\documentclass[11pt,twoside]{article}
\usepackage{atmp}
\usepackage{latexsym,amsmath,amssymb,graphicx}
\copyrightnotice{2002}{6}{643}{702}
\suppressfloats
\input xy
\xyoption{all}
 
\def\inbar{\,\vrule height1.5ex width.4pt depth0pt}
\def\bx{{\overline x}}
\def\CS {{\cal S}}
\def\IE{{\relax{\rm I\kern-.18em E}}}
\def\IL{{\relax{\rm I\kern-.18em L}}}

\def\half {{1\over 2}}

\newcommand{\CO}{{\cal O}}
\newcommand{\CU}{{\cal U}}
\newcommand{\IR}{\relax{\rm I\kern-.18em R}}
\newcommand{\IT}{\relax{\rm I\kern-.18em T}}
\newcommand{\hx}{{\widehat X}}
\newcommand{\IP}{{\relax{\rm I\kern-.18em P}}}
\newcommand{\CY}{{\cal Y}}
\newcommand{\CT}{{\cal T}}
\newcommand{\CF}{\cal{F}}
\def\IZ{\mathbb {Z}}
\def\IF{{\relax{\rm I\kern-.18em F}}}

\def\om{\overline{M}}

\def\bz{{\overline Z}}
\def\by{{\overline Y}}
\def\bcy{{\overline{\cal Y}}}
\def\inj{\hookrightarrow}

\newcommand{\ra}{\rightarrow}
\def\th{\widehat{t}}
\newcommand{\sh}{\widehat{s}}
\def\Tr{{\rm Tr}}
\def\CL {{\cal L}}
\def\ot{{1\over 2}}
\def\r{{\rangle}}
\def\ll{\langle}
\def\l{{\lambda}}
\def\ra{{\longrightarrow}}

\newdimen\tableauside\tableauside=1.0ex
\newdimen\tableaurule\tableaurule=0.4pt
\newdimen\tableaustep
\def\phantomhrule#1{\hbox{\vbox to0pt{\hrule height\tableaurule width#1\vss}}}
\def\phantomvrule#1{\vbox{\hbox to0pt{\vrule width\tableaurule height#1\hss}}}
\def\sqr{\vbox{%
  \phantomhrule\tableaustep
  \hbox{\phantomvrule\tableaustep\kern\tableaustep\phantomvrule\tableaustep}%
  \hbox{\vbox{\phantomhrule\tableauside}\kern-\tableaurule}}}
\def\squares#1{\hbox{\count0=#1\noindent\loop\sqr
  \advance\count0 by-1 \ifnum\count0>0\repeat}}
\def\tableau#1{\vcenter{\offinterlineskip
  \tableaustep=\tableauside\advance\tableaustep by-\tableaurule
  \kern\normallineskip\hbox
    {\kern\normallineskip\vbox
      {\gettableau#1 0 }%
     \kern\normallineskip\kern\tableaurule}%
  \kern\normallineskip\kern\tableaurule}}
\def\gettableau#1 {\ifnum#1=0\let\next=\null\else
  \squares{#1}\let\next=\gettableau\fi\next}

\newcommand{\CP}{\mathbf{P}}

\def\IC{{\relax\hbox{$\inbar\kern-.3em{\rm C}$}}}
\begin{document}
\title{Geometric Transitions, del Pezzo Surfaces and Open String Instantons}
\author{D.-E. Diaconescu,$^1$ B. Florea,$^2$ and A. Grassi$^3$}
\address{$^1$ Department of Physics and Astronomy,
Rutgers University,\\
Piscataway, NJ 08855-0849, USA}
\addressemail{email: duiliu@physics.rutgers.edu}
\vspace*{.2in}

\address{$^2$Mathematical Institute, University of Oxford,\\24-29 St. Giles', Oxford OX1 3LB, England}
\addressemail{email: florea@maths.ox.ac.uk}
\vspace*{.2in}
\address{$^3$ Department of Mathematics, University of
Pennsylvania,\\Philadelphia, PA 19104-6395, USA}
\addressemail{email: grassi@math.upenn.edu}
\url{hep-th/0206163}

\pagestyle{myheadings}
\markboth{GEOMETRIC TRANSITIONS, DEL PEZZO SURFACES ...}{DIACONESCU, FLOREA AND GRASSI}
\noindent
\noindent
We continue the study of a class of geometric
transitions proposed by Aganagic and Vafa which exhibit open
string instanton corrections to Chern-Simons theory. In this paper
we consider an extremal transition for a local del Pezzo model
which predicts a highly nontrivial relation between topological
open and closed string amplitudes. We show that the open string
amplitudes can be computed exactly using a combination of
enumerative techniques and Chern-Simons theory proposed by Witten
some time ago. This yields a striking conjecture relating all genus
topological amplitudes  of the local del Pezzo model
to a system of coupled Chern-Simons theories.
\newpage
\section{Introduction}
\label{intro}
In the original formulation \cite{GViii}, geometric transitions
have predicted a
remarkable relation between Chern-Simons theory on $S^3$ and
closed topological strings on the small resolution of a conifold
singularity. This correspondence has been extended to knots and links
in \cite{LMi,LMV, LMii,OV,RS} and it has been recently proven
from a linear sigma model perspective in \cite{OVii}.
A different generalization has been proposed in \cite{AViv},
where the Chern-Simons theory was corrected by open string instanton
effects. This new class of dualities yields very interesting predictions
relating topological open string amplitudes in various toric backgrounds
to certain open string expansions.
The new feature of these transitions
is a fascinating interplay of open string enumerative geometry and
Chern-Simons theory proposed by Witten in \cite{EWii}.
Open string enumerative techniques have been developed
in \cite{AAHV}-\cite{AViv},\cite{B,GJT,GZ},\cite{HV}-\cite{KL},\cite{LMi}-\cite{LM},\cite{LS}-\cite{OV},\cite{RS}-\cite{sts}.
Applying some of these results, we have successfully tested this approach
for a simple exactly soluble model in \cite{DFG}.

The question we would like to address in this paper is if one can
perform similar high precision tests of the duality in more
general toric backgrounds.
In particular, we consider the local
$dP_2$ model, which is a toric noncompact Calabi-Yau threefold
fibered over the del Pezzo surface of degree two.
This is the simplest local model containing a compact divisor
which exhibits extremal transitions. Since there is an abundance
of holomorphic curves on the del Pezzo surface, the topological closed string
amplitudes are quite complicated. So far, concrete computations
have been performed only for genus zero Gromov-Witten invariants \cite{CKYZ}.
The extremal transition in question is obtained by contracting two
$(-1,-1)$ curves on the noncompact threefold, and then smoothing out
the conifold singularities. After a somewhat technical analysis, one can
show that the resulting open string theory consists of two Chern-Simons
theories supported on two disjoint 3-spheres which are coupled by instanton
effects. Systems of this kind have been predicted by Witten in \cite{EWii}.

The main result of this paper is that the open string instanton
corrections can be summed exactly using the techniques developed
in \cite{GZ,KL,LS}. This yields a fairly simple system of
Chern-Simons theories by interpreting the instanton corrections
as Wilson loop perturbations of the Chern-Simons theories \cite{EWii}.
Then large $N$ duality predicts that the 't Hooft expansion of these
coupled Chern-Simons theories computes all topological closed string
amplitudes of the local $dP_2$ model! We show by direct computations
that this conjecture is valid up to degree four in the expansion
in terms of K\"ahler parameters.
This is very strong evidence that the conjecture is true to all
orders, but we do not have a general proof.

This paper is structured as follows. In section two we study the geometry
of the extremal transition and construct the primitive open string instantons
after deformation. Section three consists of a review of
the topological closed string theory for the local $dP_2$
model following \cite{CKYZ}.
In section four we present the main results, namely the open string
instanton expansion accompanied by Chern-Simons computations.
Here we find a precise agreement with the known genus zero Gromov-Witten
invariants, and make some higher genus predictions. Sections five
and six are devoted to open string enumerative computations
based on localization techniques as in \cite{GZ,KL,LS}.
Finally, some technical details and calculations are  presented in the
two appendixes.

{\it Acknowledgements.} During this work we have greatly benefited
from interactions with Mina Aganagic, Marcos Mari\~no and Cumrun
Vafa who were working simultaneously on a similar project
\cite{AMV}. We would like to express our special thanks to them
for sharing their ideas and insights with  us regarding the
framing dependence (section 4).

We would also like to thank Ron Donagi and Tony Pantev for
collaboration on a related project, and Bobby Acharya, Michael
Douglas, John Etnyre, Albrecht Klemm, John McGreevy and Harald
Skarke for very stimulating conversations. We owe special
thanks (and lots of tiramis\`{u}) to Corina Florea for invaluable
help with the LaTeX conversion of the original draft.
The work of D.-E. D.
has been supported by DOE grant DOE-DE-FG02-96ER40959;  A.G. is
supported in part by the  NSF Grant DMS-0074980.

\section{Geometric Transitions for Local $dP_2$ Model}

The local $dP_2$ model is a toric Calabi-Yau threefold $X$
isomorphic to the total space of the canonical bundle
$\CO(K_{dP_2})$. We have
$X=\left(\IC^5\setminus F\right)/(\IC^*)^3$ defined by the
following toric data
\begin{equation}\label{eq:toricA}
\begin{array}{rrrrrrr}
& X_0 & X_1 & X_2 & X_3 & X_4 & X_5 \cr
l_1 & -1 & 1 & -1 & 1 & 0 & 0 \cr
l_2 & -1 & 0 & 1 & -1 & 1 & 0 \cr
l_3 & -1 & 0 & 0 & 1 & -1 & 1\cr
\end{array}
\end{equation}
with disallowed locus $F=\{X_1=X_3=0\}\cup\{X_2=X_4=0\}\cup
\{X_3=X_5=0\}$.
This toric quotient can be equivalently described as a symplectic
quotient $\IC^6//U(1)^3$ with moment maps
\begin{equation}\label{eq:sympquotA}
\begin{array}{l}
|X_1|^2 + |X_3|^2 - |X_2|^2 - |X_0|^2 = \xi_1 \\
|X_2|^2 +|X_4|^2 - |X_3|^2 - |X_0|^2 = \xi_2 \\
|X_3|^2 + |X_5|^2 - |X_4|^2 - |X_0|^2 = \xi_3\\
\end{array}
\end{equation}
where $\xi_1,\xi_2, \xi_3>0$.
The toric fan of $X$ is a cone over the two dimensional polytope
represented below.

\begin{figure}[ht]
    \centering
    \scalebox{1}{\includegraphics[angle=0]{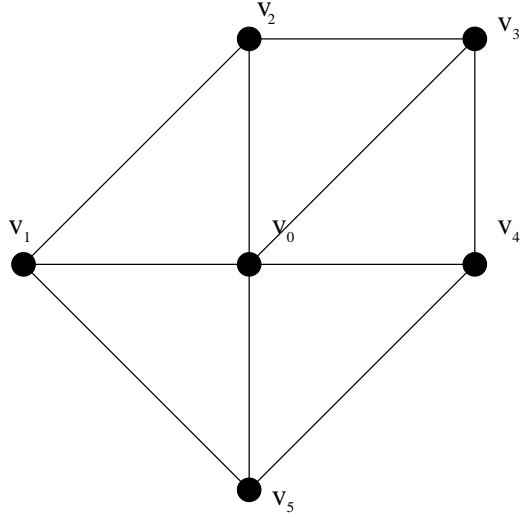}}
     \caption{A section in the toric fan of $X$. The resulting polytope
describes $\IP^2$ blown-up at two points.}\label{dptwo}

\end{figure}

There is a single compact divisor $S$ on $X$ which is the zero
section of ${  \pi:X\ra dP_2}$ defined by $X_0=0$. The Mori cone of
$X$ is generated by the curve classes $e_1, h-e_1-e_2, e_2$
corresponding the cones over $v_0v_2$, $v_0v_3$ and respectively
$v_0v_4$. One can check that these are rigid rational curves with
normal bundle $\CO_{\IP^1}(-1)\oplus \CO_{\IP^1}(-1)$. Since $X$
is a toric manifold, it can be represented as a topological
$T^2\times \IR$ fibration over $\IR^3$ \cite{MG,LV} whose
discriminant is the two dimensional planar graph represented in
fig. 1. Then the curves $e_1, h-e_1-e_2, e_2$ can be represented as
$S^1$ fibrations over certain edges of the graph as shown there.

The moment maps (\ref{eq:sympquotA}) yield the following
parameterization of the K\"ahler cone
\begin{equation}\label{eq:KclassA} J= \xi_1(h-e_1) + \xi_2 h + \xi_3 (h-e_2)
\end{equation}
where $J$ represents the restriction of the K\"ahler class to $S$.
In the following we will use alternative K\"ahler parameters
$(s_1, t, s_2)$ defined by
\begin{equation}\label{eq:KclassB}
J= -s_1 e_1 + th -s_2 e_2.
\end{equation}

We are interested in a extremal transition consisting of a
contraction of $e_1, e_2$ on $X$ followed by a smoothing of the
two nodal singularities. It turns out that the resulting singular
threefold $\hx$ can be described as a nodal hypersurface in a
toric variety $Z=\left(\IC^3\setminus\{0\}\right)\times \IC^2
/\IC^*$. The $(\IC^*)$ action is defined by
\begin{equation}\label{eq:toricB}
\begin{array}{rrrrrr}
& Z_1 & Z_2 & Z_3 & U & V \cr \IC^* & 1 & 1 & 1 & -1 & -2.\cr
\end{array}
\end{equation}
Obviously, $Z$ is isomorphic to the total space of $\CO(-1)\oplus
\CO(-2)$ over $\IP^2$. The embedding of $i:\hx\hookrightarrow Z$
is given terms of homogeneous coordinates by
\begin{equation}\label{eq:embedding}
 Z_1 =X_2X_3X_4,\quad Z_2=X_1X_2, \quad Z_3=X_4X_5,\quad U=X_0X_1X_5,
\quad V=-X_0X_3.
\end{equation}
In order to make sure this is a regular map, one has to check that
(\ref{eq:embedding}) is compatible with the toric actions
(\ref{eq:toricA}), (\ref{eq:toricB}) and that  the disallowed loci
agree. We have included the details in appendix A.1. The image of
${\hx}$ in $Z$ is given by \begin{equation}\label{eq:hyperA} UZ_1
+VZ_2Z_3=0.
\end{equation} One can easily check
that this hypersurface has exactly two nodal singularities, as
expected. The singular locus is described by
\begin{equation}\label{eq:singA}
U=0,\quad Z_1=0, \quad Z_2Z_3=0, \quad VZ_2=0, \quad VZ_3=0.
\end{equation} Since
$Z_1,Z_2,Z_3$ do not vanish simultaneously, the singular points
are $\{U=V=0, Z_1=Z_2=0\}\cup \{U=V=0, Z_1=Z_3=0\}$. Therefore we
obtain the two expected conifold singularities.

This representation of $\hx$ allows a concrete description of the
extremal transition. We can resolve the singularities by blowing
up $Z$ along the zero section $U=V=0$. The proper transform of
$\hx$ will then be isomorphic to the local $dP_2$ model $X$
discussed before. Alternatively, we can smooth out
(\ref{eq:hyperA})  by deforming the polynomial equation to
\begin{equation}\label{eq:defA}
UZ_1 +VZ_2Z_3 = \mu.
\end{equation}
We obtain a family of hypersurfaces $\CY/\Delta$ parameterized by
a complex parameter $\mu\in \Delta$, where $\Delta$ is the unit
disc. Without loss of generality we can take $\mu$ to be real and
positive, and we denote by $Y\equiv Y_\mu$ the corresponding
fiber. The transition can be very conveniently described in terms
of the $T^2$ fibration structure represented in fig. 1. The
discriminant of the torus fibration undergoes the following
sequence of transformations.
\begin{figure}[ht]
     \centering
     \scalebox{0.7}{\includegraphics[angle=0]{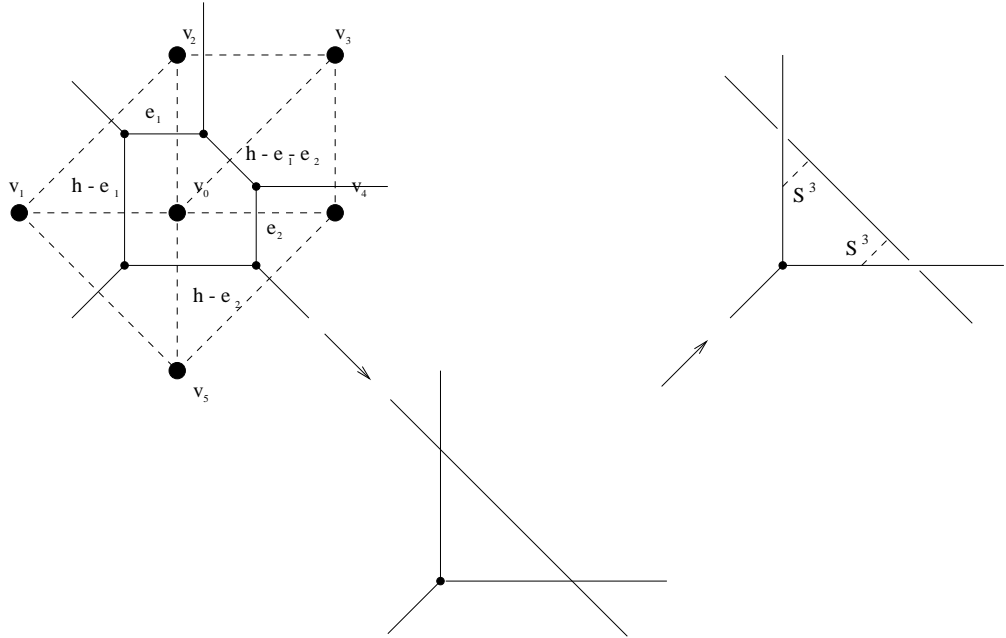}}
     \caption{Description of the extremal transition in terms of
$T^2$ fibration.}\label{transition}
\end{figure}
\noindent
As a differential manifold, $Y$ is obtained from $X$ by performing surgery
along the links of the two nodes, which are both isomorphic to
$S^2\times S^3$ \cite{HC}.
The third homology is
generated by the two vanishing cycles $L_1, L_2$ associated to the
nodal singularities, subject to the relation $[L_1]-[L_2]=0$.
Topologically, these cycles are 3-spheres which can be locally described
as fixed point sets of local antiholomorphic involutions.
We will give some details below, since this is an important point for
the rest of the paper. Let us cover $Z$ with three coordinate
patches
\begin{equation}\label{eq:locoordA}
\begin{aligned}[b]
& U_1=\{Z_1\neq 0\}:\qquad x_1={Z_2\over Z_1},\quad y_1={Z_3\over Z_1},
\quad u_1=UZ_1,\quad v_1 = VZ_1^2\\
& U_2=\{Z_2\neq 0\}:\qquad x_2={Z_1\over Z_2},\quad y_2={Z_3\over Z_2},
\quad u_2=UZ_2,\quad v_2=VZ_2^2\\
& U_3=\{Z_3\neq 0\}:\qquad x_3={Z_1\over Z_3},\quad y_3={Z_2\over Z_3},
\quad u_3=UZ_3,\quad v_3=VZ_3^2.\\
\end{aligned}
\end{equation}
In local coordinates, the hypersurface equation (\ref{eq:hyperA})
can be written
\begin{equation}\label{eq:loceqA}
\begin{aligned}[b]
&U_1:& &\qquad u_1 + v_1x_1 y_1 = \mu\cr
&U_2:& &\qquad u_2x_2 + v_2y_2 = \mu\cr
&U_3:& &\qquad u_3x_3 + v_3y_3 = \mu.\cr
\end{aligned}
\end{equation}
One can then see two conifold singularities at $\mu=0$ in the patches
$U_2, U_3$. The corresponding vanishing cycles are defined by the real sections
\begin{equation}\label{eq:vancycles}
\begin{array}{lccccc}
&U_2:& &\qquad u_2={\overline x}_2,& &\qquad v_2={\overline y}_2\\
&U_3:& &\qquad u_3={\overline x}_3,& &\qquad v_3={\overline y}_3.\\
\end{array}
\end{equation}
We show in appendix A.2. that one can choose a symplectic K\"ahler
form $\omega$ on $Y$ so that $L_1, L_2$ are lagrangian cycles.
More precisely, one can construct $\omega$ so that it is locally isomorphic
to the standard symplectic form on a deformed conifold near $L_1, L_2$.
To conclude this section, let us discuss the second homology of $Y$.
An exact sequence argument shows that
$H_2(Y,L_1\cup L_2, \IZ) \simeq H_2(Y, \IZ) =\IZ$.
One can construct certain nontrivial relative 2-cycles on $Y$
as holomorphic discs $D_1, D_2$ with boundaries on $L_1, L_2$.
Let $\Sigma_{0,1}$ be the disc
$\{|t|\leq \mu^{-1/2}\}=\{\mu^{1/2}\leq |t'|\}$
in a projective line $\IP^1$ with affine coordinates
$t, t'$. We construct a holomorphic embedding $f_1:\Sigma_{0,1}\ra Y$
given in local coordinates by
\begin{equation}\label{eq:holdiscA}
\begin{array}{lccccccccc}
&U_1:& &\qquad x_1(t)=t,& &\qquad y_1(t)= 0,& &\qquad u_1(t)=\mu,& &\qquad v_1(t) = 0\\
&U_2:& &\qquad x_2(t') = t',& &\qquad y_2(t') =0,& &\qquad u_2(t') = {\mu\over t'},& &\qquad v_2(t') =0.
\end{array}
\end{equation}
It easy to check that $f$ is well defined and it maps the boundary
$|t'|=\mu^{1/2}$ of $\Sigma_{0,1}$ to an unknot
 $\Gamma_1$ in $L_1$. We will denote the
image of $\Sigma_{0,1}$ in $Y$ by $D_1$. We can construct
similarly a disc ending on $L_2$. The embedding map is locally
given by
\begin{equation}\label{eq:holdiscB}
\begin{array}{lccccccccc}
&U_1:& &\qquad x_1(t)=0,& &\qquad y_1(t)=t,& &\qquad u_1(t)= \mu,& &\qquad v_1(t) = 0\\ 
&U_3:& &\qquad x_3(t') = t',& &\qquad y_3(t') =0,& &\qquad u_3(t') ={\mu\over t'},& &\qquad v_3(t') = 0.
\end{array}
\end{equation} To complete this discussion, note that one can also embed a holomorphic annulus $C$
in $Y$, the boundary components being mapped to $L_1, L_2$. For
this we have to use the coordinate patches $U_2, U_3$. Let
$\Sigma_{0,2}$ be the cylinder $\left\{\mu^{1/2} \leq |t|\leq
\mu^{-1/2}\right\} = \left\{\mu^{-1/2}\geq  |t'|^2 \geq
\mu^{1/2}\right\}$ in $\IP^1$ with affine coordinates $(t,t')$
(recall that $\mu$ is a positive real number inside the unit disc,
hence $\mu <1$.) We define a map $f:\Sigma_{0,2}\ra Y$ by
\begin{equation}\label{eq:holcylA}
\begin{array}{lccccccccc}
&U_2:& &\qquad x_2(t)=0,& &\qquad y_2(t)= t,& &\qquad u_2(t)=0,& &\qquad v_2(t) = {\mu \over t}\\
&U_3:& &\qquad x_3(t')=0,& &\qquad y_3(t')=t',& &\qquad u_3(t')=0,& &\qquad v_3(t')={\mu \over t'}.
\end{array}
\end{equation}
 Then the boundary component
$|t|=\mu^{1/2}$ is mapped to $L_1$, while the boundary
$|t|=\mu^{-1/2}$ is mapped to $L_2$. Note that the discs $D_1,
D_2$ and the cylinder $C$ can be in principle covered by a single
coordinate patch. We have used two coordinate patches for reasons
that will be clear in section six. Let us denote the two boundary
components of $C$ by $\Xi_1, \Xi_2$, which are again to be
regarded as knots in $L_1, L_2$. An important point for
Chern-Simons computations is that $\{\Gamma_1, \Xi_1\}$ and
respectively $\{\Gamma_2, \Xi_2\}$ are algebraic links in $L_1$,
$L_2$. This can be seen by noting that locally we can identify for
example $L_1$ to the sphere
\begin{equation}\label{eq:sphereC}
|x_2|^2+|y_2|^2 = \mu
\end{equation}
in $\IC^2$ with coordinates $(x_2, y_2)$.
Then the disc $D_1$ and
$C$ can be locally described by the equation $x_2y_2=0$ in
$\IC^2$. It is a well known fact that the intersection of this
singular curve with the sphere surrounding the origin is an
algebraic link with linking number one in the orientations induced
by the complex structure. More precisely, if we parameterize the
two boundary components as $x_2=\mu^{1/2} e^{i\theta_x}$,
$y_2=\mu^{1/2} e^{i\theta_y}$, the 1-forms $d\theta_x, d\theta_y$
define orientations of $\Gamma_1, \Xi_1$ such that the linking
number is $1$ \cite{Bd}. The same is true for $\Gamma_2, \Xi_2$ in
$L_2$. In particular this shows that $D_1, C$ and $D_2, C$ are
disconnected. Note that from now on we will fix the above
orientations for $\Gamma_1, \Xi_1, \Gamma_2, \Xi_2$. In terms of
the $T^2$ fibrations, $D_1, D_2, C$ can be represented as in fig.
3.
\begin{figure}[h]
     \centering
     \scalebox{1}{\includegraphics{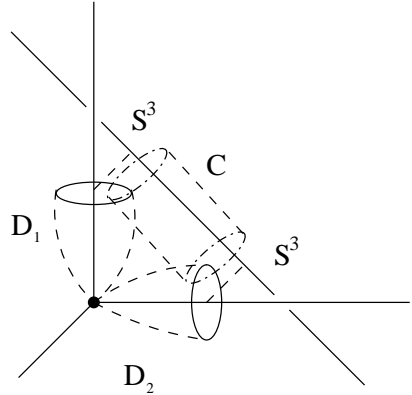}}
     \caption{Primitive open string instantons on $Y$: the linking
number of the linked boundaries is $1$.}\label{discs}
    \end{figure}
\noindent

Using this picture it is easy to see that there is a continuous family of
2-spheres interpolating between $D_1$ and $D_2$, hence the relation
$[D_1]-[D_2]=0$. Similarly, we have $[C] = [D_1]=[D_2]$, and we will
denote this relative homology class
by $\beta$.
However note that these 2-spheres are not holomorphically
embedded in $Y$. In fact we will show in section five that
there are no holomorphic curves on $Y$ and that
$\beta$ is a generator of $H_2(Y, L_1\cup L_2; \IZ)$.
Therefore $D_1, D_2, C$ constructed above are primitive open
string instantons.
Since $L_1, L_2$ are lagrangian, this
 shows that $D_1, D_2, C$ have the same symplectic area
\begin{equation}\label{eq:symparea}
t_{op}=\int_{D_1}\omega =\int_{D_2}\omega =\int_C \omega.
\end{equation}
By deforming the discs as topological spheres
away from the vanishing
cycles, one can show that classically $t=t_{op}$.
This is so because the transition leaves the symplectic form essentially
unchanged away from the singular locus.
This concludes our discussion of the extremal transitions for the
local $dP_2$ model from a geometric point of view.
The physics of the transitions
will be explored in the next sections.

\section{Closed String Amplitudes and Duality Predictions}

In the context of geometric transitions we are interested in a relation
between the topological closed string A model on $X$ and a topological
open string A model on the deformation space $Y$.
The topological open string theory on $Y$ is defined by wrapping
$N_1$ and respectively $N_2$ D-branes on the lagrangian cycles
$L_1, L_2$. The target space action of this theory consists of
two Chern-Simons theories with gauge groups $U(N_1), U(N_2)$
supported on the two cycles \cite{EWii}. We will see later
that these theories are coupled by open string instanton effects.

In order to have the right integrality properties, the topological
amplitudes must be written in terms of flat coordinates. On the
closed string side we have flat coordinates $(\sh_1,\sh_2,\th)$
corresponding to the classical coordinates $(s_1, s_2, t)$. For
simplicity, we will drop the notation $\ {\widehat{}}\ $, keeping
in mind that topological amplitudes will always be written in
terms of flat coordinates. The open string theory contains a
classical geometric parameter defined in (\ref{eq:symparea}).
Accordingly we have a flat coordinate $\th_{op}$, which will also
be denoted by $t_{op}$ from now on. As discussed above,
classically, one would predict a relation of the form
$t=\th_{op}$, but this has to be refined at quantum level, as
discussed in \cite{DFG}. Moreover, we will see later in section
four that the open string amplitudes depend in fact on three flat
coordinates corresponding to the three primitive instantons
constructed in section two.

Without going into details for the moment,
note that large $N$ duality predicts a relation between
closed and open string amplitudes of the form
\begin{equation}\label{eq:dualpredA}
{\CF}_{cl}(g_s,s_1, s_2, t)
={\CF}_{op}(g_s,\lambda_1,\lambda_2,t_{op}).
\end{equation}
Here $\lambda_1=N_1g_s$,
$\lambda_2=N_2g_s$ are the 't Hooft coupling constants of the two
Chern-Simons theories on $L_1$, $L_2$ which should be related
to the closed string parameters $(s_1,s_2)$. We will discuss
the precise relation in section four.

According to \cite{GVii}, the closed
string free energy on $X$ has the following
structure\footnote{Throughout this paper, we will consider truncated
expressions for the free energy, that is we will omit the
polynomial terms.}
\begin{equation}\label{eq:freenA}
{\CF}_{cl}(t,s_1,s_2,g_s)=\sum_{r=0}^\infty
\sum_{n=1}^\infty \sum_{{C}\in H_2(X,\IZ)} N^r_C
{1\over n}\left(2\sin{ng_s\over 2}\right)^{2r-2} {e^{-n<J,C>}}.
\end{equation}
In this expression, $N^r_C$ are the Gopakumar-Vafa invariants of $X$
which count the number of BPS states of charge $C$ and spin quantum number
$r$ in M-theory compactified on $X$; $n$ counts multicovers. In the following
we will refer to them as GV invariants.
In terms of the generators $(e_1,h-e_1-e_2,e_2)$ of the Mori cone
we have
\begin{equation}\label{eq:curveclsA}
C = d_1e_1+d(h-e_1-e_2)+d_2e_2,\qquad d,d_1,d_2\in \IZ
\end{equation}
Note that $C$ is representable by an irreducible reduced curve
only if $0\leq d_1,d_2\leq d$ or $d=0$ and $d_1=d_2=1$. Recall
that the K\"ahler class $J$ is given by (\ref{eq:KclassB}),
$J=-s_1e_1+ th-s_2e_2$. Then we can rewrite (\ref{eq:freenA}) as
\begin{eqnarray}\label{eq:freenB}
& {\CF}_{cl}(t,s_1,s_2,g_s) =
\sum_{n=1}^\infty {1\over n\left(2\sin{ng_s\over 2}\right)^2}
(e^{-ns_1}+e^{-ns_2})&\cr
&+ \sum_{r=0}^\infty \sum_{n=1}^\infty \sum_{d>0}
\sum_{0\leq d_1,d_2\leq d} N^r_{d,d_1,d_2}
{1\over n}\left(2\sin{ng_s\over 2}\right)^{2r-2}
{e^{-n(dt-(d-d_1)s_1-(d-d_2)s_2)}}.&\cr
\end{eqnarray}
Note that the first two terms represent the universal contributions of the two
exceptional curves, whose open string interpretation is well understood
\cite{GViii}. In the second series, the GV invariants are not known
for the local $dP_2$ model, except for $r=0$, when they can be computed
using mirror symmetry. This calculation has been performed in
\cite{CKYZ}.

\section{Open String Amplitudes and The Duality Map}

We now consider the open string A model defined by wrapping $N_1$
and respectively $N_2$ branes on the lagrangian cycles $L_1, L_2$
in $Y$. According to \cite{EWii}, the target space physics of this model
is captured by two Chern-Simons theories with gauge groups $U(N_1)$
and respectively $U(N_2)$ supported on the cycles $L_1, L_2$.
As explained in \cite{EWii}, the Chern-Simons theory
is in general corrected by open string instantons which give rise to Wilson
loop operators in the target space action.
The complete action can then be schematically written in the form
\begin{equation}\label{eq:CSA}
S(A_1, A_2) = S_{CS}(A_1) + S_{CS}(A_2) +
F_{inst}(g_s, t_{op}, A_1, A_2)
\end{equation}
where $A_1, A_2$ denote the two gauge fields on $L_1, L_2$.
The concrete form of the instanton expansion depends on the
details of the model. In general $A_1, A_2$ enter
$F_{inst}(g_s, t_{op}, A_1, A_2)$ via holonomy operators
associated to the boundary components of open string instantons
interpreted as knots in $L_1, L_2$.
For large volume, the instanton corrections can be treated
perturbatively from the Chern-Simons point of view. Therefore
in the large $N$ limit, the open string free energy can be written
as
\begin{equation}\label{eq:freenE}
{\CF}_{op}(t_{op}, \lambda_1, \lambda_2, g_s) =
{\CF}^{CS}_1(\lambda_1, g_s) + {\CF}^{CS}_2(\lambda_2, g_s)
+\ln \left\langle e^{F_{inst}(g_s, t_{op}, A_1, A_2)}\right\rangle
\end{equation}
where $\lambda_1=N_1g_s$ and $\lambda_2 = N_2 g_s$ denote the 't
Hooft coupling constants of the two Chern-Simons theories. In the
last term of (\ref{eq:freenE}) we have a double functional
integral over both gauge fields. Therefore the computation of the
free energy consists of two steps. First we have to find
$F_{inst}(g_s, t_{op}, A_1, A_2)$ using open string enumerative
techniques, and then compute  the Wilson line expectation values
in Chern-Simons theory. For convenience, we will denote the last
term in (\ref{eq:freenE}) by
${\CF}_{inst}(g_s,t_{op},\lambda_1,\lambda_2)$ so that
(\ref{eq:freenE}) becomes
\begin{equation}\label{eq:freenEA}
{\CF}_{op}(t_{op},\lambda_1,\lambda_2, g_s) =
{\CF}^{CS}_1(\lambda_1, g_s) + {\CF}^{CS}_2(\lambda_2, g_s)+
{\CF}_{inst}(g_s,t_{op},\lambda_1,\lambda_2).
\end{equation}

The above discussion is quite schematical since the interaction between
Chern-Simons theory and open string instantons is more subtle. According
to \cite{EWii}, the perturbative Chern-Simons expansion should be interpreted
as a sum over degenerate open string Riemann surfaces which develop
infinitely thin ribbons. The ribbons, which
 are mapped to the spheres $L_1, L_2$ as geodesic graphs, have been
interpreted in \cite{EWii} as virtual instantons at infinity.
So far there is no rigorous mathematical treatment of this type of degenerate
behavior at infinity. In particular it is not known how to actually write
down the open string amplitudes as finite dimensional integrals on a
well defined moduli space. The prescription outlined above following
\cite{EWii} is to first sum over nondegenerate instantons, i.e. Riemann surfaces
which have at worst double node singularities, and then sum over
degenerate instantons by performing the Chern-Simons path integral.

The sum over non-degenerate instantons should be in principle defined
in terms of intersection theory on some moduli space of stable open string
maps\footnote{This is a schematic discussion. Since $Y$ is noncompact,
the compactification of this moduli space is a very subtle issue.
Some aspects will be mentioned in section six.}
$\om_{g,h}(Y,L,d\beta)$, where $L=L_1\cup L_2$.
This theory has not been rigorously
developed so far, but at the level of rigor of \cite{GZ,KL,LS},
one can give a computational definition of open string amplitudes.
The main idea is to proceed by localization with respect to a torus
action induced by a torus action on $Y$ preserving $L_1, L_2$.
Although the structure of the moduli space is unknown, one can describe
in detail the structure of the fixed point loci. From this data,
we can obtain enumerative invariants essentially by adapting
on spot the known closed string techniques \cite{BF,GP,MK,LT}
in order to evaluate the contribution of
each fixed point locus.
This approach has been successfully implemented for noncompact
lagrangian cycles in \cite{GZ,KL,LS,Mii}. In that case one can fix a
flat unitary connection on the lagrangian cycles as a background field.

In the present context, since the cycles are compact, the unitary connections
become dynamical variables and one should integrate over all (gauge
equivalence classes) of such fields \cite{EWii}. This is achieved by performing
the Chern-Simons theory path integral. The coupling between finite area
instantons and Chern-Simons theory is quite subtle \cite{EWii}. As explained
there, if one had only isolated open string instantons, their effects
would be encoded in a series of holonomy (Wilson loop) operators
added to the Chern-Simons action. For each rigid isolated instanton
$D\subset Y$, one adds a term of the form $e^{-t} \Tr V$, where $V$ is
the holonomy around the boundary of $D$, which is a knot in $L$.
It is important to note that in this formula $V$ is not a flat gauge
field, therefore this operator is not invariant under deformations
of the knot. If $D$ is rigid and isolated, this is not a problem.
However, what happens if we have families of such instantons?
Then the holonomy $V$ depends on the particular member in the family,
and it is not clear how one should write the associated corrections.
A general answer to this question is not known at the present stage,
but we would like to propose an answer for situations in which
there is a torus action on $Y$ preserving the lagrangian cycles.
In such cases, one can simply use localization arguments to argue that
all nontrivial contributions to the instanton sum come from fixed
maps under this action.
Then, to each component of the fixed point locus
we associate a certain series of holonomy operators in Chern-Simons
theory, as detailed below. Because of this coupling with Chern-Simons
theory, the procedure described above should not be thought as
localization of a virtual cycle on a moduli space of maps in the usual
sense. It would certainly be desirable to have a more precise mathematical
formulation of this construction, but this is not known at the present stage.

The final step is to perform the Chern-Simons
functional integral with the instanton corrections included in order to
compute the open string free energy.
This approach has been successfully tested in a simple geometric situation in
\cite{DFG}. Note that this last step requires the choice of a framing
for each boundary $\partial D$ in order to regulate the divergences in
Chern-Simons perturbation theory. We will comment more in this point below.

In the present situation, we define an $S^1$ action on $Z$ by
\begin{equation}\label{eq:toractA}
\begin{array}{ccccccc}
& Z_1 & Z_2 & Z_3 & U & V \cr
\IC^* & \lambda_1 & \lambda_2 & 0  & -\lambda_1 & -\lambda_2.\cr
\end{array}
\end{equation}
Obviously, this action preserves $Y$ and $L_1, L_2$.
Then, a somewhat technical analysis shows
that the only primitive open string instantons left
invariant by this action are the discs $D_1, D_2$ and
the annulus $C$ constructed in the previous section.
The proof follows the lines of \cite{DFG} (section 5); one has to take a
projective completion $\by$ of $Y$ and show that the problem reduces
to finding invariant curves on $\by$ subject to certain homology
constraints. We leave the details for the next section.

Note that the two discs $D_1, D_2$ have a common origin, therefore
they form a nodal (or pinched) cylinder $D_1\cup D_2$. As discussed
at the end of section two, $D_1\cup D_2$ and $C$ are disconnected.
Therefore the fixed locus of the torus action on $\om_{g,h}(Y,L, d\beta)$
consists of two disconnected components: multicovers of $D_1\cup D_2$
and respectively multicovers of $C$. The precise structure of these
components will be discussed in detail in section five.
For now, let us note that on general grounds, an open string map to the
pinched cylinder $D_1\cup D_2$
is characterized by two degrees $d_1, d_2$ and two sets of
 winding numbers
$m_i$, $i=1, \ldots, h_1$, $n_j$, $j=1,\ldots, h_2$, where $h_1, h_2$ are
the numbers of boundary components mapped to $\Gamma_1$ and respectively
$\Gamma_2$. We have the constraints $\sum_{i=1}^{h_1} = d_1$,
$\sum_{j=1}^{h_2} n_j = d_2$. Similarly, a generic map to the cylinder $C$
is characterized by a degree $d$ and two sets of winding numbers
$m_i, n_j$, $i=1,\ldots, h_1$, $j=1,\ldots, h_2$ with $\sum_{i=1}^{h_1} m_i
= \sum_{j=1}^{h_2} n_j = d$, where $h_1, h_2$ are the numbers of boundary
components mapped to $\Xi_1, \Xi_2$.

Based on these elements, we can write down the general form of the
open string instanton expansion on $Y$. We noticed earlier that
$D_1, D_2$ and $C$ have the same symplectic area, therefore one
would be tempted to write down this expansion in terms of a single
open string K\"ahler modulus $t_{op}$. However, things are more
subtle here since the instanton expansion should be written in
terms of flat coordinates rather than classical moduli. One of the
lessons of \cite{DFG} was that in the presence of dynamical A
branes, the open string flat coordinates can receive
nonperturbative corrections generated by virtual instantons at
infinity. These corrections can be different for different open
string instantons, depending on the lagrangian cycle they end on.
For this reason, we will refine (\ref{eq:freenE}) by writing the
instanton expansion in terms of three distinct flat K\"ahler
moduli $t_1, t_2, t_c$ corresponding to $D_1, D_2, C$. We will
show later that this is in precise agreement with the duality
predictions from the closed string side. We also introduce the
following holonomy variables
\begin{equation}\label{eq:holvarA}\begin{array}{lcccc}
&U_1=\hbox{Pexp}\int_{\Xi_1} A^{(1)},&
&U_2=\hbox{Pexp}\int_{\Xi_2} A^{(2)},&\\
&V_1=\hbox{Pexp}\int_{\Gamma_1} A^{(1)},&
&V_2=\hbox{Pexp}\int_{\Gamma_2} A^{(2)}&
\end{array}\end{equation}
corresponding to the boundary components of $D_1, D_2$ and $C$.
Then the open string expansion takes the form
\begin{equation}\label{eq:instcorrA}
F_{inst}(g_s,t_1, t_2,t_c, U_1,U_2, V_1, V_2) =
F^{(1)}_{inst}(g_s,t_c,U_1,U_2) + F^{(2)}_{inst}(g_s, t_1, t_2, V_1, V_2)
\end{equation}
where\begin{equation}\label{eq:instcorrAB}\begin{aligned}
F^{(1)}_{inst}(g_s,t_c, U_1, U_2)&=&
\sum_{g=0}^\infty \sum_{h_1,h_2=0}^\infty \sum_{d=0}^\infty
\sum_{m_i\geq 0, n_j\geq 0}
i^{h_1+h_2}g_s^{2g-2+h_1+h_2} \cr
&\times& C_{g,h_1, h_2}(d;m_i,n_j)
e^{-dt_c} \prod_{i=1}^{h_1} \Tr U_1^{m_i}
\prod_{j=1}^{h_2} \Tr U_2^{n_j}\cr
\end{aligned}
\end{equation}
\begin{equation}\label{eq:instcorrAC}
\begin{aligned}
F^{(2)}_{inst}(g_s, t_1, t_2, V_1, V_2)&=&
&\sum_{g=0}^\infty \sum_{h_1,h_2=0}^\infty \sum_{d_1,d_2=0}^\infty
\sum_{m_i\geq 0, n_j\geq 0}i^{h_1+h_2}g_s^{2g-2+h_1+h_2}&\\
&\times& &F_{g,h_1, h_2}(d_1, d_2; m_i, n_j)
e^{-d_1t_1-d_2t_2}\prod_{i=1}^{h_1} \Tr V_1^{m_i}\prod_{j=1}^{h_2} \Tr V_2^{n_j}.&\\
\end{aligned}
\end{equation}
In (\ref{eq:instcorrAB}) we have a sum over multicovers of the
annulus $C$, while (\ref{eq:instcorrAC}) represents a sum over
multicovers of the two discs $D_1, D_2$, which form a nodal
cylinder. Note that the winding numbers are subject to the
constraints mentioned above, that is $\sum_{i=1}^{h_1} m_i =
\sum_{j=1}^{h_2}n_j = d$ for $C$, and $\sum_{i=1}^{h_1} m_i =d_1,\
\sum_{j=1}^{h_2}n_j = d_2$ for $D_1\cup D_2$.

As discussed earlier in this section, the coefficients
$C_{g,h_1, h_2}(d;m_i,n_j)$ as well as $F_{g,h_1, h_2}(d_1,
d_2; m_i, n_j)$ can be computed by evaluating
the contribution of the fixed loci in $\om_{g,h}(Y,L,d\beta)$.
Note however, that to each component of the fixed locus we assign a
certain holonomy operator in the Chern-Simons theory. Therefore, one
does not simply sum over all fixed loci as in standard localization
computations. This means that the contribution of each fixed component
depends on the weights of the toric action used in the localization
process. In order to obtain a physically sensible answer, we
have to make a certain choice of weights similar to the choices made
in \cite{GZ,KL,Mii}. Moreover, in our case the situation
is more complicated since we also have to make a choice of framing in
Chern-Simons theory. The two choices are in fact related, as discussed
below and in more detail in section 6.3.

Before giving the details, note that given the geometric context, there may
be many choices of weights and/or framings that result in distinct open
string expansions. At this stage we do not know if there is a
preferred choice based on certain intrinsic consistency criteria of the
open string theory. This problem is very hard, and it cannot be answered
without a better development of the mathematical formalism.
In the following we will pursue a more modest goal, namely we will try
to find a set of choices which leads to an agreement with the dual closed
string expansion. Formulated differently, we will try to find the correct
prescriptions for the duality map in this geometric situation.

For a single disc $D\subset \IC^3$ with boundary on a noncompact lagrangian
cycle $L$, it was shown in \cite{KL} that the choice of weights is equivalent to
the choice of an equivariant section of the normal bundle $N_{\partial D/L}$.
This prescription formalizes the relation between framing and toric action
found for the first time in \cite{AKV}.
In particular, the instanton expansion for $D$ depends on an integer ambiguity
$a$ which parameterizes isomorphism classes of $S^1$ equivariant sections
of the normal bundle $N_{\partial D/L}$.
For the discs $D_1, D_2$ embedded in $Y$, one can still choose the
boundary data in the form of two sections to $N_{\Gamma_1/L_1},
N_{\Gamma_2/L_2}$ which are labeled by two integers $a,b$.
Generalizing the strategy proposed by \cite{KL}, we assume that these
choices have to be compatible with the global $S^1$ action on $Y$.
By explicitly writing these conditions in local coordinates, we
show in section 6.3. that we are left with only two consistent
choices, namely $(a,b) =(0,0)$ or $(a,b)=(2,2)$. In the following we will
choose $(a,b)=(0,0)$ since in this case the instanton expansion takes a
very simple compact form. The second choice is not logically ruled out,
but leads to a very complicated formula for the instanton corrections.
We leave it for future work.

Having made this choice, the open string topological theory is still not
completely determined, since we also have to specify the framing of
the knots $\Gamma_1,\Gamma_2,\Xi_1$ and $\Xi_2$.
In principle, the equivariant sections introduced in the previous
paragraph should determine the framings of $\Gamma_1=\partial D_1$
and $\Gamma_2=\partial D_2$. However, there is a subtlety at this point
explained in detail in section 6.3. Briefly, the choice of a single section
does not determine the framing as an integer number; one also needs
a reference section which is typically provided by the geometric context.
In our case we have a natural reference section since
$\Gamma_1, \Gamma_2$ are algebraic knots. Then a short local
computation shows that the framings
of $\Gamma_1, \Gamma_2$ are $(2-a,2-b)$. Therefore
for $(a,b)=(0,0)$ we obtain framings $(2,2)$.

For the annulus $C$, the choice of framing is more subtle since
the localization computation does not require the choice of special values
of toric weights. Therefore one does not have to choose equivariant sections
on the two boundaries components $\Xi_1, \Xi_2$.
In the present context, this framing can be related to the framings
of the discs by a deformation argument detailed in section six.
The resulting values are $(1-{a\over 2}, 1-{b\over 2})=(1,1)$ for the two
boundaries of $C$.
Moreover, it is shown in \cite{AMV} that for an annulus with framings
$(p,p)$, the $p$-dependence of the amplitudes
can be absorbed in a simple shift of the open string K\"ahler parameters,
leaving the amplitudes otherwise unchanged. This allows us to choose
canonical framing without loss of generality. We are very grateful to
the authors of \cite{AMV} for explaining this to us.

The open string instanton expansions
(\ref{eq:instcorrAB}),(\ref{eq:instcorrAC}) can be determined by
adding the contributions of all fixed points of the $S^1$ action.
These are computed in section six, equations (6.30), (6.100),
(6.101) and (6.102). There is one subtle aspect at this point,
namely given the choices made so far, one has to count the
contribution of the pinched cylinder, eqn (6.102), twice in order
to match the predictions of large $N$ duality. This factor of two
does not follow directly from localization computations, and it
cannot be satisfactorily explained using our present knowledge of
moduli spaces of open string maps. In fact, since the only
criterion for introducing this factor is agreement with the closed
string dual, we should think of it as a prescription of the
duality map. A more conceptual explanation would require a much
deeper mathematical understanding of the open/closed string
duality, which is beyond the purpose of this work. We hope to
report on this aspect in the future.

To summarize this discussion, we propose the following large $N$ Chern-Simons
dual to the local $dP_2$ model
\begin{equation}\label{eq:instcorrBA}
F^{(1)}_{inst}(g_s,t_c,U_1, U_2) = -\sum_{d=1}^\infty
{e^{-dt_c}\over d} \Tr U_1^d \Tr U_2^d
\end{equation}
\begin{equation}\label{eq:instcorrBB}
\begin{aligned}
F^{(2)}_{inst}(g_s,t_1,t_2,V_1,V_2)& = \sum_{d=1}^{\infty}
{ie^{-dt_1}\over 2d\sin{dg_s\over 2}}\Tr V_1^{d}+
\sum_{d=1}^{\infty} {ie^{-dt_2}\over 2d\sin{dg_s\over 2}}\Tr V_2^d&
\cr &+{ 2}\sum_{d=1}^\infty {e^{-d(t_1+t_2)}\over d} \Tr V_1^d
\Tr V_2^d &\cr
\end{aligned}
\end{equation}
where the framings of the knots $\Gamma_1,\Gamma_2,\Xi_1,\Xi_2$
are $(2,2,0,0)$.
In the next subsection, we will present very convincing evidence for this
conjecture by computing the open string free energy up to degree four
in $e^{-t_1}, e^{-t_2}, e^{-t_c}$ and finding perfect agreement
with the closed string results.

\subsection{Chern-Simons Computations}

As discussed above, we have to evaluate
\begin{equation}\label{eq:freenEB}
{\CF}_{op}(t_1,t_2,t_c,\lambda_1,\lambda_2) =
{\CF}^{CS}_1(\lambda_1, g_s) + {\CF}^{CS}_2(\lambda_2, g_s)+
{\CF}_{inst}(g_s,t_1,t_2,t_c,\lambda_1,\lambda_2)
\end{equation}
 where
\begin{equation}\label{eq:freenEC}
{\CF}_{inst}(g_s,t_1,t_2,t_c,\lambda_1,\lambda_2) =\ln
\left\langle e^{F_{inst}(g_s,t_1,t_2,t_c,U_1,U_2,V_1,V_2)}\right\rangle.
\end{equation}
Moreover, the instanton expansion is obtained by adding
(\ref{eq:instcorrBA}) and (\ref{eq:instcorrBB})
\begin{equation}\label{eq:instcorrBC}
F_{inst}(g_s,t_1,t_2,t_c,U_1,U_2,V_1,V_2) =
F^{(1)}_{inst}(g_s,t_c,U_1,U_2) +
F^{(2)}_{inst}(g_s,t_1,t_2,V_1,V_2).\end{equation} The holonomy
variables $(U_1,V_1)$,$(U_2, V_2)$ are associated to the unknots
$(\Xi_1,\Gamma_1)$, $(\Xi_2,\Gamma_2)$ in $L_1$ and respectively
$L_2$. Each pair of knots form a link with linking number one in
the present choice of orientations. Moreover, $(\Xi_1, \Xi_2)$ are
canonically framed while $(\Gamma_1, \Gamma_2)$ have framings
$(2,2)$. This completely specifies the Chern-Simons system. The
first two terms in (\ref{eq:freenEB}) are well understood.
Performing an analytic continuation as in
\cite{MRD,GVii,GViii,VP}, we can write them in the form
\begin{equation}\label{eq:CSfreen} {\CF}^{CS}_1(\lambda_1, g_s) +
{\CF}^{CS}_2(\lambda_2, g_s) = \sum_{n=1}^\infty {1\over
n{\left(2\sin{ng_s\over 2}\right)}^2} (e^{in\l_1}+
e^{in\l_2}).\end{equation} The third term is more complicated. We
will evaluate it perturbatively up to terms of order three in
$e^{-t_1}, e^{-t_2}, e^{-t_c}$.

For a systematic approach, let us write the instanton corrections
in the form
\begin{equation}\label{eq:instcorrBD}
\begin{aligned}
F_{inst}(g_s,t_1,t_2,t_c,& U_1,U_2,V_1,V_2)\cr
& =\sum_{n=1}^\infty \left[ia_ne^{-nt_1} + ib_ne^{-nt_2}
- c_ne^{-nt_c} +2d_ne^{-n(t_1+t_2)}\right]\cr
\end{aligned}\end{equation}where
\begin{equation}\label{eq:notationB}
\begin{array}{lcccc}
& a_n = {1\over 2n\sin{ng_s\over 2}}\Tr V_1^n,&\qquad&
b_n = {1\over 2n\sin{ng_s\over 2}}\Tr V_2^n&\cr
& c_n = {1\over n} \Tr U_1^n\Tr U_2^n,&\qquad&
d_n = {1\over n} \Tr V_1^n \Tr V_2^n.&\cr
\end{array}\end{equation}
Then the first order terms are
\begin{equation}\label{eq:degoneA}
{\CF}_{inst}(g_s,t_1,t_2,t_c,\lambda_1,\lambda_2)^{(1)}
=e^{-t_1}X_{(t_1)}+ e^{-t_2}X_{(t_2)}+e^{-t_c}X_{(t_c)},\end{equation}
where we have introduced the notation
\begin{equation}\label{eq:degonebbb}
X_{(t_1)}=x_{(t_1)}=i\ll a_1\r ,~~~X_{(t_2)}=x_{(t_2)}=i\ll b_1\r
,~~~X_{(t_c)}=x_{(t_c)}=-\ll c_1\r .
\end{equation}
The expectation values in (\ref{eq:degonebbb}) can be evaluated
using the Chern-Simons techniques developed in
\cite{LMi,LMV,LMii,MV,RS}. Some extra care is needed since we have
to work with $U(N)$ Chern-Simons theory, and not $SU(N)$
\cite{MV}. Let us consider a link $\CL$ with $c$ components with
representations $R_\alpha$ and framings $p_\alpha$,
$\alpha=1,\dots, c$. The framing dependence of the expectation
value $\langle W_{R_\alpha}(\CL)\rangle$ is of the form
\begin{equation}\label{eq:framdepA}
\langle W_{R_\alpha}(\CL)\rangle_{(p_1, \ldots, p_c)} =
e^{{ig_s\over 2}\sum_{\alpha=1}^c \kappa_{R_\alpha} p_\alpha}
e^{{i\lambda\over 2} \sum_{\alpha=1}^c l_\alpha p_\alpha}
\langle W_{R_\alpha}(\CL)\rangle_{(0, \ldots, 0)}
\end{equation}
where $l_\alpha$ is the total number of boxes in the Young tableau
of $R_\alpha$, and $\kappa_{R_\alpha}$ is a group theoretic quantity
defined as follows. Let $v=1,\ldots r$ label the rows of the Young
tableau of a representation $R$, and $l_v$ denote the length of the
$v$-th row. Then we have \cite{MV}
\begin{equation}\label{eq:framdepB}
\kappa_R = l + \sum_{v=1}^r (l_v^2 -2vl_v)\end{equation} where
$l=\sum_{v=1}^r l_v$ is the total number of boxes. Applying this
formula, and taking into account the framings specified below
(\ref{eq:instcorrBC}), we have
\begin{equation}\label{eq:expvalA}
\begin{array}{lc}
& \langle a_1\rangle = (-i) e^{i\lambda_1}{e^{{1\over 2}i\lambda_1}-
e^{-{1\over 2}i\lambda_1}
\over \left(2\sin{g_s\over 2}\right)^2}\cr
& \langle b_1\rangle
= (-i) e^{i\lambda_2}{e^{{1\over 2}i\lambda_2}-e^{-{1\over 2}i\lambda_2}
\over \left(2\sin{g_s\over 2}\right)^2}\cr
& \langle c_1\rangle
= -{(e^{{1\over 2}i\lambda_1}-e^{-{1\over 2}i\lambda_1})(e^{{1\over
2}i\lambda_2}-e^{-{1\over 2}i\lambda_2})\over
\left(2\sin{g_s\over 2}\right)^2}.\cr
\end{array}\end{equation}
By direct substitution in (\ref{eq:instcorrBD}), we find
\begin{equation}\label{eq:degoneB}
\begin{array}{lc}
&{\CF}_{inst}(g_s,t_1,t_2,t_c,\lambda_1,\lambda_2)^{(1)}\cr
& ={1\over \left(2\sin{g_s\over 2}\right)^2}
\bigg[
e^{-t_1}(-e^{{1\over 2}i\lambda_1}+e^{{3\over 2}i\lambda_1})
+e^{-t_2}(-e^{{1\over 2}i\lambda_2}+e^{{3\over 2}i\lambda_2})
\cr
&\qquad\qquad\qquad
 + e^{-t_c}(e^{{1\over 2}i(\lambda_1+\lambda_2)}
-e^{{1\over 2}i(-\lambda_1+\lambda_2)}-e^{{1\over 2}i(\lambda_1-\lambda_2)}
+e^{-{1\over 2}i(\lambda_1+\lambda_2)})\bigg].\cr
\end{array}\end{equation}
Let us consider the second order terms in (\ref{eq:freenEC}). By
successively expanding the exponential and the logarithm, we
obtain
\begin{equation}\label{eq:degtwoA}
\begin{array}{lc}
&{\CF}_{inst}(g_s,t_1,t_2,t_c,\lambda_1,\lambda_2)^{(2)} =
e^{-2t_1}X_{(2t_1)}+e^{-2t_2}X_{(2t_2)}+e^{-2t_c}X_{(2t_c)}\cr
&\qquad\qquad\qquad\qquad\qquad\quad~~ 
+e^{-t_1-t_2}X_{(t_1,t_2)}+e^{-t_1-t_c}X_{(t_1,t_c)}+e^{-t_2-t_c}X_{(t_2,t_c)},\end{array}\end{equation}
where
\begin{equation}\label{eq:degtwobbb}
\begin{array}{lcccc}
&X_{(2t_i)}=x_{(2t_i)}-\ot x_{(t_i)}^2,& &i=1,2,c,&\cr
&X_{(t_i,t_j)}=x_{(t_i,t_j)}-x_{(t_i)}x_{(t_j)},& &i,j=1,2,c,~i\neq j,&
\end{array}
\end{equation}
with
\begin{equation}\label{eq:degtwoccc}
\begin{array}{lcccccc}
&x_{(2t_1)}=i\ll a_2\r -\ot\ll a_1^2\r ,&
&x_{(2t_2)}=i\ll b_2\r -\ot\ll b_1^2\r ,&
&x_{(2t_c)}=-\ll c_2\r +\ot\ll b_1^2\r ,&\cr
&x_{(t_1,t_2)}=-\ll a_1b_1\r +2\ll d_1\r ,& &x_{(t_1,t_c)}=-i\ll
a_1c_1\r ,& &x_{(t_2,t_c)}=-i\ll
b_1c_1\r. &
\end{array}\end{equation}
In order to compute the relevant expectation values, we have to
use the Frobenius formula in order to linearize quadratic
expressions in the holonomy variables. Since the formula
(\ref{eq:degtwoA}) is symmetric under the exchange of $\Gamma_1$
and $\Gamma_2$, it suffices to consider only $V_1$
\begin{equation}\label{eq:frobA}
\begin{array}{lc}
& (\Tr V_1)^2 = \Tr_{\tableau{2}} V_1 + \Tr_{\tableau{1 1}} V_1 \cr
& \Tr V_1^2 = \Tr_{\tableau{2}} V_1 - \Tr_{\tableau{1 1}} V_1\cr
\end{array}\end{equation}
where $\tableau{2},~\tableau{1 1}$ are the Young tableaux
corresponding to the symmetric and antisymmetric representations of
$U(N_1)$, respectively. Applying (\ref{eq:framdepA}) we have
\begin{equation}\label{eq:expvalB}
\begin{aligned}
\langle (\Tr V_1)^2\rangle = e^{2ig_s}e^{2i\l_1}
\langle \Tr_{\tableau{2}} V_1\rangle_0 + e^{-2ig_s} e^{2i\l_1}
\langle \Tr_{\tableau{1 1}} V_1\rangle_0\cr
\langle \Tr V_1^2 \rangle = e^{2ig_s}e^{2i\l_1}
\langle \Tr_{\tableau{2}} V_1\rangle_0 - e^{-2ig_s} e^{2i\l_1}
\langle \Tr_{\tableau{1 1}} V_1\rangle_0 \cr
\end{aligned}\end{equation}
where the subscript zero means canonical framing. Now, the expectation
value of $\Tr_R V_1$ for the unknot with the
canonical framing is given by the quantum dimension of $R$ with
quantum parameter $e^{ig_s}$ \cite{MV}. Therefore we have
\begin{equation}\label{eq:expvalBC}
\begin{array}{lc}
& \langle \Tr V_1\rangle_0 = {e^{\ot i\l_1}-e^{-\ot i\l_1} \over
e^{\ot ig_i\l}-e^{-\ot ig_i\l}}\cr & \langle \Tr_{\tableau{2}}
V_1\rangle_0 = {(e^{\ot i\l_1}-e^{-\ot i\l_1}) (e^{\ot
i\l_1}e^{\ot ig_s}-e^{-\ot i\l_1}e^{-\ot ig_s}) \over (e^{\ot
ig_s}-e^{-\ot ig_s}) (e^{ig_s}-e^{-ig_s})}\cr & \langle
\Tr_{\tableau{1 1}} V_1\rangle_0 = {(e^{\ot i\l_1}-e^{-\ot i\l_1})
(e^{\ot i\l_1}e^{-\ot ig_s}-e^{-\ot i\l_1}e^{\ot ig_s}) \over
(e^{\ot ig_s}-e^{-\ot ig_s})
\left(e^{ig_s}-e^{-ig_s}\right)}.\cr\end{array}\end{equation}
Taking also into account (\ref{eq:notationB}), after some
elementary computations, we arrive at

\begin{equation}\label{eq:degtwoC}
\begin{array}{lc}
&X_{(2t_1)}={1\over  \left(2\sin{g_s\over 2}\right)^2}
(e^{2i\l_1}-e^{3i\l_1}) +{1\over 2\left(2\sin{g_s}\right)^2}
(-e^{i\l_1}+e^{3i\l_1}).\cr\end{array}\end{equation} The result
for $X_{(2t_2)}$ can be obtained by substituting $t_1\ra t_2$ and
$\l_1\ra \l_2$ in (\ref{eq:degtwoC})
\begin{equation}\label{eq:degtwoD}
\begin{array}{lc}
&X_{(2t_2)}={1\over  \left(2\sin{g_s\over 2}\right)^2}
(e^{2i\l_2}-e^{3i\l_2})
+{1\over 2\left(2\sin{g_s}\right)^2}
(-e^{i\l_2}+e^{3i\l_2}).\cr\end{array}\end{equation}
The term $X_{(3t_c)}$ can be evaluated analogously by linearizing the quadratic
expressions in $U_1, U_2$. The computation is straightforward since
the unknots $\Xi_1, \Xi_2$ are canonically framed. We obtain
\begin{equation}\label{eq:degtwoE}
\begin{array}{lc}
& X_{(2t_c)}={1\over 2
\left(2\sin{g_s}\right)^2}(e^{i\l_1}-e^{-i\l_1})
(e^{i\l_2}-e^{-i\l_2}).\cr\end{array}\end{equation}
We now compute $X_{(t_1,t_c)}$ and $X_{(t_2,t_c)}$. These terms are
more interesting since they involve expectation values of
linked Wilson loops. We can exploit again the $\IZ/2$ symmetry which exchanges
the links $(\Gamma_1, \Xi_1)$ and $(\Gamma_2, \Xi_2)$. This means it
suffices to consider
\begin{equation}\label{eq:degtwoG}
X_{(t_1,t_c)}=
{i\over 2\sin{g_s\over 2}}\bigg[
-\langle \Tr V_1\Tr U_1\rangle
+\langle \Tr V_1\rangle \langle \Tr U_1\rangle\bigg]
\langle \Tr U_2 \rangle .
\end{equation}
As discussed at length in \cite{LMV,LMii}, in order for this
expression to have the correct integrality properties, the
expectation value of the link $\langle \Tr V_1\Tr U_1\rangle$
has to be taken with a particular normalization. For $U(N_1)$
Chern-Simons theory, this amounts to writing the expectation value
of a Hopf link $\CL$ with linking number $-1$ as \cite{LMV,LMii}
\begin{equation}\label{eq:expvalC}
\langle W(\CL)\rangle_{(0,0)}=
\left({e^{\ot i\l_1}-e^{-\ot i\l_1}\over e^{\ot ig_s}-e^{-\ot ig_s}}\right)^2
-(1-e^{-i\l_1})
\end{equation}
where the subscript $(0,0)$ means that both components are in
canonical framing. Up to normalization, this is the HOMFLY
polynomial of the Hopf link. In our case, we have a link with
linking number $+1$, which is in fact the mirror link $\CL^*$ of
$\CL$. The expectation value of the mirror link $\CL^*$ is related
to that of $\CL$ by sending $g_s\ra -g_s, \lambda \ra -\lambda$
\cite{MV}. Therefore, taking into account the framing correction,
the expression we need in (\ref{eq:degtwoG}) is
\begin{equation}\label{eq:expvalD}
\langle \Tr V_1\Tr U_1\rangle =
e^{i\l_1}
\left[\left({e^{\ot i\l_1}-e^{-\ot i\l_1}\over e^{\ot ig_s}-e^{-\ot ig_s}}\right)^2
-(1-e^{i\l_1})\right].\end{equation}
Then, a direct computation yields
\begin{equation}\label{eq:degtwoH}
X_{(t_1,t_c)}=
{1\over \left(2\sin{g_s\over 2}\right)^2}
(e^{i\l_1}-e^{2i\l_1})(e^{\ot i\l_2}-e^{-\ot i\l_2}).\end{equation}
Next, $X_{(t_2,t_c)}$ can be obtained using the exchange symmetry
as before
\begin{equation}\label{eq:degtwoI}
X_{(t_2,t_c)}=
{1\over \left(2\sin{g_s\over 2}\right)^2}
(e^{i\l_2}-e^{2i\l_2})(e^{\ot i\l_1}-e^{-\ot i\l_1}).\end{equation}
Finally, we have to evaluate $X_{(t_1,t_2)}$. Given what has been said
so far, and using the observation that
\begin{equation}\label{eq:degtwoF}
\begin{array}{lc}
& -\langle a_1b_1\rangle  +\langle a_1\rangle \langle b_1\rangle
={1\over 2\sin{g_s\over 2}}
\bigg[
\langle \Tr V_1\Tr V_2\rangle -
\langle \Tr V_1\rangle \langle \Tr V_2\rangle\bigg]
=0,\cr\end{array}\end{equation}
a straightforward computation leads to
\begin{equation}\label{eq:degtwoJ}
X_{(t_1,t_2)}= -{2 \over \left(2\sin{g_s\over
2}\right)^2}e^{i(\l_1+\l_2)} (e^{\ot i\l_1}-e^{-\ot i\l_1})(e^{\ot
i\l_2}-e^{-\ot i\l_2}).\end{equation} This concludes the
evaluation of (\ref{eq:degtwoA}). The last step is to collect all
the intermediate results
(\ref{eq:CSfreen}),(\ref{eq:degoneB}),(\ref{eq:degtwoC})-(\ref{eq:degtwoE}) and
(\ref{eq:degtwoH})-(\ref{eq:degtwoJ}) in a single formula.
Writing only terms up to order two, we have
\begin{equation}\label{eq:freenED}\begin{array}{lc}
& {\CF}_{op}(t_1,t_2,t_c,\lambda_1,\lambda_2)  \cr
& ={1\over \left(2\sin{g_s\over 2}\right)^2}
\bigg[ e^{i\l_1} + e^{i\l_2} +
e^{-t_1}(-e^{\ot i\lambda_1}+e^{{3\over 2}i\lambda_1})
+ e^{-t_2}(-e^{\ot i\lambda_2}+e^{{3\over 2}i\lambda_2})\cr
&
 + e^{-t_c}(e^{\ot i(\lambda_1+\lambda_2)}
-e^{\ot i(-\lambda_1+\lambda_2)}-e^{\ot i(\lambda_1-\lambda_2)}
+e^{-\ot i(\lambda_1+\lambda_2)})
+e^{-2t_1}(e^{2i\l_1}-e^{3i\l_1})\cr
&
+e^{-2t_2}(e^{2i\l_2}-e^{3i\l_2})
-2e^{-t_1-t_2}e^{i(\l_1+\l_2)}(e^{\ot i\l_1}-e^{-\ot i\l_1})(e^{\ot i\l_2}-e^{-\ot i\l_2})\cr
&
+e^{-t_1-t_c}(e^{i\l_1}-e^{2i\l_1})(e^{\ot i\l_2}-e^{-\ot i\l_2})
+e^{-t_2-t_c}(e^{i\l_2}-e^{2i\l_2})(e^{\ot i\l_1}-e^{-\ot i\l_1})
\bigg]\cr
&+{1\over 2\left(2\sin{g_s}\right)^2}\bigg[
e^{2i\l_1}+e^{2i\l_2} +
e^{-2t_1}(-e^{i\lambda_1}+e^{3i\lambda_1})
+ e^{-2t_2}(-e^{i\lambda_2}+e^{3i\lambda_2}) \cr
&
+ {e^{-2t_c}}(e^{{i}(\lambda_1+\lambda_2)}
-e^{{i}(-\lambda_1+\lambda_2)}-e^{{i}(\lambda_1-\lambda_2)}
+e^{-{i}(\lambda_1+\lambda_2)})\bigg].\cr\end{array}\end{equation}
We have also computed the terms of degree three and four in Appendix B,
but we will not reproduce them here.

\subsection{Comparison with Duality Predictions}

In order to test large $N$ duality, the free energy
(\ref{eq:freenED}) must be compared with its closed string
counterpart. For convenience we write below the $r=0$ (see
(\ref{eq:freenA})) closed string result of \cite{CKYZ}
\begin{equation}\label{eq:freenCA}
\begin{aligned}
{\CF}_{cl}^{(0)}(q,q_1,q_2,g_s)&=
{1\over \left(2\sin{g_s\over 2}\right)^2}
\bigg[q_1+q_2 +q(q_1^{-1}q_2^{-1}-2q_1^{-1}-2q_2^{-1}+3)
+q^2(-4q_1^{-1}q_2^{-1}\cr
&\qquad
+5q_1^{-1}+5q_2^{-1}-6)
+q^3(-6q_1^{-2}q_2^{-1}-6q_1^{-1}q_2^{-2}+7q_1^{-2}\cr
&\qquad
+7q_2^{-2}+35q_1^{-1}q_2^{-1}-32q_1^{-1}-32q_2^{-1}+27)
+q^4(-8q_1^{-3}q_2^{-1}\cr
&\qquad
-32q_1^{-2}q_2^{-2}-8q_1^{-1}q_2^{-3}+9q_1^{-3}+135q_1^{-2}q_2^{-1}
+135q_1^{-1}q_2^{-2}\cr
&\qquad
+9q_2^{-3}-110q_1^{-2}-400q_1^{-1}q_2^{-1}-110q_2^{-2}
+286q_1^{-1}+286q_2^{-1}\cr
&\qquad
-192)+\ldots\bigg]\cr
& +{1\over 2\left(2\sin{g_s}\right)^2}\bigg[
q_1^2+q_2^2+q^2(q_1^{-2}q_2^{-2}-2q_1^{-2}-2q_2^{-2}+3)\cr
&\qquad\qquad~
+q^4(-4q_1^{-2}q_2^{-2}+5q_1^{-2}+5q_2^{-2}-6)+\ldots\bigg]\cr
&+{1\over 3\left(2\sin{{3g_s\over 2}}\right)^2}\bigg[
q_1^3+q_2^3+q^3(q_1^{-3}q_2^{-3}-2q_1^{-3}-2q_2^{-3}+3)+\ldots
\bigg]\cr
&+{1\over 4\left(2\sin{2g_s}\right)^2}\bigg[
q_1^4+q_2^4+q^4(q_1^{-4}q_2^{-4}-2q_1^{-4}-2q_2^{-4}+3)+\ldots
\bigg],\cr
\end{aligned}\end{equation}
where we have introduced the notation $q=e^{-t}, q_1=e^{-s_1},
q_2=e^{-s_2}$. Note that we have written only terms up to degree
$4$ in $q$ in (\ref{eq:freenCA}).

As a first qualitative test, note that the terms in the open
string expression (\ref{eq:freenED}) satisfy the integrality
constraints of a closed string expansion \cite{GVii}. Namely, the
terms weighted by ${1\over 2\left(2\sin{g_s}\right)^2}$ have the
structure of degree two multicover contributions of the terms of
degree one. In order to perform a precise test, we have to find a
duality map relating the open string parameters $(t_1, t_2, t_c,
\l_1, \l_2)$ to the closed string parameters $(t,s_1, s_2)$. We
conjecture the following relations
\begin{equation}\label{eq:dualitymapA}
\begin{array}{lcccccc}
& \l_1=is_1,& &\l_2=is_2 & \cr
& t_1=t-{3\over 2}s_1,&  &t_2=t-{3\over2}s_2,& &t_c=t -{1\over 2}(s_1+s_2).&\cr
\end{array}\end{equation}
Note that the flat coordinates $t_1, t_2, t_c$ associated to the
three primitive instantons receive indeed different quantum
corrections, as anticipated in the paragraph preceding equation
(\ref{eq:holvarA}). This is quite sensible, given the
interpretation of these corrections in terms of degenerate
instantons proposed in \cite{DFG}.

Collecting (\ref{eq:freenED}) and the degree three and four terms
computed in Appendix B, we can write the final expression for the
free energy up to degree four
\begin{equation}\label{eq:freenEEi}
\begin{array}{lc}
&{\CF}_{op}(t_1,t_2,t_c,\lambda_1,\lambda_2)
={\CF}_{op}^{(0)}+
{\CF}_{op}^{(1)}+
{\CF}_{op}^{(2)}+{\CF}_{op}^{(3)},
\end{array}\end{equation}
where
\begin{equation}\label{eq:freenEE}\begin{array}{lc}
& {\CF}_{op}^{(0)} ={1\over \left(2\sin{g_s\over
2}\right)^2}
\bigg[e^{-s_1}+e^{-s_2}+e^{-t}(e^{s_1+s_2}-2e^{s_1}-2e^{s_2} +3)
+e^{-2t}(-4e^{s_1+s_2}\cr &\qquad\qquad
+5e^{s_1}+5e^{s_2}-6)+e^{-3t}(-6e^{2s_1+s_2}-6e^{s_1+2s_2}+7e^{2s_1}+7e^{2s_2}
\cr &\qquad\qquad +35e^{s_1+s_2}-32e^{s_1}-32e^{s_2}
+27)+e^{-4t}(-8e^{3s_1+s_2}-32e^{2s_1+2s_2}\cr
&\qquad\qquad -8e^{s_1+3s_2}+9e^{3s_1}
+135e^{2s_1+s_2}+135e^{s_1+2s_2}+9e^{3s_2}-110e^{2s_1}\cr
&\qquad\qquad -400e^{s_1+s_2}-110e^{2s_2}+286e^{s_1}+286e^{s_2}-192) \bigg]\cr &
\qquad +{1\over 2\left(2\sin{g_s}\right)^2}
\bigg[e^{-2s_1}+e^{-2s_2}+e^{-2t}(e^{2s_1+2s_2} -2e^{2s_1}
-2e^{2s_2} + 3) \cr &\qquad\qquad
+e^{-4t}(-4e^{2s_1+2s_2}+5e^{2s_1}+5e^{2s_2}-6)\bigg]\cr & \qquad
+{1\over 3\left(2\sin{{3g_s\over 2}}\right)^2}\bigg[
e^{-3s_1}+e^{-3s_2}+e^{-3t}(e^{3s_1+3s_2}-2e^{3s_1}-2e^{3s_2}+3)\bigg]
\cr & \qquad +{1\over
4\left(2\sin{2g_s}\right)^2}\bigg[e^{-4s_1}+e^{-4s_2}+e^{-4t}(e^{4s_1+4s_2}-2e^{4s_1}-2e^{4s_2}+3)
\bigg],\cr &\cr
&{\CF}_{op}^{(1)}=e^{-3t}\big[-8e^{s_1+s_2}+9e^{s_1}+9e^{s_2}-10\big]+e^{-4t}\big[9e^{2s_1+2s_2}-72e^{2s_1+s_2}\cr
&\qquad\qquad\quad
-72e^{s_1+2s_2}+68e^{2s_1}+344e^{s_1+s_2}+68e^{2s_2}-288e^{s_1}-288e^{s_2}+231\big],\cr 
&\cr &{\CF}_{op}^{(2)}=
\left(2\sin{g_s\over
2}\right)^2e^{-4t}\big[11e^{2s_1+s_2}+11e^{s_1+2s_2}-12e^{2s_1}
-112e^{s_1+s_2} -12e^{2s_2}\cr &\qquad\qquad\qquad\qquad\quad~~
+108e^{s_1}+108e^{s_2}-102\big],\cr
&\cr
&{\CF}_{op}^{(3)}=\left(2\sin{g_s\over
2}\right)^4e^{-4t}\big[13e^{s_1+s_2}-14e^{s_1}-14e^{s_2}+15\big].
\end{array}\end{equation}
Since $q=e^{-t}, q_1=e^{-s_1}, q_2=e^{-s_2}$, we see that all the
terms in ${\CF}_{op}^{(0)}$ are in perfect agreement with
(\ref{eq:freenCA}). This is a highly nontrivial test of our
conjecture. The other three expressions contain terms that have
the $g_s$ dependence of GV invariants with $r=1$, $r=2$ and $r=3$
respectively and we interpret them as duality predictions. Note
that some of these higher genus invariants can be computed using
the methods of \cite{kkv}. For example all degree three $r=1$
invariants can be computed this way obtaining perfect agreement
with (\ref{eq:freenEE}). The successful comparison obtained so far
leads us to conjecture that the free energy of the Chern-Simons
system considered in this section captures all genus topological
amplitudes of the local $dP_2$ model. 
\newpage\noindent It would be very interesting
to test this conjecture at higher order and eventually develop a
more conceptual understanding of this correspondence.

\section{Localization of Open String Maps -- Geometric Considerations}

As discussed in the previous section, the instanton expansions
(\ref{eq:instcorrBA}) and (\ref{eq:instcorrBB}) can be derived
using the open string localization techniques developed in
\cite{DFG}. In this section we carry out the first part of this
program, by finding the general structure of invariant open string
maps $f:\Sigma_{g,h}\ra Y$ subject to the homology constraint
$f_*[\Sigma_{g,h}]=d{\widetilde \beta},$ where ${\widetilde
\beta}$ is an (yet undetermined) integral generator of $H_2(Y,
L_1\cup L_2; \IZ)$. The main result is that ${\widetilde \beta}
=[D_1]=[D_2]=[C] =\beta$, where $D_1, D_2, C$ are the holomorphic
cycles constructed in section two. The sought maps are then
multicovers of $D_1, D_2, C$, as promised in section two.

We first compactify the hypersurfaces $Y_\mu$ by
taking a projective closure of the ambient variety $Z$. Recall
that $Z$ is isomorphic to the total space of $\CO(-1)\oplus \CO
(-2)$ over $\IP^2$, which can be represented as a toric variety
\begin{equation}\label{eq:toricDf}
\begin{array}{rrrrrr}
&Z_1 &Z_2 & Z_3 & U & V \cr \IC^* & 1 & 1 & 1 & -1 & -2.\cr
\end{array}\end{equation} The family $\CY/\Delta$ can be described as a
family of hypersurfaces in $Z$ determined by \begin{equation}\label{eq:famhyper}
Z_1U+VZ_2Z_3 = \mu .\end{equation} Recall also that $Y_\mu \to \IP^2$ is a
fibration with non-compact fibers and no holomorphic section (nor
multi-section).

The (relative) projective closure of $Z$ is the compact toric
variety $\bz = \left(\IC^3\setminus \{0\}\right)$ $\times
\left(\IC^3\setminus \{0\}\right) /(\IC^*)^2$ determined by
\begin{equation}\label{eq:toricD}
\begin{array}{rrrrrrr}
&Z_1 &Z_2 & Z_3 & U & V & W\cr \IC^* & 1 & 1
& 1 & -1 & -2 & 0 \cr \IC^* & 0 & 0 & 0 & 1 & 1 & 1.\cr
\end{array}\end{equation} Note
that $\bz\simeq \IP\left(\CO\oplus \CO(-1)\oplus \CO(-2)\right)$
over $\IP^2$. The Picard group of $\bz$ has rank two and is
generated by the divisor classes
\begin{equation}\label{eq:divclsA} \zeta_1:\quad Z_1=0,\qquad
\zeta_2:\quad W=0.\end{equation} By constructing the toric fan
associated to (\ref{eq:toricD}), one can check that the Mori cone
of $\bz$ is generated by the curve classes
$\eta_1=\zeta_1(\zeta_2-\zeta_1)(\zeta_2-2\zeta_1)$,
$\eta_2=\zeta_1^2(\zeta_2-2\zeta_1)$. One can also choose concrete
representatives of the form \begin{equation}\label{eq:curveclsAi}
\eta_1:\quad Z_1=U=V=0,\qquad \eta_2:\quad
Z_1=Z_2=V=0.\end{equation} The projective completion of the family
(\ref{eq:famhyper}) is a family $\bcy/\Delta$ of compact
hypersurfaces in $\bz$ given by \begin{equation}\label{eq:defC}
Z_1U +VZ_2Z_3 = \mu W.\end{equation} Let $\by$ denote a generic
fiber of this family (as noted before, we drop the subscript $\mu$
with the understanding that $\mu$ is fixed at some real positive
value). We denote by $Y{\buildrel i\over \inj} \by {\buildrel
j\over \inj} \bz$ the obvious embedding maps. The induced
fibration $\pi: \by \to \IP^2$ is a $\IP^1$ fibration; the inverse
image of a line $L$ in $\IP^2$ is a complex surface $\pi ^{-1}(L)
\simeq \IF_1$.  Moreover, $\by \simeq \IP\left(\CO(-1)\oplus
\CO(-2)\right)$, and $\by$ can be identified with $\IP^3$ blown up
at a point. It follows that $H_2(\by, \IZ) \simeq H^2(\by, \IZ)
\simeq \IZ \oplus \IZ$. As generators of $H^2(\by, \IZ)$ we can
take  the section $\sigma$ defined by $V=0$, and $\pi ^*(L)$
(defined, say, by $Z_1=0$). The second homology of $\by$ is
similarly generated by algebraic cycles. For example, one can
choose a basis consisting of the curve classes $\eta_2,
\eta_1+\eta_2$ restricted to $\by$. In particular this shows that
the pushforward map $j_*: H_2(\by, \IZ) \ra H_2(\bz, \IZ)$ is an
isomorphism.

The divisor at infinity in $\by$ is defined as the pull back of
the Cartier divisor $W=0$ on $\bz$:  $ \zeta_\infty=\by \setminus
Y$. Note that $\zeta _\infty \in |\sigma + \pi ^*(L)|$, which is
an ample divisor on $\by$. In particular any holomorphic curve on $\by$ 
must intersect $\zeta _\infty$, and $Y$ contains no holomorphic curve 
classes, as mentioned in section two.

From this point, our analysis follows \cite{DFG}. We restrict our
considerations to open string morphisms which are fixed points of
a certain torus action. Recall that there is a natural $S^1$
action (\ref{eq:toractA}) on $Z$, which can be extended such that
it preserves $\zeta _\infty$\footnote{This is not the most general
action, but it suffices for localization purposes on $\by$.}:
\begin{equation}\label{eq:torcactw}
\begin{array}{ccccccc}
& Z_1  & Z_2 & Z_3 & U & V & W \cr
&\lambda_1 & \lambda_2 & 0 & -\lambda_1 & -\lambda_2 & 0.\cr
\end{array}
\end{equation}

As in \cite{DFG}, we restrict our search to $T$-invariant open string
maps to the pair $(\by, L_1 \cap L_2)$ subject to the homology
constraint $f_*[\Sigma_{g,h}]=d{\widetilde \beta}$.
There is a subtlety here, since one might think it suffices to
consider open string maps to $Y$ instead of $\by$.
The problem is related to the compactification
of $M_{g,h}(Y,d{\widetilde \beta})$. Since $Y$ is
not compact, the correct point of view is to consider open string
maps to $(\by, L_1\cup L_2)$ subject to certain contact conditions
along $\zeta_\infty$. In our case, the order of contact should be zero,
but we cannot automatically exclude eventual compactification
effects. Therefore we will consider open string morphisms to $\by$, and
in the end show that such effects are ruled out by homology constraints.

According to \cite{GZ,KL},
the domain of an $S^1$-invariant map $f:\Sigma_{g,h}\ra \by$
has to be a nodal
bordered Riemann surface whose irreducible
components are either closed compact Riemann surfaces
or holomorphic discs. In the present situation there is an
extra possibility, namely the domain can also be an
annulus. Therefore, we can divide the problem into two parts.
The compact components of $\Sigma_{g,h}$ have to be mapped to
invariant closed curves on $\by$, which can be found using
simple toric considerations.
The second part reduces to finding
$T$-invariant maps $f:\Sigma_{0,1}\ra
\by$ and $f:\Sigma_{0,2}\ra \by$ with lagrangian boundary
conditions along $L_1$ and $L_2$.
As in \cite{DFG}, we can show that
any such $f$ can be extended to a $T$-invariant map $f: \Sigma _0 \ra
\by$ from a smooth rational curve to $\by$.
Proceeding as in section five
of \cite{DFG},  we see that the only curves on $\by$ satisfying these
conditions are
\begin{equation}\label{eq:fixedcurveA}
d_1:\qquad Z_1U = \mu W, \qquad
Z_3=V=0,
\end{equation}
\begin{equation}\label{eq:fixedcurveB}
d_2:\qquad Z_1U = \mu W, \qquad
Z_2=V=0,
\end{equation}
\begin{equation}\label{eq:fixedcurveC}
c:\qquad Z_2Z_3V = \mu W, \qquad
Z_1=U=0.
\end{equation} By writing (\ref{eq:fixedcurveA}) (resp. (\ref{eq:fixedcurveB})) in
local coordinates, it follows that $d_1$ (resp. $d_2$)
intersects $L_1$ (resp. $L_2$) along $\Gamma _1$ ($\Gamma _2$)
which divides it into two discs $D_1$ (resp. $D_2$) and
$D_1^\prime$ ($D_2^\prime$) with boundary on $L_1$ ($L_2$).
Similarly $c$ intersects both $L_1$ and $L_2$ in $\Xi _1$ and $\Xi
_2$, which divide $c$ in the cylinder $C$ and two other discs
$D_3$ and $D_4$.

For future reference, let us note at this point that there is a
family of degree two curves on $\by$ which induces similarly a family of
degree two holomorphic annuli on $Y$ with boundaries on $L_1, L_2$. In terms
of homogeneous coordinates, this family is described by
\begin{equation}\label{eq:cylfamA}
\begin{array}{lc}
&\rho Z_1^2 -\eta Z_2Z_3=0, \qquad {\overline \rho}U^2 -{\overline \eta} V=0,
\qquad UZ_1+VZ_2Z_3 = \mu W\cr
\end{array}\end{equation}
where $(\rho, \eta)$ are projective moduli. By writing these
equations in local coordinates, one can check that
(\ref{eq:cylfamA}) indeed intersects the two spheres $L_1, L_2$
along two circles. Moreover, if $(\rho,\eta) = (1,0)$,
(\ref{eq:cylfamA}) describes the cylinder $C$ with multiplicity
$2$ whereas if $(\rho,\eta)=(0,1)$ we obtain the nodal cylinder
$D_1\cup D_2$ with multiplicity $1$. Therefore $2C$ and $D_1\cup
D_2$ belong to the same moduli space of bordered Riemann surfaces
in $Y$ with boundaries on $L_1, L_2$.

Returning to our problem, we next show as in \cite{DFG} that the discs
$D_1^\prime, D_2^\prime, D_3$ and $D_4$ are in fact ruled out by
the homology constraint $f_*[\Sigma_{g,h}]=d[D]$ by computing the
homology classes  in $H_2(\by,L_1 \cap L_2;\IZ)\simeq H_2(\by,
\IZ).$  We have a commutative diagram of homology groups
\begin{equation}\label{eq:homdiag}
\xymatrix{
H_2(\by,L;\IZ) \ar[r]^{j_*} & H_2(\bz,L;\IZ) \\
H_2(\by,\IZ)\ar[r]^{j_*}\ar[u] & H_2(\bz,\IZ)\ar[u]\\}
\end{equation} in which
all the four arrows are isomorphisms. Therefore we can reduce the
problem to computing the homology classes of the discs in question
as relative cycles for $\bz$ rather than $(\by,L)$. In fact $L_1$
and $L_2$ are fillable in $\bz$;  this is realized geometrically
by deforming to $\mu=0$. In this limit, the discs become
holomorphic curves on the singular fiber $\by_0$ whose homology
classes in $H_2(\bz, \IZ)$ can be easily determined from the
algebraic equations. Note that the singularities of $\by_0$ have
no effect on this computation. In the limit $\mu=0$, $d_1$
specializes to a reducible curve with components $Z_1=U=V=0$, $V=Z_1=Z_3=0$
(class of a fiber) and $V=Z_1=Z_3=0$, which are precisely the
generators $\eta_1$ and $\eta_2$ of $H_2(\bz, \IZ)$.
The two discs $D_1^\prime, D_1$ are deformed in
this limit to these two components of $d_1$; therefore we find
\begin{equation}\label{eq:homclsA}
j_*[D_1] = \eta_1, \qquad j_*[D_1^\prime]=\eta_2.\end{equation}
By a similar reasoning
we also find
\begin{equation}\label{eq:homclsB}
j_*[D_2]= j_*[C]=\eta_1, \qquad j_*[D_2^\prime]=j_*[D_3]=
j_*[D_4]=\eta_2.\end{equation} Since $\eta_1, \eta_2$ are
generators of the Mori cone of $\bz$, using the commutative
diagram (\ref{eq:homdiag}), we conclude that
$[D_1]=[D_2]=[C]={\widetilde \beta}$ are integral generators of
$H_2(Y, L_1\cup L_2;\IZ)$. Therefore we can identify ${\widetilde
\beta}=\beta$ from now on.

Now we can complete the description of a general $S^1$-invariant map
$f:\Sigma_{g,h}\ra {\by}$ subject to the homology constraint
$f_*[\Sigma_{g,h}]=d\beta$. If the domain of the map is an annulus
$\Sigma_{0,2}$, the above analysis shows that $f:\Sigma_{0,2}
\ra \by$ has to be an invariant $d:1$ cover of $C$. If the domain is
a nodal bordered Riemann surface, the disc components have to be
mapped either to $D_1$ or $D_2$ in an invariant manner.
This leaves us with question of finding the images of the closed compact
components of $\Sigma_{g,h}$. This is a simple exercise, taking into
account the homology constraints. Note that $j_*\beta$ is identified with
$\eta_1$ under the isomorphism $H_2(\bz, L_1\cup L_2;\IZ) \simeq
H_2(\bz, \IZ)$. Since $\eta_1$ is a generator of the Mori cone of $\bz$,
it follows that any closed component of $\Sigma_{g,h}$ has to be
mapped to an invariant curve in the class $\eta_1$ lying on $\by$.
Moreover, since the image of $f$ must be connected,
this curve would have to pass through the common origin
of $D_1, D_2$ i.e. the point $P:\ Z_2=Z_3=U=V=0$. It is easy to check that
the only invariant curves on $\by$ passing through this point are
in the fiber class $\eta_2$, which is a distinct generator of the Mori
cone of $\bz$. Therefore we conclude that there are no curves
on $\by$ satisfying all the required conditions;
 all closed components of $\Sigma_{g,h}$
have to be mapped to the point $P$. This completes the description of the
fixed loci. In the next section we will evaluate the contributions
all these fixed points using localization techniques.

\section{Localization of Open String Maps -- Explicit Computations}

In this section we conclude the localization computation for open
string maps with an explicit evaluation of the contributions
of all fixed loci.
We will be using the tangent--obstruction complex technique
\cite{BF,GP,MK,LT} generalized to open string
morphisms in \cite{GZ,KL}.
Based on the results obtained so far,  the fixed
locus of the induced toric action on $\om_{g,h}(Y,L,d\beta)$
consists of two disconnected components. The first component consists
of invariant maps to the annulus $C$, while the second component
consists of invariant maps to the pinched cylinder $D_1\cup D_2$.
Given the structure of the fixed locus in the target space $Y$,
there are no other components. Let us evaluate their contributions.

\subsection{Multicovers of $C$}

Recall that the annulus $C$ is defined by an embedding
$f:\Sigma_{0,2}\ra Y$ given locally by \begin{equation}\label{eq:holcylAB}
\begin{array}{lccccccccc}
&U_2:& &\qquad x_2(t)=0,& &\qquad y_2(t)= t,& &\qquad u_2(t)=0,& &\qquad v_2(t) = {\mu \over t}\\
&U_3:& &\qquad x_3(t')=0,& &\qquad y_3(t')=t',& &\qquad u_3(t')=0,& &\qquad v_3(t')={\mu \over t'}.
\end{array}
\end{equation}
Moreover, the boundary conditions at the two ends are imposed by
the local equations of $L_1, L_2$, which are
\begin{equation}\label{eq:sphereB}
\begin{array}{lc}
& L_1:\qquad u_2={\overline x}_2,\qquad
v_2={\overline y}_2\cr & L_2:\qquad u_3={\overline x}_3,\qquad
v_3={\overline y}_3.\cr\end{array}\end{equation} Up to reparameterizations, there is a
single invariant multicover of $C$ of degree $d$, which is the
Galois cover. In local coordinates we have $f:\Sigma_{0,2}\ra Y$
given by\footnote{In order to keep the notation simple, we will denote
all open string maps generically by $f$. The meaning should be
clear from the context.}
\begin{equation}\label{eq:holcylB}
\begin{array}{lccccccccc}
&U_2:& &\qquad x_2(t)=0,& &\qquad y_2(t)= t^d,& &\qquad u_2(t)=0,& &\qquad v_2(t) = {\mu \over t^d}\\
&U_3:& &\qquad x_3(t')=0,& &\qquad y_3(t')=t'{}^d,& &\qquad u_3(t')=0,& &\qquad v_3(t')={\mu \over t'{}^d}.
\end{array}
\end{equation}
where $t,t'$ are local
coordinates on $\Sigma_{0,2}=\{\mu^{1/2d}\leq |t| \leq
\mu^{-1/2d}\}$. This means that the only nontrivial coefficients
in (\ref{eq:instcorrAB}) are $C_{0, 1,1}(d;d,d)$, the rest being
zero. These coefficients can be computed by evaluating the
contributions of the fixed loci in the moduli space of stable open
string maps along the lines of \cite{GZ,KL}.

Let us denote $\Sigma_{0,2}$ by $\Sigma$ in order to simplify the
formulas. Note that the automorphism group of this map has an
automorphism group of order $d$ generated by $(t\ra \zeta t, t'\ra
\zeta^{-1}t')$ where $\zeta$ is a $d$-th root of unity. In local
coordinates, the torus action (\ref{eq:toractA}) reads
\begin{equation}\label{eq:toractB}
\begin{array}{lccccccccc}
 U_2:& &x_2\ra e^{i\phi(\lambda_1-\lambda_2)}x_2,&
&y_2 \ra e^{-i\phi\lambda_2} y_2,&
&u_2\ra e^{i\phi(-\lambda_1+\lambda_2)}u_2,&
&v_2\ra e^{i\phi\lambda_2}v_2&\cr
 U_3:& &x_3\ra e^{i\phi\lambda_1}x_3,&
&y_3\ra e^{i\phi\lambda_2} y_3,&
&u_3\ra e^{-i\lambda_1} u_3,&
&v_3 \ra e^{-i\phi\lambda_2} v_3.&\cr
\end{array}
\end{equation} The map (\ref{eq:holcylB}) is left
invariant if we let $S^1$ act on $\Sigma$ by \begin{equation}\label{eq:toractC}
t\ra
e^{-i\phi\lambda_2/d}t,\qquad t'\ra e^{i\phi\lambda_2/d}t'.\end{equation} In
order to write down the tangent-obstruction complex for the above
map we need to introduce some more notation. Note that the tangent
bundles $T_{L_{1,2}}$ of the lagrangian cycles $L_1, L_2$ are real
subbundles of the holomorphic tangent bundle $T_{Y}$ restricted to
$L_1$ and respectively $L_2$. Pulling back to $\Sigma$ we obtain a
 triple\footnote{We denote by $f_\partial$ the restriction of
$f:\Sigma\ra Y$ to the boundary of $\Sigma$.}
 $\left(f^*(T_Y), f_\partial^*(T_{L_1}), f_\partial^*(T_{L_2})\right)$
which defines a Riemann-Hilbert bundle on $\Sigma$. We let $\CT_Y$ denote
the sheaf of germs of holomorphic sections of $f^*(T_Y)$ with boundary
values in $\left(f_\partial^*(T_{L_1}), f_\partial
^*(T_{L_2})\right)$. Similarly, the
holomorphic tangent bundle of $\Sigma$ together with the tangent bundles
to the two boundary components form a Riemann-Hilbert bundle whose associated
sheaf of holomorphic sections will be denoted by $\CT_\Sigma$.
Then the tangent-obstruction complex of $f$ takes the form
\begin{equation}\label{eq:tangobsA}
0 \ra H^0\left(\Sigma, \CT_\Sigma\right) \ra
H^0\left(\Sigma,\CT_Y\right) \ra
{\bf T}^1
 \ra H^1\left(\Sigma, \CT_\Sigma\right) \ra
H^1\left(\Sigma, \CT_Y\right) \ra {\bf T}^2 \ra 0.\end{equation}
The torus actions (\ref{eq:toractB}), (\ref{eq:toractC}) induce an
action on all terms in (\ref{eq:tangobsA}), so that we obtain a
complex of equivariant vector spaces. The contribution of the
fixed point (\ref{eq:holcylB}) is given by
\begin{equation}\label{eq:fixedpA}
{1\over |Aut(f)|} \int_{pt_{S^1}}{e({\bf T}^2)\over e({\bf T}^1)}
= {1\over d} \int_{pt_{S^1}}{e(B^m_5)e(B^m_1)\over
e(B^m_4)e(B^m_2)}\end{equation} where $B^m_i$ denotes the moving
part of the $i$-th term in the complex (\ref{eq:tangobsA}) and
$e({\ })$ is the $S^1$- equivariant Euler class. The integration
is $S^1$-equivariant integration over the fixed point locus, which
is a point.

In order to evaluate (\ref{eq:fixedpA}) we have to compute the
relevant cohomology groups and take the moving parts \cite{GP}. We
will do this computation in {\v C}ech cohomology, as in \cite{LS}.
The cylinder $\Sigma$ is covered by two open sets
\begin{equation}\label{eq:opencoverA} \CU_1=\{\mu^{1/2d}< t \leq
\mu^{-1/2d}\},\qquad \CU_2=\{\mu^{1/2d}< t'
\leq\mu^{-1/2d}\}.\end{equation} The {\v C}ech complex we are
interested in is \begin{equation}\label{eq:complexA} 0\ra
\CT_Y(\CU_1) \oplus \CT_Y(\CU_2) {\buildrel \kappa\over \ra}
\CT_Y(\CU_{12}) \ra 0\end{equation} where $\kappa(s_1, s_2) =
s_1|_{\CU_{12}}-s_2|_{\CU_{12}}$. The local sections of $\CT_Y$
over $\CU_1$, $\CU_2$ have the form
\begin{equation}\label{eq:locsectA}
\begin{array}{lc}
& s_1=\left(\sum_{n\in \IZ} \alpha_n t{}^n\right)\partial_{x_2} +
\left(\sum_{n\in \IZ} \beta_n t{}^n\right)\partial_{y_2} +
\left(\sum_{n\in \IZ} \gamma_nt{}^n\right)\partial_{u_2} + \left(\mu
\sum_{n\in \IZ} \delta_nt{}^n\right)\partial_{v_2}\cr &
s_2=\left(\sum_{n\in \IZ} \alpha'_n t'{}^n\right)\partial_{x_3} +
\left(\sum_{n\in \IZ} \beta'_n t'{}^n\right)\partial_{y_3} +
\left(\sum_{n\in \IZ} \gamma'_nt'{}^n\right)\partial_{u_3} +
\left(\mu \sum_{n\in \IZ}
\delta'_nt'{}^n\right)\partial_{v_3}\cr\end{array}\end{equation}
where the coefficients $\alpha_n, \beta_n, \gamma_n, \delta_n$,
$\alpha'_n, \beta'_n, \gamma'_n, \delta'_n$ are subject to two
types of constraints. We have constraints imposed by the boundary
conditions (\ref{eq:sphereB}) which can be written in the form
\begin{equation}\label{eq:boundcondA} \sum_{n\in \IZ} \alpha_n
r^ne^{in\theta} = \sum_{n\in \IZ} {\overline \gamma}_nr^n
e^{-in\theta},\qquad \sum_{n\in \IZ} \beta_n r^ne^{in\theta} = \mu
\sum_{n\in \IZ} {\overline \delta}  _n r^n
e^{-in\theta}\end{equation} and identical conditions for primed
coefficients. At the same time, one must impose the condition that
the infinitesimal deformations parameterized by
(\ref{eq:locsectA}) be tangent to $Y$. Recall that the local
equation of $Y$ in $U_2$ is \begin{equation}\label{eq:loceqC}
x_2u_2+y_2v_2=\mu.\end{equation} Therefore the sheaf of
holomorphic differentials on $\ Y\cap U_2 \ $ is generated by
$(dx_2, dy_2, du_2, dv_2)$ subject to the condition
\begin{equation}\label{eq:holdiffA}
x_2du_2 +u_2dx_2 +y_2dv_2+v_2dy_2=0.\end{equation} By contracting
(\ref{eq:holdiffA}) with (\ref{eq:locsectA}) and taking into
account the fact that $x_2=u_2=0$ along $C$, we find
\begin{equation}\label{eq:condA} \left(\sum_{n\in \IZ} \beta_n
t^n\right)+t^{2d} \left(\sum_{n\in \IZ}
\delta_nt^n\right)=0.\end{equation} The conditions for local
sections over $\CU_2$ are very similar. In terms of coefficients,
(\ref{eq:boundcondA}) and (\ref{eq:condA}) yield
\begin{equation}\label{eq:condB}
\alpha_n = \mu^{-n/d}{\overline
\gamma}_{-n}, \qquad \beta_n = \mu^{(d-n)/d}
{\overline \delta}_{-n},\qquad \beta_n +\delta_{n-2d}=0\end{equation} and, similarly,
\begin{equation}\label{eq:condC}
\alpha'_n = \mu^{-n/d}{\overline \gamma}'_{-n}, \qquad
\beta'_n = \mu^{(d-n)/d}{\overline \delta}'_{-n},\qquad \beta'_n
+\delta'_{n-2d}=0.\end{equation} In order to be able to compare the two
sections upon restriction, we have to write say $s_2$ in terms of
$t=1/t'$ and $\partial_{x_2},\ldots,
\partial_{v_2}$. The relevant linear transformations are
\begin{equation}\label{eq:lintransfA}
\begin{array}{lccc}
&\partial_{x_3} = y_2\partial_{x_2},& &\qquad
\partial_{y_3}=-x_2y_2\partial_{x_2}-y_2^2\partial_{y_2}
+y_2u_2\partial_{u_2}+2y_2v_2\partial_{v_2},\\
& \partial_{u_3} ={1\over y_2}\partial_{u_2},& &\qquad
\partial_{v_3}={1\over {y_2}^2} \partial_{v_2}.
\end{array}\end{equation}
Using these transformations, a straightforward computation shows that
the map $\kappa$ is given by
\begin{equation}\label{eq:cechdiffA}
\begin{array}{lc}
\kappa(s_1, s_2) = & \left(\sum_{n\in \IZ} (\alpha_n-\alpha'_{-n+d})
t^n \right)\partial_{x_2} +\left(\sum_{n\in \IZ}(\beta_n +
\beta'_{-n+2d})t^n\right)\partial_{y_2}\cr
&+\left(\sum_{n\in \IZ} (\gamma_n-\gamma'_{-n-d})t^n\right) \partial_{u_2}
+\left(\mu\sum_{n\in \IZ} (\delta_n+\delta'_{-n-2d})t^n\right)
\partial_{v_2}.\cr\end{array}\end{equation}
Let us determine the kernel and cokernel of this map. For this, let us
consider the following system of equations
\begin{equation}\label{eq:kerA}\begin{array}{lccc}
& \alpha_n -\alpha'_{-n+d} =a_n,& &\qquad
\beta_n + \beta'_{-n+2d} =b_n\\
& \gamma_n-\gamma'_{-n-d}=c_n,& &\qquad \delta_n+\delta'_{-n-2d}=d_n
\end{array}\end{equation}
where $a_n, b_n, c_n, d_n$ are coefficients of a generic section
of $\CT_Y(\CU_{12})$. Hence we have
\begin{equation}\label{eq:condD}
b_n + d_{n-2d} =0.\end{equation} Using the constraints
(\ref{eq:condB}) and (\ref{eq:condC}) we can rewrite
(\ref{eq:kerA}) in the following form
\begin{equation}\label{eq:kerB}
\begin{array}{lc}
& \alpha_n -\alpha'_{-n+d} =a_n,\qquad
\qquad \qquad \qquad \beta_n + \beta'_{-n+2d} =b_n\cr
& \mu^{n/d}\alpha_{n} -\mu^{(d-n)/d}\alpha'_{-n+d}
={\overline c}_{-n},\qquad
\mu^{(n-d)/d}\beta_n+\mu^{(d-n)/d}\beta'_{-n+2d}
={\overline d}_{-n}.\cr\end{array}\end{equation}
In order to find the kernel of $\kappa$ we have to set all $a_n,\ldots,
d_n$ to zero, and solve the resulting homogeneous equations.
Then it is easy to check that there are no nonzero
solutions except for the following cases
\begin{equation}\label{eq:kerC}\begin{array}{lc}
&2n=d,\qquad \alpha_{d/2}=\alpha'_{d/2}\cr
&n=d,\qquad \ \beta_{d}=-\beta'_{d}.\cr\end{array}\end{equation}
In all other cases, the solutions are identically zero. Note that
the first case cannot be realized unless $d$ is even.
Therefore we find that
$\hbox{Ker}(\kappa)\simeq H^0(\Sigma, \CT_Y)$ is
generated by
\begin{equation}\label{eq:cohA}\begin{array}{lc}
& \alpha_{d/2}t^{d/2}\partial_{x_2} +\beta_d t^d\partial_{y_2} +
\mu^{1/2} {\overline \alpha}_{d/2} t^{-d/2} \partial_{u_2}
+\mu{\overline \beta}_d t^{-d}\partial_{v_2},\qquad \hbox{for}\ d
\ \hbox{even}\cr & \beta_d t^d\partial_{y_2} + \mu{\overline
\beta}_d t^{-d}\partial_{v_2},\qquad \qquad \qquad\qquad \qquad
\qquad\qquad \qquad\ \ \, \hbox{for}\ d \
\hbox{odd}\cr\end{array}\end{equation} where $\beta_d
+{\overline \beta}_d=0$. We can similarly compute the cokernel of
$\kappa$. The equations (\ref{eq:kerB}) have solutions for 
any values of $a_n,\ldots, d_n$ subject to the constraints
\begin{equation}\label{eq:condE}
{\overline c}_{-d/2} = \mu^{1/2} a_{d/2},\qquad {\overline d}_{-d}
= b_d,\qquad b_d+{\overline b_d}=0.\end{equation} If these
constraints are not satisfied, the equations (\ref{eq:kerB}) have
no solutions. \\ Taking also into account (\ref{eq:condD}), it
follows that $\hbox{Coker}(\kappa)\simeq H^1(\Sigma, \CT_Y)$ is
generated by sections of the form
\begin{equation}\label{eq:cohB}\begin{array}{lc}
& a_{d/2}t^{d/2} \partial_{x_2} + b_{d} t^d \partial_{y_2}, \qquad
 \hbox{for}\ d \ \hbox{even}\cr
& b_d{t^d} \partial_{y_2}, \qquad \qquad\qquad \qquad \hbox{for}\
d \ \hbox{odd}\cr\end{array}\end{equation} where $b_d-{\overline
b}_d=0$. Note that for $d$ even we obtain one extra deformation
and one extra obstruction compared to $d$ odd. This reflects the
fact that there is a family of holomorphic cylinders of degree two
on $Y$ interpolating between $2C$ and $D_1\cup D_2$ as shown in
(\ref{eq:cylfamA}). This means that for $d$ even the invariant map
(\ref{eq:holcylB}) can be
 deformed to a $d/2:1$ cover of the holomorphic cylinders
described in (\ref{eq:cylfamA}).

In order to finish the computation of the integral in
(\ref{eq:fixedpA}), we have to determine the moving parts of
$H^{0}(\Sigma, \CT_Y)$ and $H^{1}(\Sigma, \CT_Y)$. The action of
$S^1$ on $H^0(\Sigma, \CT_Y)$ is determined by (\ref{eq:toricA})
and the explicit form of the generators (\ref{eq:cohA}),
(\ref{eq:cohB}).
 We adopt the notation conventions
of \cite{KL} for representations of $S^1$, namely the representation
$z\ra e^{iw\phi}z$, where $z\in \IC$, will be denoted by $(w)$.
The trivial real representation will be denoted by $(0)_{\IR}$.
One can define an $S^1$-equivariant isomorphism $H^0(\Sigma,
\CT_Y)\ra H^1(\Sigma, \CT_Y)$ by \begin{equation}\label{eq:equivisom}
\begin{array}{lc}
&\alpha_{d/2}t^{d/2}\partial_{x_2} +\beta_d t^d\partial_{y_2} +
\mu^{1/2} {\overline \alpha}_{d/2} t^{-d/2} \partial_{u_2}
+\mu{\overline \beta}_d t^{-d}\partial_{v_2}\ \ra \
\alpha_{d/2}t^{d/2}\partial_{x_2} +\beta_d t^d\partial_{y_2}\cr &
\beta_d t^d\partial_{y_2} + \mu{\overline \beta}_d
t^{-d}\partial_{v_2}\ \ra\ i\beta_d t^d\partial_{y_2},\cr\end{array}\end{equation} for
$d$ even and respectively odd. This shows that both $H^0(\Sigma,
\CT_Y)$ and $H^1(\Sigma, \CT_Y)$ are isomorphic to
\begin{equation}\label{eq:torrepA}\begin{array}{lcc}
& \left({\lambda_2\over 2}-\lambda_1\right) \oplus (0)_{\IR},
\qquad \hbox{for}\ d\ \hbox{even}\cr & \ \ (0)_{\IR}, \qquad\qquad
\qquad \qquad \hbox{for}\ d\
\hbox{odd}.\cr\end{array}\end{equation} Therefore we have
\begin{equation}\label{eq:fixedpB} {e(B^m_5)\over e(B^m_2)}=
1\end{equation} in (\ref{eq:fixedpA}). A similar computation shows
that $B_1^m$ and $B_4^m$ are trivial, hence
\begin{equation}\label{eq:fixedpBi} {e(B^m_1)\over
e(B^m_4)}=1.\end{equation} We conclude that contribution of the
fixed point (\ref{eq:holcylB}) to the open string topological
amplitudes is simply
\begin{equation}\label{eq:fixedpAB} C_{0,1,1}(d; d,d) = {1\over d}.\end{equation}

\subsection{ Multicovers of $D_1\cup D_2$}

We now perform a similar computation for the components of the
fixed locus which map to the pinched cylinder $D_1\cup D_2$. In
this case the structure of $S^1$ invariant maps $f:\Sigma_{g,h}\ra
Y$ is more complicated \cite{GZ,KL,LS}. Recall that the two discs
are defined locally by (\ref{eq:holdiscA}) and
(\ref{eq:holdiscB}). For convenience let us reproduce the relevant
formulas below. The common domain of the maps
$f_{1,2}:\Sigma_{0,1}\ra Y$ is the disc $\{|t|\leq
\mu^{-1/2}\}=\{\mu^{1/2}\leq |t'|\}$ in $\IP^1$ with affine
coordinates $(t,t')$. The local form of the holomorphic embeddings
is
\begin{equation}\label{eq:holdiscD}\begin{array}{lccccccccc}
&U_1:& &\qquad x_1(t)=t,& &\qquad y_1(t)= 0,& &\qquad u_1(t)=\mu,& &\qquad v_1(t) = 0\\
&U_2:& &\qquad x_2(t') = t',& &\qquad y_2(t') =0,& &\qquad u_2(t') = {\mu\over t'},& &\qquad v_2(t') =0.
\end{array}
\end{equation}
and respectively
\begin{equation}\label{eq:holdiscE}\begin{array}{lccccccccc}
&U_1:& &\qquad x_1(t)=0,& &\qquad y_1(t)=t,& &\qquad u_1(t)= \mu,& &\qquad v_1(t) = 0\\ 
&U_3:& &\qquad x_3(t') = t',& &\qquad y_3(t') =0,& &\qquad u_3(t') ={\mu\over t'},& &\qquad v_3(t') = 0.
\end{array}
\end{equation}
We will also need the defining equations for the two spheres in all coordinate
patches. For $L_1$ we have
\begin{equation}\label{eq:sphereBA}\begin{array}{lc}
& U_1:\qquad u_1x_1\bx_1=1,\qquad v_1x_1^2\bx_1={\overline y}_1\cr
& U_2:\qquad u_2={\overline x}_2,\qquad\quad\ \,  v_2={\overline y}_2\cr\end{array}\end{equation}
and for $L_2$
\begin{equation}\label{eq:sphereD}\begin{array}{lc}
& U_1:\qquad u_1y_1{\overline y}_1=1,\qquad
v_1y_1^2{\overline y}_1={\overline x}_1\cr
& U_3:\qquad u_3={\overline x}_3,\qquad \quad\ \,
v_3={\overline y}_3.\cr\end{array}\end{equation}
Recall also that the open string maps are labelled
by three integers $(g,h_1, h_2)$,
where $h_1,h_2$ denote the number of holes mapped to
$\Gamma_1=\partial D_1$ and respectively
$\Gamma_2=\partial D_2$.
The homotopy class of such a map is determined by the degrees
$d_1, d_2$ and the winding numbers $m_i$, $i=1,\ldots, h_1$, $n_j$,
$j=1,\ldots, h_2$ subject to the constraints $\sum_{i=1}^{h_1}m_i =d_1$,
$\sum_{j=1}^{h_2} n_j = d_2$. In order to describe the structure of the
corresponding fixed loci, we need to distinguish several cases.

$i)\ (g,h_1,h_2)=(0,1,0)$. The invariant maps are Galois covers $f:\Delta^1
\ra D_1$ of degree $m=d_1$ with  automorphism group $\IZ/d_1$.

$ii)\ (g,h_1,h_2)=(0,0,1)$. The invariant maps are Galois covers
$f:\Delta^2\ra D_2$ of degree $n=d_2$ with automorphism group
$\IZ/d_2$.

$iii)\ (g,h_1,h_2)=(0,2,0)$. The domain of an invariant map has to
be a pinched cylinder $\Delta^1_1\cup \Delta^1_2$. The components
are mapped as Galois covers of degrees $m_1, m_2$ to $D_1$. In
this case the automorphism group is a  product of $\IZ/m_1\times
\IZ/m_2$ with a permutation group $\CP$ which is trivial in
$m_1\neq m_2$ and it permutes the two boundary components if
$m_1=m_2$. $|Aut(f)|=|\CP|m_1m_2$ where $|\CP|=1$ if $m_1\neq
m_2$, and $|\CP|=2$ if $m_1=m_2$.

$iv)\ (g,h_1,h_2)=(0,0,2)$. The domain is similarly a pinched cylinder
$\Delta^2_1\cup \Delta^2_2$ which are again mapped as Galois covers
of degrees $(n_1,n_2)$ to $D_2$. By analogy with the previous
case we have $|Aut(f)|=|\CP|n_1n_2$, where $\CP$ is defined similarly.

$v)\ (g,h_1, h_2)=(0,1,1)$. The domain is of the form
$\Delta^1_1\cup \Delta^2_1$, and the two components are mapped to
$D_1, D_2$ as Galois covers of degrees $(m_1, n_1)=(d_1, d_2)$.
The automorphism group is simply $\IZ/m_1\times \IZ/n_1$ since permuting
the two boundary components does not preserve $f$ in this case.
Therefore $|Aut(f)|=m_1n_1$.

$vi)$ In all other cases, the domain of the map must be of the form
$\Sigma^0_g\cup \Delta_1^{1}\cup\ldots \cup \Delta^1_{h_1} \cup
\Delta^2_{1}\cup\ldots \cup \Delta^2_{h_2}$ where $(\Sigma^0_g, p_1,\ldots,
p_{h_1}, q_1,\ldots, q_{h_2})$ is a stable $h$-punctured curve
of genus $g$ and $\Delta^1_1,\ldots,\Delta^1_{h_1}, \Delta^2_{1},
\ldots, \Delta^2_{h_2}$ are holomorphic discs whose origins are
attached to $\Sigma_g$ at the punctures $(p_1,\ldots, p_{h_1})$ and
respectively $(q_1,\ldots, q_{h_2})$. The map $f$ sends $\Sigma^0_g$ to the
point $P$ which is the common origin of the discs $D_1, D_2$ in $Y$,
and maps the discs $\Delta^1_1,\ldots,\Delta^1_{h_1}$,
$\Delta^2_{1},\ldots, \Delta^2_{h_2}$ to $D_1$ and respectively $D_2$.
By $S^1$ invariance, all these maps must be Galois covers of some degrees
$m_1, \ldots , m_{h_1}$ and $n_1,\ldots, n_{h_2}$.
In this case, the automorphism group is generated by a) automorphisms
of the Galois covers and b) permutations of $(p_1, \ldots, p_{h_1})$
and respectively $(q_1, \ldots, q_{h_2})$ which leave the orders unchanged.
Note that only permutations which act on $(p_1, \ldots, p_{h_1})$ and
$(q_1, \ldots, q_{h_2})$ separately are allowed. The permutations
which exchange $p_i$ with $q_j$ are not automorphisms of $f$ even if
$m_i=n_j$ for some values of $i$ and $j$.
Therefore $Aut(f)$ is a product between $\prod_{i=1}^{h_1}
\IZ/m_i\times \prod_{j=1}^{h_2} \IZ/n_j\times \CS_{h_1}\times \CS_{h_2}$
We have accordingly $|Aut(f)| = \prod_{i=1}^{h_1} m_i \prod_{j=1}^{h_2}
n_j h_1!h_2!$.

In the following we will denote $\Sigma_{g,h}$ by $\Sigma$ and
$\Sigma_g^0$ by $\Sigma^0$ in order to simplify the equations. The
tangent-obstruction complex has the form
\begin{equation}\label{eq:tangobsB}
0\ra
Aut(\Sigma)\ra H^0(\Sigma, \CT_Y)\ra {\bf T}^1 \ra Def(\Sigma) \ra
H^1(\Sigma, \CT_Y) \ra {\bf T}_2\ra 0\end{equation}
where $Aut(\Sigma)$,
$Def(\Sigma)$ are the groups of infinitesimal automorphisms and
respectively infinitesimal deformations of $\Sigma$. In the
generic case, which is number six on the above list, we have to
evaluate
\begin{equation}\label{eq:fixedpC}
{1\over |Aut(f)|}\int_{[\om_{g,h}]_{S^1}} {e(B^m_5)e(B^m_1)\over
e(B^m_4)e(B^m_2)}\end{equation} where the notation is as before.
In the special cases (i)--(v), the fixed locus is just a point,
hence the integral should be modified accordingly. Note that in
the instanton series we have to sum over ordered $h_1$-uples
$(m_1,\ldots,m_{h_1})$ and respectively $h_2$-uples
$(n_1,\ldots,n_{h_2})$. Since the value of (\ref{eq:fixedpC}) is
invariant under a permutation of the winding numbers, we can
alternatively sum over unordered $(m_1,\ldots,m_{h_1})$,
$(n_1,\ldots,n_{h_2})$ and divide only by $\prod_{i=1}^{h_1} m_i
\prod_{j=1}^{h_2}n_j$ in (\ref{eq:fixedpC}). This is the point of
view we will take in this section.

Let us start with the generic case. Since the domain Riemann
surface is nodal, we have to use a normalization exact sequence in
order to compute the terms in (\ref{eq:tangobsB}). Let
$\CT^1_{Y_i}$, $i=1,\ldots, h_1$, $\CT^2_{Y_j}$, $j=1,\ldots, h_2$
denote the restrictions of $\CT_Y$ to the discs $\Delta^1_i$ and
respectively $\Delta_2^j$. In other words, $\CT^1_{Y_i}$ is the
sheaf associated to the Riemann-Hilbert bundle obtained by pulling
back the pair $(T_Y,T_{L_1})$ to the disc $\Delta^1_i$, and
similarly for $\CT^2_{Y_j}$. In the following we will always label
the boundary components mapped to $L_1$ by $i=1,\ldots, h_1$, and
the boundary components mapped to $L_2$ by $j=1,\ldots, h_2$. Then
we have the following exact sequence
\begin{equation}\label{eq:normseqA}
0\ra \CT_Y \ra \oplus_{i=1}^{h_1} \CT^1_{Y_i} \oplus \oplus_{j=1}^{h_2}
\CT^2_{Y_j} \oplus f_0^*(T_Y)\ra \oplus_{i=1}^{h_1}(T_Y)_{P}\oplus
\oplus_{j=1}^{h_2} (T_Y)_{P} \ra 0\end{equation}
where $f_0:\Sigma^0\ra Y$ denotes the restriction of $f$ to $\Sigma^0$.
This yields the following long exact sequence
\begin{equation}\label{eq:longseqA}\begin{aligned}
0& \ra H^0(\Sigma,\CT_Y)\ra \oplus_{i=1}^{h_1} H^0(\Delta^1_i, \CT^1_{Y_i})
\oplus \oplus_{j=1}^{h_2}
 H^0(\Delta^2_j, \CT^2_{Y_j}) \oplus H^0(\Sigma^0, f_0^*(T_Y))\cr
&\ra \oplus_{i=1}^{h_1}(T_Y)_P \oplus \oplus_{j=1}^{h_2}(T_Y)_P
\ra H^1(\Sigma, \CT_Y) \cr
&\ra \oplus_{i=1}^{h_1} H^1(\Delta^1_i, \CT^1_{Y_i}) \oplus \oplus_{j=1}^{h_2}
 H^1(\Delta^2_j, \CT^2_{Y_j}) \oplus H^1(\Sigma^0, f_0^*(T_Y))\ra 0.\cr\end{aligned}\end{equation}
We denote by $F^m_k$, $k=1,\ldots, 5$ the moving parts of the
terms in (\ref{eq:longseqA}). Then we have
\begin{equation}\label{eq:movingA}
{e(B^m_5)\over e(B^m_2)}= {e(F^m_5) e(F^m_3)\over
e(F^m_2)}.\end{equation} The other factors in (\ref{eq:fixedpC})
are
\begin{equation}\label{eq:movingAB}
{e(B_1^m)\over e(B_4^m)}={e(Aut(\Sigma)^m)\over e(Def(\Sigma)^m)}.\end{equation}
Since we are in the generic case, $Aut(\Sigma)$
consists only of rotations of the $h$ discs, which have $S^1$-weight
zero, therefore we have $e(Aut(\Sigma)^m)=1$. The moving part of
$Def(\Sigma)$ consists only of deformations of the $h$ nodes,
namely
\begin{equation}\label{eq:movingB}
Def(\Sigma)^m = \oplus_{i=1}^{h_1}
\left(T_{p_i}({\Sigma^0})\otimes T_0(\Delta^1_i)\right) \oplus
\oplus_{j=1}^{h_2} \left(T_{q_j}({\Sigma^0})\otimes T_0(\Delta^2_j)\right).\end{equation}
Therefore, in order to finish the computation we have to determine
$F^m_2,F^m_3,F^m_5$.
In order to do that, let us begin with some local considerations.
We will focus on the disc $D_1$, since $D_2$ is entirely analogous.
In the following we will need the local form of the torus action
in the coordinate patches $U_1, U_2$
\begin{equation}\label{eq:toractD}\begin{array}{lcccccccccccccccccc}
U_1:&&x_1\ra e^{i\phi(\lambda_2-\lambda_1)}x_1,& &y_1\ra
e^{-i\phi\lambda_1}y_1,& &u_1\ra u_1,& &v_1\ra
e^{i\phi(2\lambda_1-\lambda_2)}v_1,&\cr U_2:&&x_2\ra
e^{i\phi(\lambda_1-\lambda_2)}x_2,& &y_2 \ra e^{-i\phi\lambda_2}
y_2,& &u_2\ra e^{i\phi(-\lambda_1+\lambda_2)}u_2,& &v_2\ra
e^{i\phi\lambda_2}v_2.&\cr\end{array}\end{equation} We will also
need the local form of the sheaf of K\"ahler differentials on
$U_1, U_2$. Given the local equations (\ref{eq:loceqA}) we find
the following relations
\begin{equation}\label{eq:holdiffB}\begin{array}{lc}
& U_1:\qquad  du_1+ y_1v_1dx_1+ x_1v_1dy_1+x_1y_1dv_1=0\cr
& U_2:\qquad  x_2du_2 +u_2dx_2 +y_2dv_2+v_2dy_2=0.\cr\end{array}\end{equation}

Let us now compute the {\v C}ech cohomology groups
$H^*(\Delta^1_i,\CT^1_{Y_i})$ and $H^*(\Delta^2_j,\CT^2_{Y_j})$. We
consider as before a disc $\Delta$ of the form $\{|t| \leq
{\mu}^{-1/2d}\}=\{\mu^{1/2d}\leq |t'|\}$ in $\IP^1$ with affine
coordinates $(t,t')$. A Galois cover of degree $d$ $f:\Delta\ra
D_1$ is given in local coordinates by
\begin{equation}\label{eq:invmapB}
\begin{array}{lccccccccc}
&U_1:& &\qquad x_1(t)=t^d,& &\qquad y_1(t)= 0,& &\qquad u_1(t)=\mu,& &\qquad v_1(t) = 0\\
&U_2:& &\qquad x_2(t') = t'{}^d,& &\qquad y_2(t') =0,& &\qquad u_2(t') = {\mu\over t'{}^d},& &\qquad v_2(t') =0.
\end{array}
\end{equation} 
Note that (\ref{eq:invmapB})
is left invariant if we let $S^1$ act on the domain $\Delta$ by
\begin{equation}\label{eq:toractDA}
t\ra e^{i\phi(\lambda_1-\lambda_2)/d},\qquad t'\ra
e^{i\phi(\lambda_2-\lambda_1)/d}.\end{equation} We cover the disc $\Delta$ by
two open sets \begin{equation}\label{eq:opnecoverB}
\CU_1=\{0\leq |t| <
{\mu}^{-1/2d}\},\qquad \CU_2=\{\mu^{1/2d} \leq |t| <
(\mu+\epsilon^2)^{1/2d}\}\end{equation} and construct the {\v C}ech complex
\begin{equation}\label{eq:cechB}
0\ra \CT_Y(\CU_1)\oplus \CT_Y(\CU_2) {\buildrel \kappa
\over \ra} \CT_Y(\CU_{12}) \ra 0.\end{equation} The generic sections in
$\CT_Y(\CU_1), \CT_Y(\CU_2)$ have the form \begin{equation}\label{eq:locsectC}\begin{array}{l}
 s_1 = \left(\sum_{n=0}^{\infty}\alpha_n t^n\right)
\partial_{x_1}+ \left(\sum_{n=0}^\infty \beta_n t^n\right)
\partial_{y_1} + \left(\mu \sum_{n=0}^\infty \gamma_n t^n\right)
\partial_{u_1} + \left(\sum_{n=0}^\infty \delta_n t^n\right)
\partial_{v_1}\cr  s_2 = \left(\sum_{n\in \IZ}\alpha'_n
t'{}^n\right) \partial_{x_2}+ \left(\sum_{n\in \IZ}\beta'_n
t'{}^n\right) \partial_{y_2} + \left(\mu\sum_{n\in \IZ}\gamma'_n
t'{}^n\right) \partial_{u_2} + \left(\sum_{n\in \IZ}\delta'_n
t'{}^n\right) \partial_{v_2}.\cr\end{array}\end{equation} Note that we sum over $n\geq 0$
for sections in $\CT(\CU_1)$. In order to have a uniform notation,
we can extend these sums to $n \in \IZ$, with the convention that
$\alpha_n, \ldots, \delta_n$ are zero for $n<0$.

The coefficients $\alpha_n,\beta_n,\ldots, \delta'_n$ are again
subject to two types of constraints. The boundary conditions at $|t'|=
{\mu}^{1/2d}$ take the form
\begin{equation}\label{eq:boundcondB}
\sum_{n\in \IZ} \alpha'_n r^n e^{in\theta'} =\mu
\sum_{n\in \IZ} {\overline \gamma'}_n r^n e^{-in\theta'},\qquad
\sum_{n\in \IZ} \beta'_n r^n e^{in\theta'} =
\sum_{n\in \IZ} {\overline \delta'}_n r^n e^{-in\theta'}\end{equation}
where $r=\mu^{1/2d}$.
This yields the following relations between
coefficients
\begin{equation}\label{eq:boundcondC}
\alpha'_n = \mu^{(d-n)/d} {\overline \gamma'}_{-n},\qquad \beta'_n
= \mu^{-n/d} {\overline \delta'}_{-n}.\end{equation} Obviously,
the un-primed coefficients are not subject to boundary conditions.
Next, we want the infinitesimal deformations (\ref{eq:locsectC})
to be tangent to $Y$. By evaluating (\ref{eq:holdiffB}) along
$D_1$, and contracting with (\ref{eq:locsectA}), we obtain
\begin{equation}\label{eq:condF}
\sum_{n\in \IZ} \alpha'_n t'{}^n + t'{}^{2d}\sum_{n\in\IZ}\gamma'_n t'{}^n =0,\qquad
\sum_{n=0}^\infty \gamma_n t^n =0.\end{equation}
The resulting relations between coefficients can be written as
\begin{equation}\label{eq:condG}
\alpha'_n + \gamma'_{n-2d} =0, \qquad \gamma_n =0.\end{equation}
In order to compute the kernel and cokernel of $\kappa$, we have to
rewrite $s_2$ in terms of $t$ and $\partial_{x_1}, \ldots, \partial_{v_1}$.
The relevant linear transformations are
\begin{equation}\label{eq:lintransfB}\begin{array}{lccc}
& \partial_{x_2}=-x_1^2 \partial_{x_1}-x_1y_1\partial_{y_1}
+u_1x_1\partial_{u_1}+ 2x_1v_1\partial_{v_1},& &\qquad
\partial_{y_2}=x_1\partial_{y_1},\\
&\partial_{u_2}={1\over x_1} \partial_{u_1},& &\qquad 
\partial_{v_2}={1\over x_1^2} \partial_{v_1}.\end{array}\end{equation}
Then, by direct computations, one can check that the map $\kappa$
takes the form
\begin{equation}\label{eq:cechdiffB}\begin{aligned}
\kappa(s_1,s_2)= &\left(\sum_{n\in \IZ} (\alpha_n+
\alpha'_{-n+2d})t^n \right)\partial_{x_1}+ \left(\sum_{n\in \IZ}
(\beta_n- \beta'_{-n+d})t^n\right)\partial_{y_1}\cr
&+\left(\sum_{n\in \IZ} (\delta_n
-\delta'_{-n-2d})t^n\right)\partial_{v_1}.\cr\end{aligned}\end{equation}
Again, in order to determine the kernel and cokernel of $\kappa$ we
have to consider the following system of equations
\begin{equation}\label{eq:kerD}\begin{array}{lccc}
& \alpha_n+\alpha'_{-n+2d} = a_n,& &\qquad
\beta_n- \beta'_{-n+d}=b_n \\
& & &\qquad 
\delta_n -\delta'_{-n-2d} =d_n\cr\end{array}\end{equation}
where $a_n, b_n, c_n, d_n$ are coefficients in the Laurent expansion
of an arbitrary section of $\CT_Y(\CU_{12})$. Therefore they are subject to
the constraints
\begin{equation}\label{eq:condH}
c_n=0.\end{equation} In order to find the kernel, we set
$a_n,\ldots, d_n$ to zero and solve for the coefficients in the
left hand side of equations (\ref{eq:condD}). Combining
(\ref{eq:boundcondC}) and (\ref{eq:condG}) we have
\begin{equation}\label{eq:condHA}
\alpha'_n + \mu^{(d-n)/d}{\overline \alpha'}_{-n+2d}=0,\qquad
\beta'_n-\mu^{-n/d}{\overline \delta'}_{-n}=0.\end{equation}
Consider the equation in the first column of (\ref{eq:condD}).
Since $\alpha_n =0$ for $n<0$, using the first equation in
(\ref{eq:condHA}), we find that $\alpha'_n=0$ for $n<0$ or $2d<n$.
This leaves only $2d+1$ nonzero coefficients
$\alpha_n=-\alpha'_{-n+2d}$, $0\leq n\leq 2d$ subject to the
relations
\begin{equation}\label{eq:condJ}
\alpha_n +\mu^{(d-n)/d}{\overline \alpha}_{-n+2d} =
0.\end{equation} Next, using (\ref{eq:kerD}) and the second
equation in (\ref{eq:condHA}) we can find the relations
\begin{equation}\label{eq:condJA}
\beta_n =\mu^{(n-d)/d}{\overline \delta}_{-n-d},\qquad
\beta_{-n-d} =\mu^{-(n+2d)/2}{\overline \delta}_n.\end{equation}
Since $\beta_n, \delta_n =0$ for $n<0$, it follows that the only
solutions are $\beta_n = \delta_n =0$.
Therefore the kernel of $\kappa$ is generated by sections of the form
\begin{equation}\label{eq:kerE}\begin{array}{lc}
& \left[\sum_{n=0}^{d-1}(\alpha_n t^n -\mu^{(n-d)/d}{\overline \alpha}_n
t^{2d-n}) + \alpha_dt^d\right] \partial_{x_1}\cr\end{array}\end{equation}
where $\alpha_d +{\overline \alpha_d}=0$.

Now let us determine the cokernel of $\kappa$, which is generated
by local sections with coefficients $a_n, \ldots, d_n$ for which
the equations (\ref{eq:kerD}) have no solutions. Let us first
analyze the equation in the first column of (\ref{eq:kerD}). Using
(\ref{eq:condHA}), we find that
\begin{equation}\label{eq:cokerC}\begin{array}{lc}
& \alpha_n =0, \qquad \alpha'_{n} =-\mu^{(d-n)/d}{\overline
a}_n,\qquad \qquad \qquad \qquad\qquad \ \hbox{for}\ n<0\cr &
\alpha_{n}+\alpha'_{-n+2d} = a_n,\qquad \alpha'_n
-\mu^{(d-n)/d}{\overline \alpha'}_{-n+2d} =0, \qquad \hbox{for} \
0\leq n\leq 2d\cr & \alpha_n =\mu^{(n-d)/d}{\overline a}_{2d-n}
-a_n, \qquad \alpha'_n =a_{2d-n}, \qquad \qquad \quad \ \,
\hbox{for}\ 2d<n.\cr\end{array}\end{equation} It is clear that
these equations have solutions for any values of $a_n$ therefore
we do not obtain any obstructions from the first equation in
(\ref{eq:condD}). We have to perform a similar analysis for the
second set of equations in (\ref{eq:condD}). Exploiting again the
fact that $\beta_n, \delta_n =0$ for $n<0$, one can show that
\begin{equation}\label{eq:cokerD}
\beta'_n =-b_{-n+d},\qquad \delta'_{-n} = -d_{n-2d}\end{equation}
for $d< n< 2d$. Substituting (\ref{eq:cokerD}) in the second
equation of (\ref{eq:condHA}) we obtain the following condition on
$b_n, d_n$
\begin{equation}\label{eq:cokerE}
b_{-n-d}=\mu^{-(n+2d)/d}{\overline d}_n\end{equation}
for $-d<n<0$. Therefore the cokernel of $\kappa$ is $(d-1)$-dimensional
and generated by sections of the form
\begin{equation}\label{eq:cokerL}
\left(\sum_{n=-d+1}^{-1} d_n t^n \right)\partial_{v_1}.\end{equation}
In particular, if $d=1$ there are no obstructions.

Collecting the results obtained so far, we can determine the
contribution of a single disc to the integrand in
(\ref{eq:fixedpC}). Recall that we are using the notation
conventions of \cite{KL} for representations of $S^1$, namely the
representation $z\ra e^{iw\phi}z$ is  denoted by $(w)$. The real
trivial representation is denoted by $(0)_{\IR}$. Then, taking
into account (\ref{eq:toractD}) and (\ref{eq:toractDA}),
$H^0(\Delta, \CT_Y)$ computed in (\ref{eq:kerE}) is
$S^1$-isomorphic to\footnote{There is a subtlety here related to
the choice of signs for the weights of the toric action on
$H^{0,1}(\Delta, \CT_Y)$, which can be traced to the choice of
orientations. We made this choice so that the signs agree with
\cite{KL}.}
\begin{equation}\label{eq:equivdefA}
\left(\lambda_1-\lambda_2\right)\oplus
\left((\lambda_1-\lambda_2){d-1\over d}\right)\oplus \ldots \oplus
\left((\lambda_1-\lambda_2){1\over d}\right) \oplus
(0)_{\IR}.\end{equation} Similarly, the obstruction group
$H^1(\Delta, \CT_Y)$ computed in (\ref{eq:cokerL}) is
$S^1$-isomorphic to
\begin{equation}
\begin{aligned}\label{eq:equivobsA}
\left(\lambda_2-2\lambda_1 +{1\over d}(\lambda_1-\lambda_2)\right)
&\oplus \left(\lambda_2-2\lambda_1 +{2\over
d}(\lambda_1-\lambda_2)\right) \oplus\ldots \cr \oplus  &
\left(\lambda_2-2\lambda_1 +{d-1\over
d}(\lambda_1-\lambda_2)\right).
\end{aligned}
\end{equation}
The above formulas have been derived for $f:\Delta \ra D_1$ a
Galois cover of $D_1$. We can perform entirely analogous
computations for Galois covers of $D_2$. In fact, one can simply
find the $S^1$ action on the corresponding deformation and
obstruction groups by exchanging $x_2\leftrightarrow x_3$ and
$y_2\leftrightarrow y_3$ in the above computations. In that case,
$H^0(\Delta, \CT_Y)$ is $S^1$-isomorphic to
\begin{equation}\label{eq:equivdefB}
\left(\lambda_1\right) \oplus
\left({d-1\over d}\lambda_1 \right) \oplus \ldots \oplus
\left({1\over d}\lambda_1 \right) \oplus (0)_{\IR}\end{equation}
and $H^1(\Delta, \CT_Y)$ is $S^1$-isomorphic to
\begin{equation}\label{eq:equivdefC}
\left(\lambda_2-2\lambda_1+{1\over d}\lambda_1\right) \oplus
\left(\lambda_2-2\lambda_1+{2\over d}\lambda_1\right)\oplus
\ldots\oplus
\left(\lambda_2-2\lambda_1+{d-1\over d}\lambda_1\right).\end{equation}

Using these formulas, we can complete the computation of the
moving parts of the terms $F_2, F_3, F_5$ in (\ref{eq:longseqA}).
Recall that $\sum_{i=1}^{h_1} m_i=d_1$, $\sum_{j=1}^{h_2}n_j=d_2$,
where $(d_1, d_2)$ are the degrees of the map $f:\Sigma^0 \cup
\cup_{i=1}^{h_1}\Delta^1_i \cup \cup_{j=1}^{h_2}\Delta^2_j\ra Y$
with respect to the two discs $D_1, D_2$. Moreover, $\Sigma^0$ is
mapped to the point $P:Z_2=Z_3=0$ on $Y$, hence
$H^0(\Sigma^0,f_0^*(T_Y))\simeq
H^0(\Sigma_0,\CO_{\Sigma^0})\otimes T_P(Y)$. Then we have
\begin{equation}\label{eq:movingC}\begin{array}{lcc} & e(F_2^m)= H^{d_1+d_2+3}
\lambda_1(\lambda_1-\lambda_2)(\lambda_2-2\lambda_1)
(\lambda_1-\lambda_2)^{d_1} \lambda_1^{d_2}\cr &\qquad
\qquad\times\prod_{i=1}^{h_1} \prod_{k=0}^{m_i-1}
\left({m_i-k\over m_i}\right) \prod_{j=1}^{h_2}
\prod_{l=0}^{n_j-1} \left({n_j-l\over n_j}\right)\cr 
\cr
& e(F_3^m)=
H^{3(h_1+h_2)}
(\lambda_1(\lambda_1-\lambda_2)(\lambda_2-2\lambda_1))^{h_1+h_2}\cr
\cr
& e(F_5^m)=
c_g(\IE^*((\lambda_1-\lambda_2)H))c_g(\IE^*(\lambda_1H))
c_g(\IE^*((\lambda_2-2\lambda_1)H))H^{d_1+d_2-h_1-h_2}\cr
&\qquad\qquad \times\prod_{i=1}^{h_1}\prod_{k=1}^{m_i-1}
\left(\lambda_2-2\lambda_1 +{k\over
m_i}(\lambda_1-\lambda_2)\right)
\prod_{j=1}^{h_2}\prod_{l=1}^{n_j-1} \left(\lambda_2-2\lambda_1
+{l\over n_j} \lambda_1\right)\cr\end{array}\end{equation} where
$H\in H^2(BS^1)\simeq H^2_{S^1}({pt})$ is the generator of the
equivariant cohomology ring of a point, and $\IE$ denotes the
Hodge bundle on the moduli space of stable pointed curves
$\om_{g,h}$. Note that we are using the orientation conventions of
\cite{KL}. In the second equation in (\ref{eq:movingC}), the
expressions of the form $c_g(\IE^*(\eta H))$ with
$\eta=\lambda_1-\lambda_2, \lambda_1, \lambda_2-2\lambda_1$ denote
\begin{equation}\label{eq:equivhodgeA}
c_g(\IE^*(\eta H))= (\eta H)^g -c_1(\IE)(\eta H)^{g-1} +c_2(\IE)
(\eta H)^{g-2} + \ldots + (-1)^g c_g(\IE)\end{equation} in the
equivariant cohomology ring $H^*_{S^1}(\om_{g,h})$. The last
ingredients we need are the terms in equation (\ref{eq:movingAB}).
Since we are in the generic case, $Aut(\Sigma)$ consists only of
rotations of the $h$ discs
$\Delta^1_1,\ldots,\Delta^1_{h_1},\Delta^2_1, \ldots,
\Delta^2_{h_2}$ which are generated by $t\partial t$ over $\IR$.
Therefore $Aut(\Sigma) \simeq (0)_\IR$, and the moving part is
trivial. We are left with the moving part of $Def(\Sigma)$, which
is given by (\ref{eq:movingB}). This yields
\begin{equation}\label{eq:movingF} e(Def(\Sigma)^m) =
\prod_{i=1}^{h_1}\left({\lambda_1-\lambda_2\over m_i}H -\psi_i
\right) \prod_{j=1}^{h_2}\left({\lambda_1\over n_j}H -
\psi_j\right)\end{equation} where $\psi_i=c_1(\IL_i),\ i=1,\ldots,
h_1$, $\psi_j=c_1(\IL_j),\ j=1,\ldots, h_2$ are the first Chern
classes of the tautological line bundles $\IL_i, \IL_j$ over
$\om_{g,h}$ associated to the marked points $p_i, q_j$.

Now we can collect all the results obtained so far, and write down
an integral expression for the coefficients $F_{g,h_1, h_2}$ in
(\ref{eq:instcorrAC})
\begin{equation}\label{eq:inscorrC}\begin{array}{lc}
&F_{g,h_1,h_2}(d_1,d_2; m_i,n_j)= {1\over
\prod_{i=1}^{h_1}m_i\prod_{j=1}^{h_2}n_j}
\int_{\left[\om_{g,h}\right]_{S^1}} {e(F_5^m)e(F^m_3)\over
e(F^m_2) e(Def(\Sigma)^m)}\cr & =
{\left[\lambda_1(\lambda_1-\lambda_2)(\lambda_2-2\lambda_1)\right]^
{h_1+h_2-1}\over (\lambda_1-\lambda_2)^{d_1}\lambda_1^{d_2}}
\prod_{i=1}^{h_1}
{\prod_{k=1}^{m_i-1}((\lambda_2-2\lambda_1)m_i+k(\lambda_1
-\lambda_2))\over (m_i-1)!}\cr &~\times\prod_{j=1}^{h_2}
{\prod_{l=1}^{n_j-1}((\lambda_2-2\lambda_1) n_j +l\lambda_1)\over
(n_j-1)!}\cr &~\times\int_{\left[\om_{g,h}\right]_{S^1}}
{c_g(\IE^*((\lambda_1-\lambda_2)H))c_g(\IE^*(\lambda_1H))
c_g(\IE^*((\lambda_2-2\lambda_1)H))H^{2(h_1+h_2)-3}\over
\prod_{i=1}^{h_1}\left({(\lambda_1-\lambda_2)}H -m_i\psi_i \right)
\prod_{j=1}^{h_2}\left(\lambda_1H - n_j\psi_j\right)}.
\cr\end{array}\end{equation} This is the result for the generic
case. Note that this formula takes values in the fraction field of
the cohomology ring $H^*(BS^1)$ and in general it need not be a
multiple the unit element. In order to obtain a physically
meaningful answer, we have to impose certain conditions on the
weights $\lambda_1, \lambda_2$ whose origin has been explained in
\cite{KL}. We will discuss this specialization of the toric action
after treating the special cases (i)-(v).

$i)\ (g,h_1, h_2) =(0,1,0)$. In this case, the map is of the form
$f:\Delta^1\ra D_1$ given in local coordinates by
(\ref{eq:invmapB}) with $d=d_1$. Therefore we can directly use
equations (\ref{eq:equivdefA}) and (\ref{eq:equivobsA}) obtaining
\begin{equation}\label{eq:specialA}
{e(B^m_5)\over e(B^m_2)} =
{d_1^{d_1-1}H^{-1} \over (d_1-1)!(\lambda_1-\lambda_2)^{d_1}}
\prod_{k=1}^{d_1-1} \left(\lambda_2-2\lambda_1+{k\over d_1}
(\lambda_1-\lambda_2)\right).\end{equation}
The automorphism group $Aut(\Delta^1)$ is in this case nontrivial and
generated by $t'\partial_{t'}, \partial_{t'}$ which span
$(0)_\IR \oplus \left({\lambda_1-\lambda_2\over d_1}\right)$. This
yields
\begin{equation}\label{eq:specialB}
e(B_1^m) = {\lambda_1-\lambda_2\over d_1}H.\end{equation}
The overall contribution to the multicover formula is
\begin{equation}\label{eq:specialC}\begin{aligned}
F_{0,1,0}(d_1,0;d_1,0)& = {1\over d_1}\int_{{pt}_{S^1}}
{e(B^m_5)e(B^m_1)\over e(B^m_2)}\cr
& = {1\over d_1^2(d_1-1)!} {1\over (\lambda_1-\lambda_2)^{d_1-1}}
\prod_{k=1}^{d_1-1} \left((\lambda_2-2\lambda_1)d_1+{k}
(\lambda_1-\lambda_2)\right).
\cr\end{aligned}\end{equation}

$ii)\ (g,h_1, h_2) = (0,0,1)$. This case is analogous to the previous
one. We just have to exchange $D_1\leftrightarrow D_2$, and the final result
is
\begin{equation}\label{eq:specialD}
F_{0,0,1}(0,d_2; 0,d_2) = {1\over d_2^2(d_2-1)!}
{1\over \lambda_1^{d_2-1}}
{\prod_{l=1}^{d_2-1} \left((\lambda_2-2\lambda_1)d_2+{l}
\lambda_1\right)}.\end{equation}

$iii)\ (g,h_1,h_2)=(0,2,0)$. In this case we have a map $f:\Delta^1_1 \cup
\Delta^1_2 \ra D_1$ which maps the two components of the domain to
$D_1$ with degrees $m_1, m_2$, $m_1+m_2=d_1$. Since the domain is nodal,
we have to consider again a normalization exact sequence of the form
\begin{equation}\label{eq:normseqB}
0\ra \CT_Y \ra \CT^1_{Y_1}\oplus \CT^1_{Y_2}\ra (T_Y)_{P}\ra 0\end{equation}
which yields a long exact sequence
\begin{equation}\label{eq:longseqB}\begin{aligned}
0 &\ra H^0(\Sigma,\CT_Y)\ra H^0(\Delta^1_1, \CT^1_{Y_1}) \oplus
 H^0(\Delta^1_2, \CT^1_{Y_2}) \ra (T_Y)_P\cr
& \ra H^1(\Sigma, \CT_Y) \ra H^1(\Delta^1_1, \CT^1_{Y_1}) \oplus
 H^1(\Delta^1_2, \CT^1_{Y_2}) \ra 0.\cr\end{aligned}\end{equation}
We denote the terms of (\ref{eq:longseqB}) by $F_1,\ldots, F_5$ as
before, so that
\begin{equation}\label{eq:specialE}
F_{0,2,0}(d_1,0;m_1,m_2,0,0) = {1\over m_1m_2} \int_{{pt}_{S^1}}
{e(F^m_5)e(F^m_3)\over e(F^m_2)}{e(Aut(\Sigma)^m)\over
e(Def(\Sigma)^m)}.\end{equation} The contributions of the terms in
(\ref{eq:longseqB}) have been evaluated before. The automorphism
group is again trivial from an equivariant point of view since we
have only rotations of the two discs. The moving part of
$Def(\Sigma)$ consists of deformations of the node, which are
parameterized by $T_0(\Delta^1_1)\otimes T_0(\Delta^1_2)$.
Therefore
\begin{equation}\label{eq:specialF}
e(Def(\Sigma)^m) = \left({1\over m_1}+{1\over m_2}\right)
(\lambda_1-\lambda_2)H.\end{equation}
Collecting all the factors, we arrive at the following expression
\begin{equation}\label{eq:specialG}\begin{aligned}
F_{0,2,0}(d_1,0;m_1,m_2,0,0) = {1\over d_1}
{\lambda_1(\lambda_2-2\lambda_1)\over (\lambda_1-\lambda_2)^{d_1}}
\prod_{i=1}^2 {\prod_{k=1}^{m_i-1}\left((\lambda_2-2\lambda_1)m_i +
k(\lambda_1-\lambda_2)\right)\over (m_i-1)}.\end{aligned}\end{equation}

$iv)\ (g,h_1,h_2)=(0,0,2)$. This case can be obtained from the
previous one by exchanging $D_1\leftrightarrow D_3$. We have a map
$f:\Delta^2_1 \cup \Delta^2_2\ra D_2$ with degrees $n_1, n_2$,
$n_1+n_2=d_2$. Going through the steps
(\ref{eq:normseqB})-(\ref{eq:specialG}) yields the following
formula
\begin{equation}\label{eq:specialH} F_{0,0,2}(0,d_2;0,0,n_1,n_2) =
{1\over d_2} {(\lambda_1-\lambda_2)(\lambda_2-2\lambda_1)\over
\lambda_1^{d_2}} \prod_{j=1}^2{\prod_{l=1}^{n_j-1}
\left((\lambda_2-2\lambda_1)n_j+ l\lambda_1\right)\over
(n_j-1)!}.\end{equation}

$v)\ (g,h_1,h_2)=(0,1,1)$. We have a map $f:\Delta^1_1\cup
\Delta^2_1\ra Y$ mapping the two components of the domain onto
$D_1, D_2$ with degrees $d_1, d_2$. Then one has to use a
normalization sequence similar to (\ref{eq:normseqB}) and evaluate
the contributions of the terms as before. Therefore we have
\begin{equation}\label{eq:normseqC}
0\ra \CT_Y \ra \CT^1_{Y_1}\oplus \CT^2_{Y_2} \ra (T_Y)_P\ra 0\end{equation}
which yields a long exact sequence
\begin{equation}\label{eq:longseqC}\begin{aligned}
0&\ra H^0(\Sigma, \CT_Y) \ra  H^0(\Delta^1_1,\CT^1_{Y_1})\oplus
H^0(\Delta^2_1,\CT^2_{Y_2}) \ra (T_Y)_P\cr &\ra H^1(\Sigma, \CT_Y)
\ra  H^1(\Delta^1_1,\CT^1_{Y_1})\oplus
H^1(\Delta^2_1,\CT^2_{Y_2})\ra 0.\cr\end{aligned}\end{equation}
Denoting again the terms of (\ref{eq:longseqC}) by $F_1, \ldots,
F_5$, we have
\begin{equation}\label{eq:movingG}
{e(B^m_5)\over e(B_m^2)} =
{e(F_5^m) e(F_3^m) \over e(F_2^m)}.\end{equation}
In order to complete the computation, we also need to evaluate
\begin{equation}\label{eq:movingH}
{e(B^m_1)\over e(B^m_4)} = {e(Aut(\Sigma)^m) \over e(Def(\Sigma)^m)}.\end{equation}
This is analogous to cases (iii) and (iv) above. The only nontrivial
moving part consists of deformations of the node of the domain, which
are parameterized by $T_0(\Delta^1_1)\otimes T_0(\Delta^2_1)$.
Hence we are left with
\begin{equation}\label{eq:movingI}
e(Def(\Sigma)^m)= {\l_1-\l_2\over d_1} +{\l_1\over d_2}.\end{equation}
Collecting all the intermediate results, we obtain
\begin{equation}\label{eq:specialI}\begin{aligned}
F_{0,1,1}(d_1,d_2;d_1,d_2) & = {1\over d_1d_2}
\int_{{pt}_{S^1}}{e(F_5^m) e(F_3^m) \over
e(F_2^m)e(Def(\Sigma)^m)}\cr & ={1\over (d_1-1)!(d_2-1)!}
{\lambda_2-2\lambda_1\over (d_2(\lambda_1-\lambda_2) +
d_1\lambda_1) (\lambda_1-\lambda_2)^{d_1-1}\lambda_1^{d_2-1}}\cr
&~\times \prod_{k=1}^{d_1-1}\left((\lambda_2-2\lambda_1)d_1
+k(\lambda_1-\lambda_2) \right)
\prod_{l=1}^{d_2-1}\left((\lambda_2-2\lambda_1)d_2 + l\lambda_1
\right).\cr\end{aligned}\end{equation} This concludes our
localization computations for open string morphisms. In principle,
instanton corrections (\ref{eq:instcorrAC}) can be obtained by
adding add the contributions (\ref{eq:inscorrC}),
(\ref{eq:specialC}), (\ref{eq:specialD}), (\ref{eq:specialG}),
(\ref{eq:specialH}), and (\ref{eq:specialI}). Besides being very
complicated, the resulting expression would be a homogeneous
rational function of $\lambda_1, \lambda_2$ rather than a number.
This is a common problem with open string localization
computations \cite{KL,GZ} which can be solved by making a specific
choice of the toric action.

\subsection{Choice of Toric Action}

As discussed in detail in section four, in order to obtain a
physically meaningful answer,
one has to choose specific values for the weights $\lambda_1,
\lambda_2$. This approach is similar to that of \cite{GZ,KL,Mii},
except that in our case the motivation for these choices is somewhat
different. In those references, there is an ambiguity in the
definition of the virtual fundamental class due to the fact that
the open string moduli spaces have boundaries. In our situation,
we do not really have a virtual class in the standard sense since
the contribution of each fixed locus is a formal series of operators
in Chern-Simons theory. Therefore we have to choose weights in order
to make sense of this series of instanton corrections for each
fixed component. Moreover, the choice of weights has to be
correlated to the choice of framings in Chern-Simons theory.

We will follow the approach of \cite{KL}, choosing two sections of the
real normal bundles $N^1_{\IR}$, $N^2_{\IR}$.
We will work
in the coordinate patch $U_1$, which covers both discs. The
boundaries $\Gamma_1, \Gamma_2$ of the two discs can be
parameterized by
\begin{equation}\label{eq:boundariesA}\begin{array}{lc} & \Gamma_1:\qquad
x_1=\mu^{-1/2} e^{i\theta_1}, \qquad y_1=0, \qquad u_1=\mu, \qquad
v_1=0\cr & \Gamma_2:\qquad x_1=0,\qquad y_1=\mu^{-1/2}
e^{i\theta_2},\qquad u_1=\mu,\qquad v_1=0.\cr\end{array}\end{equation}
The sections of
the real normal bundles must be of the form
\begin{equation}\label{eq:framA}\begin{array}{lc} &
N^1_{\IR}:\qquad(y_1,v_1) = \left(e^{i(a-1)\theta_1},
\mu^{3/2}e^{-ia\theta_1}\right) \cr & N^2_{\IR}:\qquad(x_1,v_1) =
\left(e^{i(b-1)\theta_2}, \mu^{3/2}e^{-ib\theta_2}\right)
\cr\end{array}\end{equation} where $a,b\in \IZ$. At this point,
one might be tempted to conclude that the sections
(\ref{eq:framA}) determine the framing of the two knots $\Gamma_1,
\Gamma_2$ in $L_1, L_2$. In fact, we have to be more careful here,
since we need more data in order to determine the framing as an
integer number. Note that the sections (\ref{eq:framA}) can be
regarded as sections of the circle bundles associated to
$N_{\IR}^1, N_{\IR}^2$, which are topologically trivial. The group
of homotopy classes of sections of an $S^1$ bundle over $S^1$ is
isomorphic to $\IZ$, but one has to choose an isomorphism in order
to associate an integer number to such a class. More concretely,
the choice of such an isomorphism is equivalent to the choice of a
trivialization of the circle bundle which plays the role of
reference section. In our case, we can obtain natural
trivializations of $N_{\IR}^1, N_{\IR}^2$ exploiting the fact that
$\Gamma_1, \Gamma_2$ are algebraic knots. By changing coordinates
to $(x_2,y_2,u_2,v_2)$ and respectively $(x_3,y_3,u_3,v_3)$ the
spheres $L_1,L_2$ can be identified with the canonical sphere in
$\IC^2$, as in equation (\ref{eq:sphereC}). Moreover, the boundary
components of $D_1, D_2$ are now parameterized by
\begin{equation}\label{eq:boundariesB}\begin{array}{lc}
& \Gamma_1:\qquad x_2=\mu^{1/2} e^{i\theta'_1}, \qquad
y_2=0,\qquad u_2=\mu^{1/2} e^{-i\theta'_1}, \qquad v_2=0\cr &
\Gamma_2:\qquad x_3=\mu^{1/2} e^{i\theta'_2},\qquad y_3=0,\qquad
u_3=\mu^{1/2} e^{-i\theta'_2},\qquad
v_3=0\cr\end{array}\end{equation} where $\theta'_1 = -\theta_1$,
$\theta'_2=-\theta_2$. In the new coordinates, the sections
(\ref{eq:framA}) read
\begin{equation}\label{eq:framAB}\begin{array}{lc} & N^1_{\IR}:\qquad (y_2,v_2) =
\left(\mu^{1/2}e^{i(2-a)\theta'_1},
\mu^{1/2}e^{-i(2-a)\theta'_1}\right)\cr &  N^2_{\IR}:\qquad
(x_2,v_2) = \left(\mu^{1/2}e^{i(2-b)\theta'_2},
\mu^{1/2}e^{-i(2-b)\theta'_2}\right).\cr\end{array}\end{equation}
In this form, one can choose canonical reference sections for
$N^1_{\IR}, N^2_{\IR}$ determined by $\partial_{y_2}$ and respectively
$\partial_{x_2}$. With respect to these sections, the framings of
the two knots are  $(2-a,2-b)$.

Following the strategy proposed in \cite{KL}, we require that the
sections (\ref{eq:framA}) be equivariant. Then the toric action is
fixed, since we have the following constraints on the weights
\begin{equation}\label{eq:weightsA}\begin{array}{lccc} 
&-\lambda_1 =
(a-1)(\lambda_2-\lambda_1),& &\qquad 2\lambda_1 -\lambda_2 =
-a(\lambda_2-\lambda_1)\\ 
& \lambda_2-\lambda_1 =
-(b-1)\lambda_1,& &\qquad 2\lambda_1-\lambda_2 = b\lambda_1.\end{array}\end{equation}
Using the equations in the first column, we easily find
\begin{equation}\label{eq:weightsB}
\lambda_1(\lambda_2-\lambda_1)(ab-a-b)=0.\end{equation}
If we
choose either $\lambda_1=0$ or $\lambda_2-\lambda_1=0$, one can
check that the remaining equations imply $\lambda_1=\lambda_2=0$.
This is not an acceptable solution, therefore we are left with the
equation
\begin{equation}\label{eq:framB} ab=a+b,\qquad a,b\in \IZ.\end{equation} This has two
solutions, namely $(a,b)=(0,0)$ or $(a,b)=(2,2)$. The first case,
$(a,b)=(0,0)$ implies $2\lambda_1-\lambda_2=0$, while the second
case implies $\lambda_2=0$.

At this point, we do not have any further selection criteria,
hence we cannot rule any solution out. However, the first
solution, $(a,b)=(0,0)$ is more convenient since it yields the
simple closed form for the instanton expansion
(\ref{eq:instcorrBA}), (\ref{eq:instcorrBB}). In the following we
will make this choice, and finish the derivation of
(\ref{eq:instcorrBA}), (\ref{eq:instcorrBB}). The second solution
should also lead to consistent results, but it is much harder to
do explicit calculations. We will not pursue this problem here.

If $(a,b)=(0,0)$, we have to impose the relation
$2\lambda_1-\lambda_2=0$ in (\ref{eq:inscorrC}),
(\ref{eq:specialC}), (\ref{eq:specialD}), (\ref{eq:specialG}),
(\ref{eq:specialH}), (\ref{eq:specialI}) and collect the results.
Let us start with (\ref{eq:inscorrC}). Because of the factor
$(\lambda_2-2\lambda_1)^{h_1-h_2-1}$ we can obtain a nonzero
answer only if $h_1+h_2=1$, that is from bordered Riemann surfaces
with one boundary component. Some care is needed in this argument,
since in principle we could get other contributions if the
expansion of the integrand in (\ref{eq:inscorrC}) yields terms
with negative powers of $(\lambda_2-2\lambda_1)$. In order to show
that such terms are absent, we can rewrite (\ref{eq:inscorrC}) as
\begin{equation}\label{eq:instcorrD}
\begin{aligned}
F_{g,h_1,h_2}&(d_1,d_2;m_i,n_j)=
 {(\lambda_2-2\lambda_1)^{h_1+h_2-1}
\over (\lambda_1-\lambda_2)^{d_1+1}\lambda_1^{d_2+1}}
\prod_{i=1}^{h_1} {\prod_{k=1}^{m_i-1}((\lambda_2-2\lambda_1)m_i+k(\lambda_1
-\lambda_2))\over (m_i-1)!}\cr
&\times
\prod_{j=1}^{h_2} {\prod_{l=1}^{n_j-1}((\lambda_2-2\lambda_1) n_j
+l\lambda_1)\over (n_j-1)!}\cr
&\times\int_{\left[\om_{g,h}\right]_{S^1}}
c_g(\IE^*((\lambda_1-\lambda_2)H))c_g(\IE^*(\lambda_1H))
c_g(\IE^*((\lambda_2-2\lambda_1)H))H^{(h_1+h_2)-3}\cr
&\times
\prod_{i=1}^{h_1}\left[\sum_{k=0}^{\infty}
\left({m_i\psi_iH^{-1}\over \lambda_1-\lambda_2}\right)^k\right]
\prod_{j=1}^{h_2}\left[\sum_{l=0}^\infty
\left({n_j\psi_jH^{-1}\over \lambda_1}\right)^l
\right].
\cr\end{aligned}\end{equation}
All terms in this expressions contain negative powers of
$\lambda_1, (\lambda_1-\lambda_2)$ but not $(\lambda_2-2\lambda_1)$.
Therefore we can conclude that imposing the relation $\lambda_2-2\lambda_1=0$
leaves only terms with $h_1+h_2=1$. This leads to a significant simplification
of the instanton expansion. The nontrivial contributions are
\begin{equation}\label{eq:instcorrE}
\begin{array}{lc}
F_{g,1,0}(d_1,0;d_1,0)  =
  {1\over \lambda_1^2}
\int_{\left[\om_{g,h}\right]_{S^1}} & (-1)^g
c_g(\IE^*(-\lambda_1H))c_g(\IE^*(\lambda_1H)) c_g(\IE)H^{-2} \cr &
\times\left[\sum_{k=0}^{\infty} \left(-{d_1\psi H^{-1}\over
\lambda_1}\right)^k\right].\cr\end{array}\end{equation} \begin{equation}\label{eq:instcorrF}\begin{array}{lc}
F_{g,0,1}(0,d_2;0,d_2) = {1\over \lambda_1^2}
\int_{\left[\om_{g,h}\right]_{S^1}} & (-1)^g
c_g(\IE^*(-\lambda_1H))c_g(\IE^*(\lambda_1H))c_g(\IE)H^{-2} \cr &
\times\left[\sum_{l=0}^\infty \left({d_2\psi H^{-1}\over
\lambda_1}\right)^l \right].\cr\end{array}\end{equation} Now we can finish the
computation as in \cite{KL} using the relation in \cite{FP}
\begin{equation}\label{eq:hodgerel}
c_g(\IE^*(\lambda_1H))c_g(\IE^*(-\lambda_1H))=
(-1)^g(\lambda_1H)^g.\end{equation} Then the final result reads \begin{equation}\label{eq:instcorrG}
F_{g,1,0}(d_1,0;d_1,0)=d_1^{2g-2}b_g, \qquad
F_{g,0,1}(0,d_2;0,d_2)=d_2^{2g-2}b_g\end{equation} where $b_g$ are the
Bernoulli numbers, and $g\geq 1$.

The contributions of special cases $(i)-(iv)$ be computed easily
by direct evaluation \begin{equation}\label{eq:instcorrH}\begin{array}{lc}
 &
F_{0,1,0}(d_1,0;d_1,0) = {1\over d_1^2} \cr & F_{0,0,1}(0,d_2;
0,d_2) = {1\over d_2^2} \cr & F_{0,2,0}(d_1,0;m_1,m_2,0,0)=
F_{0,0,2}(0,d_2; 0,0,n_1,n_2) = 0.\cr\end{array}\end{equation}
This leaves us with (\ref{eq:specialI}) corresponding to the {\it
fifth} case. Here we have again a factor of
$(\lambda_2-2\lambda_1)$ which gives a zero answer unless we have
similar factors in the denominator. The only monomial in the
denominator which can produce powers of $(\lambda_2-2\lambda_1)$
is $d_2(\lambda_1-\lambda_2) + d_1\lambda_1$. If $d_1=d_2=d$ this
reduces to $d(2\lambda_1-\lambda_2)$ cancelling the effect of the
similar factor in the denominator. Therefore we obtain
\begin{equation}\label{eq:instcorrI} F_{0,1,1}(d,d;d,d) = -{1\over
d}.\end{equation} It is easy to check that all other amplitudes
are zero.

In order to complete the description of the Chern-Simons system,
we have to specify the framing of the knots. The framing of
$\Gamma_1, \Gamma_2$ has been discussed below equation
(\ref{eq:framAB}). Since we have fixed $a=b=0$, we have framings
$(2,2)$. On the other hand, the framing of $\Xi_1, \Xi_2$ is not
fixed by the localization computation. The result
(\ref{eq:fixedpAB}) is true for any values of $(\lambda_1,
\lambda_2)$. This is actually an important consistency check of
the formalism, since one can see from (\ref{eq:toractB}) that once
we fix $(a,b)=(0,0)$ or $(a,b)=(2,2)$, there is no
$S^1$-equivariant choice of framings of $\Xi_1, \Xi_2$. Had such a
choice been necessary, we would have found an inconsistency
between the localization computations for $C$ and $D_1\cup D_2$.
However, we can determine the framing of $\Xi_1, \Xi_2$ using the
following deformation argument. As noted in the paragraph
containing equation (5.11) and also below (6.25), if $d$ is even,
the $d:1$ cover of $C$ and the $d/2:1$ cover of $D_1, D_2$ belong
to the same component of the moduli space $\om_{0,2}(Y,L,d\beta)$.
A natural assumption is that the homotopy class of the framing is
preserved under deformations of maps which do not change the
isotopy type of the knots in $L_1, L_2$. In the present case, this
shows that the framing of $\Xi_1, \Xi_2$ have to be equal half the
framing of $\Gamma_1, \Gamma_2$, that is $\left(1-{a\over 2},
1-{b\over 2}\right)$. This concludes the computation of open string
instanton corrections.

\appendix
\section{Some Geometric Facts}

In this appendix we elaborate on some technical points which have
been used without proof in section two. First, we would like to
check that the map $i:\hx\ra Z$ defined in (\ref{eq:embedding}) is
a well defined toric morphism. The second issue we would like to
address is the construction of a suitable symplectic K\"ahler form
of the deformed hypersurface $Y_\mu$ so that $L_1, L_2$ are
lagrangian cycles.

\subsection{The Toric Embedding}

Recall that the map $i:\hx\ra Z$ is defined in terms of
homogeneous coordinates by
\begin{equation}\label{eq:embeddingi}
Z_1 = X_2X_3X_4,\quad Z_2=X_1X_2, \quad Z_3=X_4X_5,\quad
U=X_0X_1X_5, \quad V=-X_0X_3.\end{equation} In order to check that
this is a well defined toric morphism, we have to show that it is
compatible with the toric actions and with the disallowed loci.
Note first that the monomials (\ref{eq:embeddingi}) transform
under the toric action (\ref{eq:toricA}) as
\begin{equation}\label{eq:toricC}
\begin{array}{ccccccc}
Z_1 & Z_2 & Z_3 & U & V\cr
0 & 0 & 0 & 0 & 0\cr
1 & 1 & 1 & -1 & -2\cr
0 & 0 & 0 & 0 & 0,\cr
\end{array}\end{equation}
which is indeed compatible with (\ref{eq:toricB}). In order to
check the disallowed loci, note that $\hx$ is defined by the toric
data (\ref{eq:toricA}), the K\"ahler parameters $\xi_1, \xi_3$
being set to zero in (\ref{eq:sympquotA}). This corresponds to a
partial triangulation of the polytope in fig. 1. obtained by
erasing the simplexes $v_0v_2$ and $v_0v_4$. Therefore we have to
show that given the moment map equations (\ref{eq:sympquotA}) with
$\xi_1=\xi_3=0$ and $\xi_2>0$, the monomials $Z_1,Z_2,Z_3$ cannot
vanish simultaneously. This is elementary, but we include the
details below for completeness. Let us first rewrite the equations
(\ref{eq:sympquotA}) in the form
\begin{equation}\label{eq:sympquotB}\begin{array}{lcc}
|X_2|^2 +|X_4|^2 - |X_3|^2 - |X_0|^2& = &\xi_2 \cr
|X_1|^2+|X_4|^2-2|X_0|^2& = &\xi_2\cr
|X_2|^2+|X_5|^2-2|X_0|^2& = &\xi_2\cr
|X_1|^2 + |X_3|^2 +|X_5|^2 - 3|X_0|^2& = &\xi_2.\cr\end{array}\end{equation}
The last equation is a linear combination of the first three, but it will
be needed in this particular form below.
Next, note that for $Z_2$ and $Z_3$ to vanish simultaneously, one of the
following conditions must be realized

\smallskip

\centerline{$a)$ $X_1=X_4=0$}

\smallskip

\centerline{$b)$ $X_1=X_5=0$}

\smallskip

\centerline{$c)$ $X_2=X_4=0$}

\smallskip

\centerline{$d)$ $X_2=X_5=0$}

It suffices to prove on a case by case basis that if any of these
conditions is realized, $Z_1$ cannot vanish. Note that $a)$,
$d)$ and $c)$ are immediately excluded by (\ref{eq:sympquotB}).
We are left with $b)$. If $X_1=X_5=0$, it follows from
(\ref{eq:sympquotB}) that $X_2, X_3, X_4$ are not allowed to
vanish, therefore $Z_1$ is not allowed to vanish. It is also easy
to check that any two of $Z_1,Z_2,Z_3$ are allowed to vanish at
the same time. The monomials $U,V$ are allowed to vanish
independently of $Z_1,Z_2,Z_3$ since they are multiples of $X_0$.
We conclude that the map (\ref{eq:embeddingi}) is well defined.

\subsection{The Symplectic Form}

Another claim made in section two is that we can
choose a symplectic K\"ahler form $\omega$ on $Y$
which agrees locally near the cycles $L_1, L_2$
with the standard symplectic form on a deformed conifold.
The following construction is a refinement of standard symplectic
surgery techniques \cite{EGH,JE}.

Since the argument is local, it suffices
to consider only one cycle, say $L_1$.
Since $Z$ is toric, it can be represented as a symplectic quotient
$\IC^5//U(1)$ with level sets
\begin{equation}\label{eq:sympquotC}
|Z_1|^2 + |Z_2|^2 + |Z_3|^2 +|U|^2 + |V|^2 = \xi\end{equation}
where $\xi\in \IR_+$. Therefore $Z$ is endowed with a
symplectic
K\"ahler form $\omega_0$ obtained by descent from the invariant form
\begin{equation}\label{eq:invform}
\Omega = {i\over 2} \left(dZ_1\wedge d{\overline Z}_1+ dZ_2\wedge
d{\overline Z}_2+dZ_3\wedge d{\overline Z}_3+ dU\wedge d{\overline
U}+ dV\wedge d{\overline V}\right).\end{equation} In the patch
$Z_3\neq 0$, we can rewrite (\ref{eq:sympquotB}) in terms of local
coordinates as
\begin{equation}\label{eq:sympquotD}
(|x_2|^2 + |y_2|^2 + 1)|Z_3|^6 +|u_2|^2|Z_3|^2 +
|v_2|^2 = \xi |Z_3|^4,\end{equation}
which can be interpreted as an equation of degree three in $|Z_3|^2$.
At this point, it is convenient to introduce polar coordinates
\begin{equation}\label{eq:polcoordA}
x_2 = r_x e^{i\theta_x},\qquad y_2 = r_y e^{i\theta_y}, \qquad u_2
= r_u e^{i\theta_u}, \qquad v_2 = r_v e^{i\theta_v}.\end{equation}
By solving for $|Z_3|^2$ in (\ref{eq:sympquotD}), we obtain
$|Z_3|^2 =F(r_x,r_y,r_u,r_v)$ for some real positive function $F$.
Then we can take a local transversal slice for the $U(1)$ action
on $\IC^5$ of the form
\begin{equation}\label{eq:locsliceA}\begin{array}{lc}
Z_1 = F(r_x,r_y,&r_u,r_v)r_xe^{i\theta_x},~~ Z_2 =
F(r_x,r_y,r_u,r_v)r_ye^{i\theta_y},~~ Z_3 =
F(r_x,r_y,r_u,r_v)r_x,\cr & U=
F(r_x,r_y,r_u,r_v)r_ue^{i\theta_u},~~ V =
F(r_x,r_y,r_u,r_v)r_ve^{i\theta_v}.\end{array}\end{equation}
Substituting (\ref{eq:locsliceA}) in (\ref{eq:invform}) we find
that $\omega_0$ has the local form
\begin{equation}\label{eq:locformA}
\omega_0|_{U_2} = \half d\left[F(r_x,r_y,r_u,r_v)(r_x^2 d\theta_x
+r_y^2d\theta_y+r_u^2d\theta_u+r_v^2d\theta_v)\right].\end{equation}
On the other hand the standard symplectic form $\omega_c$ can be written
\begin{equation}\label{eq:stformA}
\omega_c = {c\over 2} d\left(r_x^2d\theta_x + r_y^2d\theta_y+r_u^2d\theta_u
+r_v^2 d\theta_v\right)\end{equation}
where $c$ is a positive real constant.
Note that in polar coordinates the cycle $L_2$ is determined by
\begin{equation}\label{eq:polcoordB}
r_x=r_u,\quad\theta_x+\theta_u=0,\qquad
r_y=r_v, \quad \theta_y+\theta_v=0,\qquad
r_x^2+r_u^2=\mu.\end{equation}
Now let us consider the polycylinder $C_\epsilon(r)$ defined by
\begin{equation}\label{eq:polcyl}
\mu^{1/2} + r -\epsilon\leq r_x,r_y,r_u,r_v\leq
\mu^{1/2} + r +\epsilon.\end{equation}
In the following argument, we will also need the polydiscs
$\Delta(r-\epsilon)=\{0\leq r_x,r_y,r_u,r_v\leq
\mu^{1/2} + r -\epsilon\}$, $\Delta(r+\epsilon) =\{0\leq
r_x,r_y,r_u,r_v\leq\mu^{1/2} + r+\epsilon\}$.
The main idea is to construct a symplectic K\"ahler form $\omega$ on
$Z$ which interpolates smoothly between $\omega_c$ and $\omega_0$
over $C_\epsilon(r)$. Let $\rho:\IR_+\ra [0,1]$ be a decreasing
interpolating smooth function such that
\begin{displaymath}\label{eq:stepfct}
\rho(t) =
\left\{\begin{array}{lcccc}
&1,&&\hbox{for}\
0\leq t\leq \mu^{1/2} + r -\epsilon&\cr
&0,&&\hbox{for}\ \mu^{1/2} + r +\epsilon \leq t.\hfill&\end{array}\right.\end{displaymath}
Then we define the form
\begin{equation}\label{eq:sympformA}\begin{array}{lcc}
&\omega= \omega_0 \cr &+\half
d\left[(c-F(r_x,r_y,r_u,r_v))(\rho(r_x)r_x^2 d\theta_x
+\rho(r_y)r_y^2d\theta_y+\rho(r_u)r_u^2d\theta_u
+\rho(r_v)r_v^2d\theta_v)\right].\end{array}\end{equation} It is
straightforward to check that $\omega$ agrees with $\omega_c$ in
the polydisc $\Delta(r-\epsilon)$ and with $\omega_0$ on the
complement of $\Delta(r+\epsilon)$. Since $\omega$ is also closed,
in order to complete the argument, we have to check it is a
K\"ahler form, that is $\omega(\psi, J\psi)\geq 0$ for any tangent
vector $\psi$ to $Z$. It is clear that it suffices to check that
$\omega$ is positive definite over $C_\epsilon(r)$. A
straightforward computation shows that
\begin{equation}\label{eq:sympformBi}\begin{array}{lcc}
&\omega|_{C_\epsilon(r)} =
 \rho(r_x) r_xdr_x\wedge d\theta_x +
\rho(r_y) r_ydr_y\wedge d\theta_y +
\rho(r_u) r_u dr_u\wedge d\theta_u+
\rho(r_v) r_v dr_v\wedge d\theta_v \cr
&+ (1-\rho(r_x))d(r_x^2F)\wedge d\theta_x
+ (1-\rho(r_y))d(r_y^2F)\wedge d\theta_y
+ (1-\rho(r_u))d(r_u^2F)\wedge d\theta_u\cr
&+ (1-\rho(r_v))d(r_v^2F)\wedge d\theta_v+ (c-F)\big[\rho'(r_x)r_x^2dr_x\wedge d\theta_x +
\rho'(r_y)r_y^2dr_y\wedge d\theta_y\cr &+
\rho'(r_u)r_u^2dr_u\wedge d\theta_u +\rho'(r_v)r_v^2dr_v\wedge d\theta_v \big].\cr\end{array}\end{equation}
Moreover, the complex structure $J$ on $U_2$ is given locally by
\begin{equation}\label{eq:cpxstrA}
J(\partial_{r_x}) = {1\over r_x} \partial_{\theta_x},\qquad
J(\partial_{\theta_x}) = -r_x \partial_{r_x}\end{equation} and
similar expressions for the other coordinates. Then one can check
that the terms in the first three lines of (\ref{eq:sympformBi})
are positive definite since $\omega_0, \omega_c$ must be positive
definite. Therefore it suffices to check that the form defined by
the last two lines of (\ref{eq:sympformBi}) is positive
semi-definite. At this point, recall that we have chosen $\rho(t)$
a decreasing function, hence $\rho'(t)\leq 0$. Given the explicit
form of the complex structure, it is easy to check that the form
in question is positive semi-definite if $F(r_x,r_y,r_u,r_v)-c
\leq 0$. Since the polycylinder is compact, this can be achieved
throughout $C_\epsilon(r)$ if we choose $c$ sufficiently small.
Therefore, after a rescaling of the standard symplectic form
$\omega_c$ by a sufficiently small positive constant, $\omega$ is
a symplectic K\"ahler form on $Z$. By construction $\omega$ agrees
with $\omega_c$ in a neighborhood of $L_1$, hence $L_1$ is a
lagrangian cycle. It is now obvious that we can perform the same
construction for $L_2$.

\section{Degree $3$ and $4$ Chern-Simons Computation}

Let us consider now the third order terms in (\ref{eq:freenEC}).
By successively expanding the exponential and the logarithm, we
obtain
\begin{equation}\label{eq:thirdorder}\begin{array}{lc}
{\cal F}_{inst}&(g_s,t_1,t_2,t_c,\lambda_1,\lambda_2)^{(3)}
=e^{-3t_1}X_{(3t_1)}+e^{-3t_2}X_{(3t_2)}+e^{-3t_c}X_{(3t_c)}\cr
&+e^{-2t_1-t_2}X_{(2t_1,t_2)}
+e^{-t_1-2t_2}X_{(t_1,2t_2)}
+e^{-t_1-t_2-t_c}X_{(t_1,t_2,t_c)}+e^{-2t_1-t_c}X_{(2t_1,t_c)}\cr
&+e^{-2t_2-t_c}X_{(2t_2,t_c)}+e^{-t_1-2t_c}X_{(t_1,2t_c)}+e^{-t_2-2t_c}X_{(t_2,2t_c)},
\end{array}\end{equation}
where
\begin{equation}\label{eq:thexes}\begin{array}{lc}
&X_{(3t_i)}=x_{(3t_i)}-x_{(2t_i)}x_{(t_i)}+{1\over 3}x_{(t_i)}^3,~~~i=1,2,c,\cr
&X_{(2t_i,t_j)}=x_{(2t_i,t_j)}-x_{(2t_i)}x_{(t_j)}-x_{(t_i,t_j)}x_{(t_i)}+x_{(t_i)}^2x_{(t_j)} ~~~i,j=1,2,c,
~i\neq j,\cr
&X_{(t_1,t_2,t_c)}=x_{(t_1,t_2,t_c)}-x_{(t_1,t_2)}x_{(t_c)}-x_{(t_1,t_c)}x_{(t_2)}-x_{(t_2,t_c)}x_{(t_1)}
+2x_{(t_1)}x_{(t_2)}x_{(t_c)},
\end{array}\end{equation}
with $x_{(\cdots )}$ defined as in (\ref{eq:degonebbb}), (\ref{eq:degtwoccc}) and
\begin{equation}\label{eq:thexesthreei}\begin{array}{lc}
&x_{(3t_1)}=i\ll a_3\r -\ll a_1a_2\r -{i\over 6}\ll a_1^3\r
,~~~x_{(3t_2)}=i\ll b_3\r -\ll b_1b_2\r -{i\over 6}\ll b_1^3\r ,\cr
&x_{(3t_c)}=-\ll c_3\r +\ll c_1c_2\r -{1\over 6}\ll c_1^3\r ,~~~
x_{(2t_1,t_2)}=-\ll a_2b_1\r +2i\ll a_1d_1\r -{i\over 2}\ll a_1^2b_1\r
,\cr
&x_{(t_1,2t_2)}=-\ll a_1b_2\r +2i\ll b_1d_1\r -{i\over 2}\ll
a_1b_1^2\r ,~~~x_{(t_1,t_2,t_c)}=\ll a_1b_1c_1\r -2\ll c_1d_1\r ,\cr
&x_{(2t_1,t_c)}=-i\ll a_2c_1\r +\ot\ll a_1^2c_1\r ,~~~x_{(2t_2,t_c)}
=-i\ll b_2c_1\r +\ot\ll b_1^2c_1\r ,\cr
&x_{(t_1,2t_c)}=-i\ll a_1c_2\r +{i\over 2}\ll a_1c_1^2\r ,~~~x_{(t_2,2t_c)}=-i\ll b_1c_2\r 
+{i\over 2}\ll b_1c_1^2\r.
\end{array}\end{equation}
First, we evaluate $X_{(3t_1)}$. In
doing so, we have to use again the Frobenius formula in order to
linearize cubic expressions in the holonomy variables. We have
\begin{equation}\label{eq:tracesii}\begin{array}{lc}
&\left(\Tr V_1\right)^3=\Tr_{\tableau{3}}V_1+2\Tr_{\tableau{2
1}}V_1+ \Tr_{\tableau{1 1 1}}V_1\cr &\Tr V_1\Tr
V_1^2=\Tr_{\tableau{3}}V_1-\Tr_{\tableau{1 1 1}}V_1\cr &\Tr
V_1^3=\Tr_{\tableau{3}}V_1-\Tr_{\tableau{2 1}}V_1+ \Tr_{\tableau{1
1 1}}V_1 .\end{array}\end{equation} Applying (\ref{eq:framdepA})
we have
\begin{equation}\label{eq:tracesexpval}\begin{array}{lc}
&\left<\left(\Tr V_1\right)^3\right>=e^{{3\over 2}i\l_1}e^{3ig_s}\left<\Tr_{\tableau{3}}V_1\right>_0
+2e^{3ig_s}\left<\Tr_{\tableau{2 1}}V_1\right>_0+
e^{-{3\over 2}i\l_1}e^{3ig_s}\left<\Tr_{\tableau{1 1 1}}V_1\right>_0\cr
&\left<\Tr V_1\Tr V_1^2\right>=e^{{3\over 2}i\l_1}e^{3ig_s}\left<\Tr_{\tableau{3}}V_1\right>_0
-e^{-{3\over 2}i\l_1}e^{3ig_s}\left<\Tr_{\tableau{1 1 1}}V_1\right>_0\cr
&\left<\Tr V_1^3\right>=e^{{3\over 2}i\l_1}e^{3ig_s}\left<\Tr_{\tableau{3}}V_1\right>_0
-e^{3ig_s}\left<\Tr_{\tableau{2 1}}V_1\right>_0+
e^{-{3\over 2}i\l_1}e^{3ig_s}\left<\Tr_{\tableau{1 1 1}}V_1\right>_0
.\end{array}\end{equation}
The expectation values in the canonical framing are given by
\begin{equation}\label{eq:zerotracesresthree}\begin{array}{lc}
&\left<\Tr_{\tableau{3}}V_1\right>_0={(y-y^{-1})(yx-y^{-1}x^{-1})(yx^2-y^{-1}x^{-2})\over
(x-x^{-1})(x^2-x^{-2})(x^3-x^{-3})},\cr
&\left<\Tr_{\tableau{2 1}}V_1\right>_0={(x+x^{-1})(y-y^{-1})(yx-y^{-1}x^{-1})(yx^{-1}-y^{-1}x)\over
(x-x^{-1})(x^2-x^{-2})(x^3-x^{-3})},\cr
&\left<\Tr_{\tableau{1 1 1}}V_1\right>_0={(y-y^{-1})(yx^{-1}-y^{-1}x)(yx^{-2}-y^{-1}x^2)\over
(x-x^{-1})(x^2-x^{-2})(x^3-x^{-3})},
\end{array}\end{equation}
where $x=e^{\ot ig_s}$ and $y=e^{\ot i\l_1}$. Using
(\ref{eq:tracesii})-(\ref{eq:zerotracesresthree}), a
straightforward computation gives
\begin{equation}\label{eq:thexesi}
X_{(3t_1)}={e^{{5\over 2}i\lambda_1}\over (2{\rm sin}{g_s\over 2})^2}
(e^{i\lambda_1}-1)(2e^{i\lambda_1}-1)+{1\over 3(2{\rm
sin}{3g_s\over 2})^2}(-e^{{3\over 2}i\lambda_1}+e^{{9\over
2}i\lambda_1})+e^{{7\over 2}i\lambda_1}-e^{{9\over 2}i\lambda_1}.\end{equation}
By symmetry, we can therefore write
\begin{equation}\label{eq:thexesii}
X_{(3t_2)}={e^{{5\over 2}i\lambda_2}\over (2{\rm sin}{g_s\over 2})^2}
(e^{i\lambda_2}-1)(2e^{i\lambda_2}-1)+{1\over 3(2{\rm
sin}{3g_s\over 2})^2}(-e^{{3\over 2}i\lambda_2}+e^{{9\over
2}i\lambda_2})+e^{{7\over 2}i\lambda_2}-e^{{9\over 2}i\lambda_2}.\end{equation}
A similar computation gives
\begin{equation}\label{eq:thexesiii}
X_{(3t_c)}={1\over 3(2{\rm sin}{3g_s\over 2})^2}(e^{{3\over
2}i\lambda_1}- e^{-{3\over 2}i\lambda_1})(e^{{3\over
2}i\lambda_2}- e^{-{3\over 2}i\lambda_2}).\end{equation} Next,
$X_{(2t_1,t_2)}$, $X_{(t_1,2t_2)}$ and $X_{(t_1,t_2,t_c)}$ can be
easily computed using (\ref{eq:expvalA}), (\ref{eq:expvalB}),
(\ref{eq:expvalBC}), (\ref{eq:expvalD}) and the exchange symmetry.
The results read
\begin{equation}\label{eq:thexesiiitwo}\begin{array}{lc}
&X_{(2t_1,t_2)}=2{e^{i({3\over 2}\lambda_1+\lambda_2)}\over (2{\rm
sin}{g_s\over 2})^2}(e^{{1\over 2}i\lambda_1}-
e^{-{1\over 2}i\lambda_1})(e^{{1\over 2}i\lambda_2}-
e^{-{1\over 2}i\lambda_2})(3e^{i\lambda_1}-1)\cr
&~~~~~~~~~~~-2e^{{5\over
2}i\lambda_1+i\lambda_2}(e^{{1\over 2}i\lambda_1}
-e^{-{1\over 2}i\lambda_1})(e^{{1\over 2}i\lambda_2}-
e^{-{1\over 2}i\lambda_2}),\cr
&X_{(t_1,2t_2)}=2{e^{i(\lambda_1+{3\over 2}\lambda_2)}\over (2{\rm
sin}{g_s\over 2})^2}(e^{{1\over 2}i\lambda_1}-
e^{-{1\over 2}i\lambda_1})(e^{{1\over 2}i\lambda_2}-
e^{-{1\over 2}i\lambda_2})(3e^{i\lambda_2}-1)\cr
&~~~~~~~~~~~-2e^{{5\over
2}i\lambda_2+i\lambda_1}(e^{{1\over 2}i\lambda_1}
-e^{-{1\over 2}i\lambda_1})(e^{{1\over 2}i\lambda_2}-
e^{-{1\over 2}i\lambda_2}),\cr
&X_{(t_1,t_2,t_c)}={e^{\ot i(\lambda_1+\lambda_2)}\over (2{\rm
sin}{g_s\over 2})^2}(e^{{1\over 2}i\lambda_1}-
e^{-{1\over 2}i\lambda_1})(e^{{1\over 2}i\lambda_2}-
e^{-{1\over
2}i\lambda_2})(5e^{i\lambda_1+i\lambda_2}-2e^{i\lambda_1}-2e^{i\lambda_2})\cr
&~~~~~~~~~~~~-2e^{{3\over 2}i\lambda_1+{3\over 2}i\lambda_2}(e^{{1\over 2}i\lambda_1}-
e^{-{1\over 2}i\lambda_1})(e^{{1\over 2}i\lambda_2}-
e^{-{1\over 2}i\lambda_2})
.\end{array}\end{equation}
Next, in order to evaluate $X_{(2t_1,t_c)}$ and $X_{(2t_2,t_c)}$ we
need to compute expectation values of the form $\langle (\Tr V_i)^2\Tr
U_i\rangle$, $\langle \Tr V_i^2\Tr U_i\rangle$ for the Hopf links
$(\Gamma_i,\Xi_i)$ with $i=1,2$. Again, due to the exchange symmetry,
it is sufficient to compute the above expectation values for the first
link only. After taking into account the framing correction we have
\begin{equation}\label{eq:thelinksii}\begin{array}{lc}
&\langle (\Tr V_1)^2\Tr U_1\rangle
=e^{2ig_s}e^{2i\l_1}\ll\Tr_{\tableau{2}}V_1\Tr U_1\r_0+e^{-2ig_s}e^{2i\l_1}\ll\Tr_{\tableau{1
1}}V_1\Tr U_1\r_0,\\
&\langle \Tr V_1^2\Tr U_1\rangle
=e^{2ig_s}e^{2i\l_1}\ll\Tr_{\tableau{2}}V_1\Tr U_1\r_0-e^{-2ig_s}e^{2i\l_1}\ll\Tr_{\tableau{1
1}}V_1\Tr U_1\r_0.
\end{array}\end{equation}
Now, for a Hopf link with linking number $-1$ we have
\begin{equation}\label{eq:Hopflinkcan}
\ll\Tr_{R_1}V_1\Tr_{R_2}V_2\r_0 =e^{{ig_s\over
2}(k_{R_1}+k_{R_2})}\sum_{\rho\in R_1\otimes R_2}e^{-{ig_s\over 2}k_{\rho}}{\rm
dim}_q\rho ,
\end{equation}
where the sum is after all the representations $\rho$ occurring in
the decomposition of the tensor product of $R_1$ and $R_2$ and
${\rm dim}_q\rho$ is the quantum dimension of the representation
$\rho$ \cite{MV} . Now, using (\ref{eq:expvalA}),
(\ref{eq:expvalB}), (\ref{eq:expvalBC}), (\ref{eq:expvalD}),
(\ref{eq:thelinksii}) and (\ref{eq:Hopflinkcan}) and keeping in
mind that in our case the linking number is $+1$ we obtain after a
simple computation
\begin{equation}\label{eq:thexesiiii}\begin{array}{lc}
&X_{(2t_1,t_c)}
={e^{2i\lambda_1}\over(2{\rm sin}{g_s\over
2})^2}(3e^{i\lambda_1}-1)(e^{{1\over 2}i\lambda_1}-
e^{-{1\over 2}i\lambda_1})(e^{{1\over 2}i\lambda_2}-
e^{-{1\over 2}i\lambda_2})\cr
&\qquad\quad~~ -e^{3i\lambda_1}
(e^{{1\over 2}i\lambda_1}-e^{-{1\over 2}i\lambda_1})
(e^{{1\over 2}i\lambda_2}-e^{-{1\over 2}i\lambda_2}).
\end{array}\end{equation}
By the exchange symmetry, we have
\begin{equation}\label{eq:thexesiiiii}\begin{array}{lc}
&X_{(2t_2,t_c)}={e^{2i\lambda_2}\over(2{\rm sin}{g_s\over
2})^2}(3e^{i\lambda_2}-1)(e^{{1\over 2}i\lambda_1}-
e^{-{1\over 2}i\lambda_1})(e^{{1\over 2}i\lambda_2}-
e^{-{1\over 2}i\lambda_2})\cr
&\qquad\quad~~ -e^{3i\lambda_2}
(e^{{1\over 2}i\lambda_1}-e^{-{1\over 2}i\lambda_1})
(e^{{1\over 2}i\lambda_2}-e^{-{1\over 2}i\lambda_2}).
\end{array}\end{equation}
Finally, in order to compute $X_{(t_1,2t_c)}$ and $X_{(t_2,2t_c)}$ we
need to evaluate expectation values of the form
$\langle \Tr V_i(\Tr U_i)^2\rangle$, $\langle \Tr V_i\Tr U_i^2\rangle$
for the Hopf links $(\Gamma_i,\Xi_i)$ with $i=1,2$. Taking into
account the framing correction we have, for one of the links,
\begin{equation}\label{eq:thelinksiiv}\begin{array}{lc}
&\langle \Tr V_1(\Tr U_1)^2\rangle
=e^{i\l_1}(\ll\Tr V_1\Tr_{\tableau{2}} U_1\r_0+\ll\Tr V_1\Tr_{\tableau{1
1}} U_1\r_0),\cr
&\langle \Tr V_1\Tr U_1^2\rangle
=e^{i\l_1}(\ll\Tr V_1\Tr_{\tableau{2}} U_1\r_0-\ll\Tr V_1\Tr_{\tableau{1
1}} U_1\r_0).
\end{array}\end{equation}
Now, using (\ref{eq:expvalA}), (\ref{eq:expvalB}),
(\ref{eq:expvalBC}), (\ref{eq:expvalD}), (\ref{eq:Hopflinkcan})
and (\ref{eq:thelinksiiv}) and the exchange symmetry, we obtain
after another direct computation
\begin{equation}\label{eq:thexesiiiiii}\begin{array}{lc}
&X_{(t_1,2t_c)}=-{e^{i(2\lambda_1-{1\over 2}\lambda_2)}\over (2{\rm sin}{g_s\over
2})^2}(e^{{1\over 2}i\lambda_1}-e^{-{1\over 2}i\lambda_1})(e^{{1\over 2}i\lambda_2}-
e^{-{1\over 2}i\lambda_2}),\cr
&X_{(t_2,2t_c)}=-{e^{i(-{1\over 2}\lambda_1+2\lambda_2)}\over (2{\rm sin}{g_s\over
2})^2}(e^{{1\over 2}i\lambda_1}-e^{-{1\over 2}i\lambda_1})(e^{{1\over 2}i\lambda_2}-
e^{-{1\over 2}i\lambda_2}).
\end{array}\end{equation}
Let us consider now the fourth order terms in (\ref{eq:freenEC}).
By successively expanding the exponential and the logarithm, we
obtain
\begin{equation}\label{eq:fourthorder}\begin{array}{lc}
{\cal F}_{inst}&(g_s,t_1,t_2,t_c,\lambda_1,\lambda_2)^{(4)}=
e^{-4t_1}X_{(4t_1)}+e^{-4t_2}X_{(4t_2)}+e^{-4t_c}X_{(4t_c)}\cr
&+e^{-3t_1-t_2}X_{(3t_1,t_2)}+e^{-2t_1-2t_2}X_{(2t_1,2t_2)}+e^{-t_1-3t_2}X_{(t_1,3t_2)}
+e^{-3t_1-t_c}X_{(3t_1,t_c)}\cr
&+e^{-2t_1-t_2-t_c}X_{(2t_1,t_2,t_c)}+e^{-t_1-2t_2-t_c}X_{(t_1,2t_2,t_c)}+
e^{-3t_2-t_c}X_{(3t_2,t_c)}\cr
&+e^{-2t_1-2t_c}X_{(2t_1,2t_c)}+e^{-t_1-t_2-2t_c}X_{(t_1,t_2,2t_c)}+e^{-2t_2-2t_c}X_{(2t_2,2t_c)}\cr
&+e^{-t_1-3t_c}X_{(t_1,3t_c)}+e^{-t_2-3t_c}X_{(t_2,3t_c)},
\end{array}\end{equation}
where
\begin{equation}\label{eq:fourthorderi}\begin{array}{lc}
&X_{(4t_i)}=x_{(4t_i)}-x_{(3t_i)}x_{(t_i)} -\ot
x_{(2t_i)}^2+x_{(2t_i)}x_{(t_i)}^2-{1\over 4}x_{(t_i)}^4,~~~i=1,2,c,\cr
&X_{(3t_i,t_j)}=x_{(3t_i,t_j)}-x_{(3t_i)}x_{(t_j)}-x_{(2t_i,t_j)}x_{(t_i)}
-x_{(2t_i)}x_{(t_i,t_j)}+2x_{(2t_i)}x_{(t_i)}x_{(t_j)}\cr
&\qquad\quad~~-x_{(t_i)}^3x_{(t_j)}+x_{(t_i)}^2x_{(t_i,t_j)},~~~i,j=1,2,c,~
i\neq j,\cr
&X_{(2t_i,2t_j)}=x_{(2t_i,2t_j)}-x_{(2t_i)}x_{(2t_j)}-x_{(2t_i,t_j)}x_{(t_j)}-x_{(t_i,2t_j)}x_{(t_i)}
+x_{(2t_i)}x_{(t_j)}^2\cr
&\qquad\qquad +x_{(2t_j)}x_{(t_i)}^2-\ot x_{(t_i,t_j)}^2-{3\over
2}x_{(t_i)}^2x_{(t_j)}^2+2x_{(t_i,t_j)}x_{(t_i)}x_{(t_j)},\cr
&\qquad\qquad {\rm for}~~i,j=1,2,c,~i\neq j,\cr
&X_{(2t_i,t_j,t_k)}=x_{(2t_i,t_j,t_k)}-x_{(2t_i,t_j)}x_{(t_k)}-x_{(2t_i,t_k)}x_{(t_j)}
-x_{(2t_i)}x_{(t_j,t_k)}+2x_{(2t_i)}x_{(t_j)}x_{(t_k)}\cr
&\qquad\qquad~-x_{(t_i,t_j,t_k)}x_{(t_i)}+x_{(t_i)}^2x_{(t_j,t_k)}
-3x_{(t_i)}^2x_{(t_j)}x_{(t_k)}-x_{(t_i,t_j)}x_{(t_i,t_k)}\cr
&\qquad\qquad~+2x_{(t_i,t_j)}x_{(t_i)}x_{(t_k)}+2x_{(t_i,t_k)}x_{(t_i)}x_{(t_j)},\cr
&\qquad\qquad {\rm for}~~i,j,k=1,2,c,~i\neq j,~j\neq k,~k\neq i.
\end{array}\end{equation}
with
\begin{equation}\label{eq:fourthorderiii}\begin{array}{lc}
&x_{(4t_1)}=i\ll a_4\r-\ll a_1a_3\r -\ot\ll a_2^2\r-{i\over 2}\ll
a_1^2a_2\r+{1\over 24}\ll a_1^4\r ,\cr &x_{(4t_2)}=i\ll b_4\r-\ll
b_1b_3\r -\ot\ll b_2^2\r-{i\over 2}\ll b_1^2b_2\r+{1\over 24}\ll
b_1^4\r ,\cr &x_{(4t_c)}=-\ll c_4\r+\ll c_1c_3\r +\ot\ll
c_2^2\r-{1\over 2}\ll c_1^2c_2\r+{1\over 24}\ll c_1^4\r ,\cr
&x_{(3t_1,t_2)}=-\ll a_3b_1\r +2i\ll a_2d_1\r -i\ll a_1a_2b_1\r
-\ll a_1^2d_1\r +{1\over 6}\ll a_1^3b_1\r ,\cr
&x_{(2t_1,2t_2)}=-\ll a_2b_2\r +2\ll d_2\r -{i\over 2}\ll
a_1^2b_2\r -{i\over 2}\ll a_2b_1^2\r +2\ll d_1^2\r +{1\over 4}\ll
a_1^2b_1^2\r -2\ll a_1b_1d_1\r ,\cr &x_{(t_1,3t_2)}=-\ll a_1b_3\r
+2i\ll b_2d_1\r -i\ll a_1b_1b_2\r -\ll b_1^2d_1\r +{1\over 6}\ll
a_1b_1^3\r ,\cr &x_{(3t_1,t_c)}=-i\ll a_3c_1\r +\ll a_1a_2c_1\r
+{i\over 6}\ll a_1^3c_1\r ,\cr & x_{(3t_2,t_c)}=-i\ll b_3c_1\r
+\ll b_1b_2c_1\r +{i\over 6}\ll b_1^3c_1\r ,\cr
&x_{(2t_1,t_2,t_c)}=\ll a_2b_1c_1\r-2i\ll a_1c_1d_1\r +{i\over
2}\ll a_1^2b_1c_1\r ,\cr &x_{(t_1,2t_2,t_c)}=\ll a_1b_2c_1\r-2i\ll
b_1c_1d_1\r +{i\over 2}\ll a_1b_1^2c_1\r ,\cr
&x_{(2t_1,2t_c)}=-i\ll a_2c_2\r +{i\over 2}\ll a_2c_1^2\r+\ot\ll
a_1^2c_2\r -{1\over 4}\ll a_1^2c_1^2\r ,\cr
&x_{(t_1,t_2,2t_c)}=\ll a_1b_1c_2\r -2\ll c_2d_1\r -\ot\ll
a_1b_1c_1^2\r +\ll c_1^2d_1\r ,\cr &x_{(2t_2,2t_c)}=-i\ll b_2c_2\r
+{i\over 2}\ll b_2c_1^2\r+\ot\ll b_1^2c_2\r -{1\over 4}\ll
b_1^2c_1^2\r ,\cr &x_{(t_1,3t_c)}=-i\ll a_1c_3\r +i\ll a_1c_1c_2\r
-{i\over 6}\ll a_1c_1^3\r ,\cr & x_{(t_2,3t_c)}=-i\ll b_1c_3\r
+i\ll b_1c_1c_2\r -{i\over 6}\ll b_1c_1^3\r
\end{array}\end{equation}
and the other $x_{(\cdots )}$ defined as in (\ref{eq:degonebbb}),
(\ref{eq:degtwoccc}) and (\ref{eq:thexesthreei}).\ First, we
evaluate $X_{(4t_1)}$. In doing so, we have to use again the
Frobenius formula in order to linearize quartic expressions in the
holonomy variables. 
We have
\begin{equation}\label{eq:tracesexpvalfour}\begin{array}{lc}
&\left(\Tr V\right)^4=\Tr_{\tableau{4}}V+3\Tr_{\tableau{3 1}}V+
2\Tr_{\tableau{2 2}}V+3\Tr_{\tableau{2 1
1}}V+ \Tr_{\tableau{1 1 1 1}}V ,\cr
&\left(\Tr V\right)^2\Tr V^2=\Tr_{\tableau{4}}V+\Tr_{\tableau{3 1}}V-\Tr_{\tableau{2 1
1}}V-\Tr_{\tableau{1 1 1 1}}V ,\cr
&\Tr V^2\Tr V^2=\Tr_{\tableau{4}}V-\Tr_{\tableau{3 1}}V+
2\Tr_{\tableau{2 2}}V-\Tr_{\tableau{2 1
1}}V+ \Tr_{\tableau{1 1 1 1}}V,\cr
&\Tr V\Tr
V^3=\Tr_{\tableau{4}}V-\Tr_{\tableau{2
2}}V+\Tr_{\tableau{1 1 1 1}}V ,\cr
&\Tr V^4=\Tr_{\tableau{4}}V-\Tr_{\tableau{3 1}}V+\Tr_{\tableau{2 1
1}}V-\Tr_{\tableau{1 1 1 1}}V.
\end{array}\end{equation}
The expectation values in the canonical framing are given by
\begin{equation}\label{eq:zerotracesresfour}\begin{array}{lc}
&\left<\Tr_{\tableau{4}}V_1\right>_0={(y-y^{-1})(yx-y^{-1}x^{-1})(yx^2-y^{-1}x^{-2})(yx^3-y^{-1}x^{-3})\over
(x-x^{-1})(x^2-x^{-2})(x^3-x^{-3})(x^4-x^{-4})},\cr
&\left<\Tr_{\tableau{3
1}}V_1\right>_0={(y-y^{-1})(yx-y^{-1}x^{-1})(yx^2-y^{-1}x^{-2})(yx^{-1}-y^{-1}x)
\over (x-x^{-1})^2(x^2-x^{-2})(x^4-x^{-4})},\cr
&\left<\Tr_{\tableau{2
2}}V_1\right>_0={(y-y^{-1})^2(yx-y^{-1}x^{-1})(yx^{-1}-y^{-1}x)\over
(x-x^{-1})(x^2-x^{-2})^2(x^3-x^{-3})},\cr
&\left<\Tr_{\tableau{2 1
1}}V_1\right>_0={(y-y^{-1})(yx-y^{-1}x^{-1})(yx^{-1}-y^{-1}x)(yx^{-2}-y^{-1}x^2) \over
(x-x^{-1})^2(x^2-x^{-2})(x^4-x^{-4})},\cr
&\left<\Tr_{\tableau{1 1 1 1}}V_1\right>_0={(y-y^{-1})(yx^{-1}-y^{-1}x)(yx^{-2}-y^{-1}x^2)(yx^{-3}-y^{-1}x^3)\over (x-x^{-1})(x^2-x^{-2})(x^3-x^{-3})(x^4-x^{-4})},
\end{array}\end{equation}
where $x=e^{\ot ig_s}$ and $y=e^{\ot i\l_1}$. We also have
\begin{equation}\label{eq:theks}
k_{\tableau{4}}=12,~~k_{\tableau{3 1}}=4,~~k_{\tableau{2
2}}=0,~~k_{\tableau{2 1 1}}=-4,~~k_{\tableau{1 1 1 1}}=
-12.\end{equation} Now, using (\ref{eq:tracesexpvalfour}) ,
(\ref{eq:zerotracesresfour}) and (\ref{eq:theks}) we obtain
\begin{equation}\label{eq:thexesfoura}\begin{array}{lc}
&X_{(4t_1)}={e^{3i\lambda_1}\over (2{\rm sin}{g_s\over 2})^2}(e^{i\l_1}-1)(-7e^{2i\l_1}+6e^{i\l_1}-1)
+{1\over 2(2{\rm sin}g_s)^2}(e^{4i\l_1}-e^{6i\l_1})\cr
&\qquad~~+{1\over 4(2{\rm
sin}{2g_s})^2}(-e^{2i\l_1}+e^{6i\l_1})+e^{4i\l_1}(e^{i\l_1}-1)(11e^{i\l_1}-5)\cr
&\qquad~~+\left(2{\rm sin}{g_s\over 2}\right)^2e^{4i\l_1}(e^{i\l_1}-1)(-6e^{i\l_1}+1)+\left(2{\rm sin}{g_s\over 2}\right)^4(-e^{5i\l_1}+e^{6i\l_1}),
\end{array}\end{equation}
and, by the exchange symmetry,
\begin{equation}\label{eq:thexesfourb}\begin{array}{lc}
&X_{(4t_2)}={e^{3i\lambda_2}\over (2{\rm sin}{g_s\over 2})^2}(e^{i\l_2}-1)(-7e^{2i\l_2}+6e^{i\l_2}-1)
+{1\over 2(2{\rm sin}g_s)^2}(e^{4i\l_2}-e^{6i\l_2})\cr
&\qquad~~+{1\over 4(2{\rm
sin}{2g_s})^2}(-e^{2i\l_2}+e^{6i\l_2})+e^{4i\l_2}(e^{i\l_2}-1)(11e^{i\l_2}-5)\cr
&\qquad~~+\left(2{\rm sin}{g_s\over 2}\right)^2e^{4i\l_2}(e^{i\l_2}-1)(-6e^{i\l_2}+1)+\left(2{\rm sin}{g_s\over 2}\right)^4(-e^{5i\l_2}+e^{6i\l_2}).
\end{array}\end{equation}
Another straightforward calculation yields
\begin{equation}\label{eq:thexesfourc}\begin{array}{lc}
X_{(4t_c)}={1\over 4(2{\rm sin}{2g_s})^2}(e^{2i\lambda_1}-
e^{-2i\lambda_1})(e^{2i\lambda_2}-
e^{-2i\lambda_2}).\end{array}\end{equation} Now, using
(\ref{eq:expvalA}), (\ref{eq:expvalB}), (\ref{eq:expvalBC})
(\ref{eq:expvalD}), (\ref{eq:tracesexpval}) and
(\ref{eq:zerotracesresthree}),  we get
\begin{equation}\label{eq:thexesfourd}\begin{array}{lc}
&X_{(3t_1,t_2)}={2e^{i(2\l_1 +\l_2)}\over(2{\rm sin}{g_s\over 2})^2}
(e^{\ot i\l_1}-e^{-\ot i\l_1})
(e^{\ot i\l_2}-e^{-\ot i\l_2})(-13e^{2i\l_1}+9e^{i\l_1}-1)\cr
&\qquad~~~~~+4e^{i(3\l_1+\l_2)}(e^{\ot i\l_1}-e^{-\ot
i\l_1})(e^{\ot i\l_2}-e^{-\ot
i\l_2})(8e^{i\l_1}-1)\cr
&\qquad~~~~~+2\left(2{\rm sin}{g_s\over 2}\right)^2e^{i(3\l_1+\l_2)}(e^{\ot i\l_1}-e^{-\ot
i\l_1})(e^{\ot i\l_2}-e^{-\ot
i\l_2})(-7e^{i\l_1}+1)\cr
&\qquad~~~~~+2\left(2{\rm sin}{g_s\over 2}\right)^4e^{i(4\l_1 +\l_2)}
(e^{\ot i\l_1}-e^{-\ot i\l_1})(e^{\ot i\l_2}-e^{-\ot
i\l_2}),\end{array}\end{equation}
and, by the exchange symmetry,
\begin{equation}\label{eq:thexesfourdii}\begin{array}{lc}
&X_{(t_1,3t_2)}={2e^{i(\l_1+2\l_2)}\over(2{\rm sin}{g_s\over 2})^2}
(e^{\ot i\l_1}-e^{-\ot i\l_1})
(e^{\ot i\l_2}-e^{-\ot i\l_2})(-13e^{2i\l_2}+9e^{i\l_2}-1)\cr
&\qquad~~~~~+4e^{i(\l_1+3\l_2)}(e^{\ot i\l_1}-e^{-\ot
i\l_1})(e^{\ot i\l_2}-e^{-\ot
i\l_2})(8e^{i\l_2}-1)\cr
&\qquad~~~~~+2\left(2{\rm sin}{g_s\over 2}\right)^2e^{i(\l_1+3\l_2)}(e^{\ot i\l_1}-e^{-\ot
i\l_1})(e^{\ot i\l_2}-e^{-\ot
i\l_2})(-7e^{i\l_2}+1)\cr
&\qquad~~~~~+2\left(2{\rm sin}{g_s\over 2}\right)^4e^{i(\l_1 +4\l_2)}
(e^{\ot i\l_1}-e^{-\ot i\l_1})(e^{\ot i\l_2}-e^{-\ot
i\l_2}).\cr
\end{array}\end{equation}
\newpage
\noindent Next, a similar computation gives
\begin{equation}\label{eq:thexesfourdi}\begin{array}{lc}
&X_{(2t_1,2t_2)}\cr &={2e^{{3\over 2}i(\l_1+\l_2)}\over(2{\rm
sin}{g_s\over 2})^2}(e^{\ot i\l_1}-e^{-\ot i\l_1})(e^{\ot
i\l_2}-e^{-\ot i\l_2})(-18e^{i(\l_1
+\l_2)}+9e^{i\l_1}+9e^{i\l_2}-4)\cr & -{2e^{2i(\l_1 +\l_2)}\over
2(2{\rm sin}{g_s})^2}(e^{ i\l_1}-e^{- i\l_1})(e^{
i\l_2}-e^{-i\l_2})\cr & +e^{{3\over 2}i(\l_1+\l_2)} (e^{\ot
i\l_1}-e^{-\ot i\l_1})(e^{\ot i\l_2}-e^{-\ot i\l_2})(45e^{i(\l_1
+\l_2)}-14e^{i\l_1}-14e^{i\l_2}+3)\cr & +\left(2{\rm sin}{g_s\over
2}\right)^2e^{{3\over 2}i(\l_1+\l_2)} (e^{\ot i\l_1}-e^{-\ot
i\l_1})(e^{\ot i\l_2}-e^{-\ot i\l_2})(-20e^{i(\l_1
+\l_2)}+3e^{i\l_1}+3e^{i\l_2}-1)\cr & +3\left(2{\rm sin}{g_s\over
2}\right)^4e^{{3\over 2}i(\l_1+\l_2)} (e^{\ot i\l_1}-e^{-\ot
i\l_1})(e^{\ot i\l_2}-e^{-\ot i\l_2}).
\end{array}\end{equation}
Now we compute $X_{(3t_1,t_c)}$ and $X_{(3t_2,t_c)}$. Using
(\ref{eq:expvalA}), (\ref{eq:expvalB}), (\ref{eq:expvalBC}),
(\ref{eq:expvalD}), (\ref{eq:tracesexpval}),
(\ref{eq:zerotracesresthree}) and (\ref{eq:Hopflinkcan}) and the
fact that the linking number is $+1$ we obtain
\begin{equation}\label{eq:thexesfourg}\begin{array}{lc}
&X_{(3t_1,t_c)}={e^{{5\over 2}i\l_1}\over(2{\rm sin}{g_s\over 2})^2}
(e^{\ot i\l_1}-e^{-\ot i\l_1})
(e^{\ot i\l_2}-e^{-\ot i\l_2})(-13e^{2i\l_1}+9e^{i\l_1}-1)\cr
&\qquad~~~~~+2e^{{7\over 2}i\l_1}(e^{\ot i\l_1}-e^{-\ot
i\l_1})(e^{\ot i\l_2}-e^{-\ot
i\l_2})(8e^{i\l_1}-1)\cr
&\qquad~~~~~+\left(2{\rm sin}{g_s\over 2}\right)^2
e^{{7\over 2}i\l_1}(e^{\ot i\l_1}-e^{-\ot
i\l_1})(e^{\ot i\l_2}-e^{-\ot
i\l_2})(-7e^{i\l_1}+1)\cr
&\qquad~~~~~+\left(2{\rm sin}{g_s\over 2}\right)^4e^{{9\over 2}i\l_1}
(e^{\ot i\l_1}-e^{-\ot i\l_1})(e^{\ot i\l_2}-e^{-\ot
i\l_2}),
\end{array}\end{equation}
and, by the exchange symmetry,
\begin{equation}\label{eq:thexesfourh}\begin{array}{lc}
&X_{(3t_2,t_c)}={e^{{5\over 2}i\l_2}\over(2{\rm sin}{g_s\over 2})^2}
(e^{\ot i\l_1}-e^{-\ot i\l_1})
(e^{\ot i\l_2}-e^{-\ot i\l_2})(-13e^{2i\l_2}+9e^{i\l_2}-1)\cr
&\qquad~~~~~+2e^{{7\over 2}i\l_2}(e^{\ot i\l_1}-e^{-\ot
i\l_1})(e^{\ot i\l_2}-e^{-\ot
i\l_2})(8e^{i\l_2}-1)\cr
&\qquad~~~~~+\left(2{\rm sin}{g_s\over 2}\right)^2
e^{{7\over 2}i\l_2}(e^{\ot i\l_1}-e^{-\ot
i\l_1})(e^{\ot i\l_2}-e^{-\ot
i\l_2})(-7e^{i\l_2}+1)\cr
&\qquad~~~~~+\left(2{\rm sin}{g_s\over 2}\right)^4e^{{9\over 2}i\l_2}
(e^{\ot i\l_1}-e^{-\ot i\l_1})(e^{\ot i\l_2}-e^{-\ot
i\l_2}).
\end{array}\end{equation}\newpage
\noindent Another direct computation gives
\begin{equation}\label{eq:thexesfouri}\begin{array}{lc}
&X_{(2t_1,t_2,t_c)}={e^{ i(\l_1+\ot\l_2)}\over(2{\rm sin}{g_s\over
2})^2} (e^{\ot i\l_1}-e^{-\ot i\l_1}) (e^{\ot i\l_2}-e^{-\ot
i\l_2})\cr
&\times(-29e^{i(2\l_1+\l_2)}+18e^{2i\l_1}+19e^{i(\l_1+\l_2)}-8e^{i\l_1}-2e^{i\l_2})\cr
& +e^{i(2\l_1+\ot\l_2)}(e^{\ot i\l_1}-e^{-\ot i\l_1}) (e^{\ot
i\l_2}-e^{-\ot
i\l_2})(33e^{i(\l_1+\l_2)}-12e^{i\l_1}-12e^{i\l_2}+2)\cr &
+2\left(2{\rm sin}{g_s\over 2}\right)^2
e^{i(2\l_1+\ot\l_2)}(e^{\ot i\l_1}-e^{-\ot i\l_1}) (e^{\ot
i\l_2}-e^{-\ot i\l_2})(-7e^{i(\l_1+\l_2)}+e^{i\l_1}+e^{i\l_2})\cr
& +2\left(2{\rm sin}{g_s\over 2}\right)^4e^{i(3\l_1+{3\over
2}\l_2)}(e^{\ot i\l_1}-e^{-\ot i\l_1}) (e^{\ot i\l_2}-e^{-\ot
i\l_2}),
\end{array}\end{equation}

\noindent and, by the exchange symmetry,
\begin{equation}\label{eq:thexesfourii}\begin{array}{lc}
&X_{(t_1,2t_2,t_c)}={e^{ i(\ot\l_1+\l_2)}\over(2{\rm sin}{g_s\over
2})^2} (e^{\ot i\l_1}-e^{-\ot i\l_1}) (e^{\ot i\l_2}-e^{-\ot
i\l_2})\cr
&\times(-29e^{i(\l_1+2\l_2)}+18e^{2i\l_2}+19e^{i(\l_1+\l_2)}-8e^{i\l_2}-2e^{i\l_1})\cr
& +e^{i(\ot\l_1+2\l_2)}(e^{\ot i\l_1}-e^{-\ot i\l_1}) (e^{\ot
i\l_2}-e^{-\ot
i\l_2})(33e^{i(\l_1+\l_2)}-12e^{i\l_1}-12e^{i\l_2}+2)\cr &
+2\left(2{\rm sin}{g_s\over 2}\right)^2
e^{i(\ot\l_1+2\l_2)}(e^{\ot i\l_1}-e^{-\ot i\l_1}) (e^{\ot
i\l_2}-e^{-\ot i\l_2})(-7e^{i(\l_1+\l_2)}+e^{i\l_1}+e^{i\l_2})\cr
&+2\left(2{\rm sin}{g_s\over 2}\right)^4e^{i({3\over
2}\l_1+3\l_2)}(e^{\ot i\l_1}-e^{-\ot i\l_1}) (e^{\ot
i\l_2}-e^{-\ot i\l_2}).
\end{array}\end{equation}
Next, we compute $X_{(2t_1,2t_c)}$ and $X_{(2t_2,2t_c)}$. Using
(\ref{eq:expvalA}), (\ref{eq:expvalB}), (\ref{eq:expvalBC}),
(\ref{eq:expvalD}), (\ref{eq:thelinksii}), (\ref{eq:Hopflinkcan})
and (\ref{eq:thelinksiiv}) we obtain
\begin{equation}\label{eq:thexesfourj}\begin{array}{lcc}
&X_{(2t_1,2t_c)} \cr &\quad ={e^{\ot
i(5\l_1-\l_2)}\over(2{\rm sin}{g_s\over 2})^2} (e^{\ot
i\l_1}-e^{-\ot i\l_1})(e^{\ot i\l_2}-e^{-\ot
i\l_2})(-2e^{i(\l_1+\l_2)}+7e^{i\l_1}+e^{i\l_2}-3)\cr
&\quad +{1\over(2{\rm
sin}{g_s})^2}(e^{2i\l_1}-e^{4i\l_1})(e^{i\l_2}-e^{-i\l_2})\cr
&\quad +e^{\ot i(5\l_1-\l_2)}(e^{\ot i\l_1}-e^{-\ot
i\l_1})(e^{\ot i\l_2}-e^{-\ot
i\l_2})(e^{i(\l_1+\l_2)}-5e^{i\l_1}+1)\cr &\quad
+\left(2{\rm sin}{g_s\over 2}\right)^2e^{\ot i(7\l_1-\l_2)}
(e^{\ot i\l_1}-e^{-\ot i\l_1})(e^{\ot i\l_2}-e^{-\ot i\l_2}),
\end{array}\end{equation}
\newpage
\noindent and, by the exchange symmetry,
\begin{equation}\label{eq:thexesfourk}\begin{array}{lcc}
&X_{(2t_2,2t_c)} \cr &\quad={e^{\ot i(-\l_1+5\l_2)}\over(2{\rm
sin}{g_s\over 2})^2} (e^{\ot i\l_1}-e^{-\ot i\l_1})(e^{\ot
i\l_2}-e^{-\ot
i\l_2})(-2e^{i(\l_1+\l_2)}+7e^{i\l_2}+e^{i\l_1}-3)\cr &\quad
+{1\over(2{\rm
sin}{g_s})^2}(e^{2i\l_2}-e^{4i\l_2})(e^{i\l_1}-e^{-i\l_1})\cr
&\quad +e^{\ot i(-\l_1+5\l_2)}(e^{\ot i\l_1}-e^{-\ot
i\l_1})(e^{\ot i\l_2}-e^{-\ot
i\l_2})(e^{i(\l_1+\l_2)}-5e^{i\l_2}+1)\cr &\quad +\left(2{\rm
sin}{g_s\over 2}\right)^2e^{\ot i(-\l_1+7\l_2)} (e^{\ot
i\l_1}-e^{-\ot i\l_1})(e^{\ot i\l_2}-e^{-\ot i\l_2}).
\end{array}\end{equation}
After a computation similar to (\ref{eq:thexesfouri}) we get
\begin{equation}\label{eq:thexesfourl}\begin{array}{lc}
&X_{(t_1,t_2,2t_c)}={e^{\ot i(\l_1+\l_2)}\over(2{\rm sin}{g_s\over
2})^2}(e^{\ot i\l_1}-e^{-\ot i\l_1})(e^{\ot i\l_2}-e^{-\ot
i\l_2})\cr
&\qquad\qquad~\times(-2e^{i(\l_1+\l_2)}-2e^{i(\l_1-\l_2)}-2e^{i(-\l_1+\l_2)}+5e^{i\l_1}+5e^{i\l_2}-2)\cr
&\qquad\qquad~ -4e^{{3\over 2}i(\l_1+\l_2)}
(e^{\ot i\l_1}-e^{-\ot i\l_1})(e^{\ot i\l_2}-e^{-\ot i\l_2})(e^{\ot
i(\l_1-\l_2)}+e^{\ot i(-\l_1+\l_2)}).
\end{array}\end{equation}
Finally, a computation similar to (\ref{eq:thexesfourg}) yields
\begin{equation}\label{eq:thexesfourm}\begin{array}{lc}
&X_{(t_1,3t_c)}=-{e^{ i({5\over 2}\l_1-\l_2)}\over(2{\rm sin}{g_s\over 2})^2}
(e^{\ot i\l_1}-e^{-\ot i\l_1})(e^{\ot i\l_2}-e^{-\ot
i\l_2}),\cr
&X_{(t_2,3t_c)}=-{e^{ i(-\l_1+{5\over 2}\l_2)}\over(2{\rm sin}{g_s\over 2})^2}
(e^{\ot i\l_1}-e^{-\ot i\l_1})(e^{\ot i\l_2}-e^{-\ot
i\l_2}).
\end{array}\end{equation}

\bibliographystyle{alpha}

\end{document}